\RequirePackage[thmmarks]{ntheorem}
\makeatletter
\renewtheoremstyle{plain} 
  {\item[\hskip\labelsep \theorem@headerfont ##1\ \textup{##2}\theorem@separator]} 
  {\item[\hskip\labelsep \theorem@headerfont ##1\ \textup{##2}\ (##3)\theorem@separator]}
\makeatother

\let\latexdocument\document
\let\latexarabic\arabic

\documentclass[lineno]{biometrika}

\let\document\latexdocument
\let\arabic\latexarabic

\def\rm{}

\usepackage{amsfonts,amsmath,amssymb}
\usepackage{enumerate}
\usepackage{times}
\usepackage{bm}
\usepackage{natbib}
\usepackage[percent]{overpic}
\usepackage{float}
\usepackage{graphicx}
\usepackage[margin=0pt]{subfig}
\usepackage[plain,noend]{algorithm2e}
\usepackage{lipsum}
\usepackage{bm}
\usepackage{natbib}
\usepackage[plain,noend]{algorithm2e}
\usepackage[utf8]{inputenc}
\usepackage{amssymb}
\usepackage[justification=centering]{caption}
\usepackage{mathrsfs}

\usepackage{graphicx}
\usepackage{mathtools}
\usepackage{booktabs}
\usepackage{amstext} % for \text macro
\usepackage{epstopdf}
\usepackage{array} 
\usepackage{booktabs}
\usepackage{caption}
\usepackage{epstopdf}
\usepackage{xcolor}
\usepackage{comment}

\makeatletter
\renewcommand{\algocf@captiontext}[2]{#1\algocf@typo. \AlCapFnt{}#2} % text of caption
\def\@algocf@capt@plain{top}
\renewcommand{\algocf@makecaption}[2]{%
  \addtolength{\hsize}{\algomargin}%
  \sbox\@tempboxa{\algocf@captiontext{#1}{#2}}%
  \ifdim\wd\@tempboxa >\hsize%     % if caption is longer than a line
    \hskip .5\algomargin%
    \parbox[t]{\hsize}{\algocf@captiontext{#1}{#2}}% then caption is not centered
  \else%
    \global\@minipagefalse%
    \hbox to\hsize{\box\@tempboxa}% else caption is centered
  \fi%
  \addtolength{\hsize}{-\algomargin}%
}
\makeatother

\def\T{{ \mathrm{\scriptscriptstyle T} }}

\DeclareMathOperator{\tr}{tr}

\newcommand{\black}[1]{{\leavevmode\color{black}{#1}}}

 \newcommand\independent{\protect\mathpalette{\protect\independenT}{\perp}}
\def\independenT#1#2{\mathrel{\rlap{$#1#2$}\mkern2mu{#1#2}}}

\newcommand{\calS}{\mathcal S}
\newcommand{\calG}{\mathcal G}
\newcommand{\calM}{\mathcal M}
\newcommand{\calP}{\mathcal P}
\newcommand{\calL}{\mathcal L}
\newcommand{\calV}{\mathcal V}
\newcommand{\calT}{\mathcal T}

\newcommand{\calD}{\mathcal D}
\newcommand{\calF}{\mathcal F}
\newcommand{\calA}{\mathcal A}

\newcommand{\covm}{M(\calL,\calL)}

\newcommand{\bs}{ {\boldsymbol s} }

\newcommand{\eps}{\epsilon}
\newcommand{\sigs}{\sigma^2}
\newcommand{\taus}{\tau^2}

\newcommand{\given}{\,|\,}

\newcommand{\calU}{{\cal U}}

\newcommand{\tF}{\widetilde{F}}
\newcommand{\tf}{\widetilde{f}}

\usepackage{xr}
\makeatletter
\newcommand*{\addFileDependency}[1]{% argument=file name and extension
	\typeout{(#1)}
	\@addtofilelist{#1}
	\IfFileExists{#1}{}{\typeout{No file #1.}}
}
\makeatother

\begin{document}

\jname{Biometrika}
%% The year, volume, and number are determined on publication
\jyear{}
\jvol{}
\jnum{}
%% The \doi{...} and \accessdate commands are used by the production team
%\doi{10.1093/biomet/asm023}
\accessdate{}

%% These dates are usually set by the production team
\received{}
\revised{}

%% The left and right page headers are defined here:
\markboth{D. Dey et~al.}{Biometrika style}

\title{Graphical Gaussian Process Models for Highly Multivariate Spatial Data}
  \author{Debangan Dey}
  \affil{Department of Biostatistics, Johns Hopkins Bloomberg School of Public Health}
  
  \author{Abhirup Datta}
  \affil{Department of Biostatistics, Johns Hopkins Bloomberg School of Public Health  \email{abhidatta@jhu.edu}}
  
    \author{Sudipto Banerjee}
  \affil{Department of Biostatistics, University of California Los Angeles}

\maketitle

\begin{abstract}
For multivariate spatial Gaussian process (GP) models, customary specifications of cross-covariance functions do not exploit relational inter-variable graphs to ensure process-level conditional independence among the variables. This is undesirable, especially for highly multivariate settings, where popular cross-covariance functions such as the multivariate Mat\'ern suffer from a curse of dimensionality as the number of parameters and floating point operations scale up in quadratic and cubic order, respectively, in the number of variables. We propose a class of multivariate {\em Graphical Gaussian Processes} using a general construction called {\em stitching} that crafts cross-covariance functions from graphs and ensures process-level conditional independence among variables. For the Mat\'ern family of functions, stitching yields a multivariate GP whose univariate components are Mat\'ern GPs, and which conforms to process-level conditional independence as specified by the graphical model. For highly multivariate settings and decomposable graphical models, stitching offers massive computational gains and parameter dimension reduction. We demonstrate the utility of the graphical Mat\'ern GP to jointly model highly multivariate spatial data using simulation examples and an application to air-pollution modelling.
\end{abstract}


\section{Introduction}
\label{sec:intro}
Multivariate spatial data abound in the natural and environmental sciences for studying features of the joint distribution of multiple spatially dependent variables \citep[see, for example,][]{wackernagel2013multivariate,creswikle11,ban14}. The objectives are to estimate associations over spatial locations for each variable and those among the variables. Let $y(s)$ be a $q\times 1$ vector of spatially-indexed dependent outcomes within any location $s \in \calD \subset \mathbb{R}^d$ with $d=2$ or $3$. A multivariate spatial regression model on our spatial domain $\calD$ specifies a univariate spatial regression model for each outcome as
\begin{equation}\label{eqn:mgp}
y_i(s) = x_i(s)^{\T}\beta_i + w_i(s) + \epsilon_i(s)\;,\quad i=1,2,\ldots,q,\; s \in \calD
\end{equation}
where $y_i(s)$ is the $i$-th element of $y(s)$, $x_i(s)$ is a $p_i\times 1$ vector of predictors, $\beta_i$ is the $p_i\times 1$ vector of slopes, each $w_i(s)$ is a spatial process and $\epsilon_i(s) \stackrel{ind}{\sim} N(0,\tau^2_i)$ is the random noise in outcome $i$. We customarily assume that $w(s)=(w_1(s),w_2(s),\ldots,w_q(s))^{\T}$ is a multivariate Gaussian process (GP)
specified by a zero mean and a cross-covariance function that introduces dependence over space and among the $q$ variables. %We assume the mean to be zero and focus on 
The cross-covariance is a matrix-valued function $C=(C_{ij}): {\cal D}\times {\cal D} \mapsto \mathbb{R}^{q \times q}$ with %for any pair of locations $(s,s')$ to the $q\times q$ matrix 
%$C(s,s')=
$C_{ij}(s,s')%$ %with $(i,j)$-th element $C_{ij}(s,s')
= \mbox{Cov}(w_i(s),w_j(s'))$ for any pair of locations $(s,s')$. Cross-covariance functions must ensure that for any finite set of locations $\calS = \{s_1,\ldots,s_n\}$,  %Therefore, a cross-covariance function must satisfy the following two conditions: (i) $C(s,s') = C(s',s)^{\T}$ for any two $s$ and $s'$; and (ii) $\sum_{i,j=1}^n a_i^{\T}C(s_i,s_j)a_j$ for any set of $n$ distinct locations $\{s_1, s_2, \ldots, s_n\}$ and any set of $n$ non-zero vectors $a_i\in \Re^q\setminus\{0\}$. These conditions are equivalent to the covariance matrix for $w = (w(s_1),w(s_2),\ldots,w(s_n))^{\T}$, i.e., 
the $nq\times nq$ matrix $C(\calS,\calS) = (C(s_i,s_j))$ is positive definite (p.d.). %\black{With a valid process specification, we cast (\ref{eqn:mgp}) into a hierarchical model and compute the posterior distribution derived from \begin{equation}\label{eqn:mgp_hierarchical}
%p(\beta, \tau, \theta) \times N(w(\calS)\given 0, C_{\theta}(\calS,\calS))\times \prod_{j=1}^n N(y(s_j)\given X(s_j)\beta + w(s_j), D_{\tau})\;,
%\end{equation}
%where $X(s_j) = \mbox{diag}(x_1(s_j)^{\T}, x_2(s_j)^{\T},\ldots, x_q(s_j)^{\T})$ is $q\times (\sum_{i=1}^q p_i)$, $\beta = (\beta_1^{\T}, \beta_{2}^{\T}\ldots,\beta_q^{\T})^{\T}$, $w(\calS) = (w(s_1)^{\T}, w(s_{2})^{\T}\ldots,w(s_n)^{\T})^{\T}$, $D_{\tau} = \mbox{diag}(\tau^2_1, \tau^2_2,\ldots,\tau^2_q)$, $\theta$ is a collection of parameters defining the cross-covariance function and $p(\beta, \tau, \theta)$ is the prior on model parameters.} 

Valid classes of cross-covariance functions have been comprehensively reviewed in \cite{genton2015cross}. %, where various characterizations and constructions are discussed with the advantages and disadvantages of each. %The linear model of coreginalization (LMC) \citep{goulard1992linear,schmidt2003bayesian} constructs a multivariate random field using linear combinations of $r$ univariate random fields, where choosing $r \ll q$ produces low-rank or spatial factor models. Computational benefits ensue; one needs a small number, $r$, of univariate processes to ensure non-negative definiteness and, hence, avoids working with $q \gg r$ processes. Using a similar framework, computationally efficient approaches has been proposed for large $n$ \citep{ren2013hierarchical,taylor2019spatial}. A major drawback of LMC is that the approach lacks interpretability due to the usage of latent processes. 
%Another The LMC builds linear combination of latent spatial processes, thereby endowing each $w_j(s)$ with the same smoothness (the smoothness of the roughest latent process). This is implausible in most applications because different spatial variables typically exhibit very different degrees of smoothness. An alternate approach constructs cross-covariances by convolving univariate processes with kernel functions \citep{ver1998constructing,ver2004flexible,majumdar2007multivariate}. However, barring certain very special cases, the resulting cross-covariance functions are analytically intractable, hence less interpretable, and may require cumbersome numerical integration for estimating process parameters. Some of the aforementioned difficulties are obviated by a conditional approach developed in \cite{cressie2016multivariate}, where the univariate GPs are specified sequentially, conditional on the previous GPs assuming some ordering of the $q$ variables. %that the variables modeled as a multivariate GP conform to some ordering, which is typically unrealistic especially in highly multivariate applications involving many variables.  Other notable approaches include \cite{apanasovich2010cross} who considered using latent dimensions to embed all the variables in a larger dimensional space and use standard covariance functions on this augmented space. This embedding restricts all pairwise-correlations among the different variables to be positive, unless combined with LMC. 
Of particular interest are multivariate Mat\'ern cross-covariance functions \citep{gneiting2010matern, apanasovich2012valid}, where the marginal covariance functions for each $w_i(s)$ and the cross-covariance functions between $w_i(s)$ and $w_j(s')$ are Mat\'ern functions. In its most general form, the multivariate Mat\'ern is appealing as it ensures that each univariate process is a Mat\'ern GP with its own range, smoothness and spatial variance although the parameters need to be constrained to ensure positive-definiteness of the cross-covariance function. % $C(\calS,\calS)$ is p.d.

\black{Our current focus is the increasingly commonplace %to inference from (\ref{eqn:mgp_hierarchical}) in the 
\emph{highly-multivariate} setting with a large number of dependent outcomes (e.g., $q \sim 10^2+$) at each spatial location. %Such settings are increasingly commonplace in the environmental and physical sciences. 
While substantial attention has been accorded to spatial data with massive number of locations (large $n$) \citep[see, e.g.,][for a review]{heaton2019case}, the highly multivariate setting fosters separate computational issues. Likelihoods for popular cross-covariance functions, such as the multivariate Mat\'ern, involve $O(q^2)$ parameters, and $O(q^3)$ floating point operations (flops). % whose dimensionality is quadratic in $q$, and involve the computations involving the resulting likelihood are cubic in $q$. %For example, a dataset on $20$ variables would involve at least $200$ parameters.
Optimizing over or sampling from high-dimensional parameter spaces is inefficient even for modest values of $n$. Illustrations of multivariate Mat\'ern models %of the aforementioned approaches 
have typically been restricted to applications with $q \leq 5$.}

  %The parsimonious Mat\'ern \citep{gneiting2010matern} imposes equality of the spatial range for all variables. \cite{apanasovich2012valid} laid out more general sufficient conditions for the parameters, yielding a very broad class of multivariate Mat\'ern covariances. 

In non-spatial settings, Gaussian graphical models are extensively used as a dimension-reduction tool %to represent the joint distribution of several variables for many types of non-spatial data, most applications in spatial settings have been concerned with reducing the computational burden for large $n$ by replacing the complete graph between locations with nearest-neighbor graphs \citep{nngp}. 
to parsimoniously model conditional dependencies in highly multivariate data. %For spatial data, graphical models  % have also recently been used have primarily been used to achieve scalability of GP likelihoods with respect to the number of locations ($n$) using nearest neighbor graphs \citep{nngp, katzfuss2021, pbf2020}. \black{However, popular spatial cross-covariance functions like the multivariate Mat\'ern are not endowed with % $C$ (or $C(\calS,\calS)$) with 
any exploitable graphical structure for scalable computation, nor do they adhere to posited conditional independence relations among the outcomes as are often introduced in high-dimensional outcomes \citep{coxwermuth1996}. 
Our innovation here is to develop multivariate GPs that conform to {\em process-level conditional independence} posited by an inter-variable graph over $q$ dependent outcomes while attending to scalability considerations for large $q$.

%However, all current multivariate Mat\'ern models  
%The positive-definiteness condition puts constraints on parameters for a multivariate Mat\'ern GP  where both the univariate covariance functions for each variable, and the cross-covariance functions between each pair of variables are members of the Mat\'ern family. \cite{gneiting2010matern} first formulated Multivariate Mat\'ern model with a simpler assumptions (parsimonius model) and \cite{apanasovich2012valid} laid out more general sufficient coditions for the parameters. But, the full specification of this model r
%require estimating $O(q^2)$ cross-covariance parameters and the multivariate Gaussian likelihood involves the inverse and determinant of the dense $nq \times nq$ covariance matrix, which is prohibitive if $n$ or $q$ is large. %These drawbacks %of the multivariate Mat\'ern model prohibits its usage  become prohibitive in the analysis of highly multivariate spatial data. 
 %Exceptions include the spatial factor models that have modelled with $q\sim 10$, but suffer from the drawbacks of the LMC and have often required inflexible restrictions such as ordering of range parameters to identify process parameters from the data \citep{ren2013hierarchical}. 

%We address the %high-dimensional multivariate, henceforth referred to as \emph{highly-multivariate} setting (large $q$) with tens to hundreds of variables measured at each spatial location, which is %. Such settings are becoming increasingly commonplace in the environmental and physical sciences. % where inference is sought on large numbers of dependent outcomes. 

To adapt graphical models to multivariate spatial process-based settings, we generalize notions of process-level conditional independence for discrete time-series \citep{dahlhaus2000graphical, dahlhaus2003causality}  to continuous spatial domains. We define multivariate {\em graphical Gaussian Processes (GGPs)} that satisfy process-level conditional independence as specified by an inter-variable graph. We focus on GGPs with properties deemed critical for handling multivariate spatial data. Specifically, we seek to retain the flexibility to model and interpret spatial properties of the random field for each variable separately. Except for the multivariate Mat\'ern, most other multivariate covariance functions fail to retain this property. 

We address and resolve challenges in constructing spatial processes that retain marginal properties and are also GGP. For example, while the existing multivariate Mat\'ern models preserve the univariate marginals as Mat\'ern GPs, we show (Section~\ref{sec:mat}) that no parametrisation of the multivariate Mat\'ern yields a GGP. On the other hand, the literature on graphical multivariate discrete time-series models, hitherto, have not attempted to preserve marginal properties and have benefited from the regular discrete setting of equispaced time-points, in both non-parametric \citep{dahlhaus2000graphical, dahlhaus2003causality,eichler2008testing} and parametric \citep{eichler2012fitting} analysis. We  resolve both of these challenges for irregular spatial data.
%In spatial contexts, it is important to ensure that the univariate components are endowed with distributions that facilitate the interpretation of the spatial properties of each univariate surface (e.g., Mat\'ern GP).%  as most.    developed graphical multivariate time-series but did not attempt to preserve univariate properties, focusing primarily on non-parametric estimation and testing of the process-level conditional independence.

Our development relies upon the seminal work of \cite{dempster1972covariance} on {\em covariance selection}, which ensures the existence of  multivariate distributions that retain univariate marginals while satisfying conditional-independence relations specified by an inter-variable graph. While covariance selection can facilitate approximate likelihood-based inference for graphical VAR models \citep{eichler2012fitting} by exploiting the expansion of the inverse spectral density matrix of VAR(p) models in terms of the inverse covariance matrices over finite (p) time-lags, such finite-lag representations do not typically hold for spatial covariance functions over ${\calD} \subset \mathbb{R}^d$. 

One of our key contributions here is to identify the construction of a marginal-retaining GGP as a {\em process-level covariance selection} problem. We use covariance selection on the spectral density matrix to prove existence, uniqueness and information-theoretic optimality of a marginal retaining GGP. % that also exactly retains cross-covariances between all variable pairs included in the graphical model. %As this GGP does not yield a tractable covariance functions or likelihood, 
We subsequently introduce a novel practicable method to approximate this optimal GGP by \emph{stitching} GPs together using an inter-variable graph. Stitching relies on the orthogonal decomposition of a GP into a fixed-rank predictive process \citep{banerjee2008gaussian} on a finite set of locations and a residual process. We show how to endow the predictive process with the desired conditional-independence structure via covariance selection, and %. Stitching first exploits covariance selection to endow the multivariate predictive process with . Subsequently, 
use componentwise-independent residual processes to create a well defined multivariate GP that exactly preserves (i) dependencies modelled by the graph; and (ii) the marginal distributions on the entire domain. Stitching with Mat\'ern GPs yields a {\em multivariate graphical Mat\'ern GP} with a tractable likelihood for irregular spatial data such that (i) each outcome process is endowed with the original Mat\'ern GP; (ii) we retain process-level conditional independence modelled by the graph; (iii) cross-covariances for variable pairs included in the graph are exactly or approximately Mat\'ern.

We also demonstrate computational scalability with respect to $q$. We show that for decomposable graphical models, stitching facilitates drastic dimension-reduction of the parameter space and fast likelihood evaluations by obviating large matrix operations. Additionally, stitching harmonizes graphical models with parallel computing to employ a chromatic Gibbs sampler for delivering efficient fully model-based Bayesian inference. We also show how our framework can adapt to (i) deliver inference for an unknown inter-variable graph; (ii) model spatial time-series; and (iii) model multivariate spatial factor models.%; and (iv) accommodate non-stationary processes.} %Ensuring positive definiteness of the covariance matrix requires complex constraints on high-dimensional spaces, which exacerbates the problem of scalability. %A sparse graphical model between the variables can potentially drastically reduce the dimensionality and complexity of the parameter space. 

\section{Method}\label{sec:meth}
\subsection{Process-level conditional independence and Graphical Gaussian Processes}\label{sec:exist}
%We will derive a \emph{Graphical Gaussian process}, henceforth GGP, from a given cross-covariance function and an inter-variable graph $\calG_\calV = \{\calV,E_\calV\}$, where $\calV$ is the set of variable-indices and $E_\calV$ is the set of edges. A key feature of our proposed GGP is that it will retain the same univariate spatial covariance functions for each variable and the same cross-covariance functions for pairs of variables $(i,j) \in E$. Furthermore, the GGP {\em conforms} to the graph $\calG_V$ in the following sense.  This relates to the notion of conditional independence of two processes given all the other processes.
We define {\em process-level conditional independence} for a multivariate GP $w(\cdot)=(w_1(\cdot),\ldots,w_q(\cdot))^\T$ over $\calD$. %as required in (ii). 
We adapt the analogous definition for multivariate discrete time-series in \cite{dahlhaus2000graphical} to a continuous-space paradigm. Let $\calV=\{1,\ldots,q\}$, $B\subset \calV$ and $w_{B}(\calD) = \{w_k(s) : k\in B,\; s\in \calD\}$. %, where each $w_k(\cdot)$ is a spatial process . 
Two processes $w_{i}(\cdot)$ and $w_{j}(\cdot)$ are conditionally independent given the processes $\{w_k(\cdot) \given k \in V \setminus \{i,j\}\}$ if $\mbox{Cov}(z_{iB}(s), z_{jB}(s'))=0$ for all $s,s'\in \calD$ and $B = \calV \setminus \{i,j\}$, where $z_{kB}(s) = w_k(s) - \mbox{E}[w_k(s)\given \sigma(\{w_j(s'): j\in B,\; s'\in \calD\})]$, where $\sigma(\cdot)$ is the usual $\sigma$-algebra generated by its argument. % denotes the ``residual'' process. 
%Conditional independence boils down to zeros in the spectral density matrix for stationary processes (see, e.g., Theorem~2.4 in \cite{dahlhaus2000graphical}). If $F(\omega)=(f_{ij}(\omega))$ is the $q \times q$ spectral density matrix corresponding to the cross-covariance functions $C_{ij}$ of a $q\times 1$ stationary Gaussian process on some domain $\calD$, then two processes $w_i(\cdot)$ and $w_j(\cdot)$ are conditionally independent if and only if $f^{ij}(\omega) = 0$ for almost all $\omega$, where %$f^{ij}(\omega)$ is the $(i,j)$th element of $F(\omega)^{-1}=(f^{ij}(\omega))$. The result is analogous to graphical multivariate Gaussian models, where the absence of an edge between two variables implies a zero in the corresponding entry in the precision matrix. %\cite{dahlhaus2000graphical} described the concept using each time series component as a vertex in a graph and an edge determined by the partial correlation defined for the time series.
Let $\calG_\calV = (\calV,E_\calV)$ be a graph, where $E_\calV$ is a pre-specified set of edges among pairs of variables. %Having defined process-level conditional independence, 
We now define a \emph{Graphical Gaussian Process} (GGP) with respect to (or conforming to) $\calG_\calV$ as follows.
\begin{definition}\label{def: gp-cond-ind}
[Graphical Gaussian Process] A $q\times 1$ GP $w(\cdot)$ is a Graphical Gaussian Process (GGP) with respect to a graph $\calG_\calV = (\calV,E_\calV)$ when the univariate GPs $w_i(\cdot)$ and $w_j(\cdot)$ are conditionally independent for every $(i,j) \notin E_\calV$. We denote such a process as GGP$(\calG_\calV)$.
\end{definition}
Any collection of $q$ independent GPs will trivially constitute a GGP with respect to any graph $\calG_\calV$. More pertinent is the ability of a GGP to approximate a full (non-graphical) GP. This is particularly relevant for inference because the full GP is computationally impracticable for large $q$. Theorem~\ref{th:exists} shows that given a graph $\calG_\calV$ and a multivariate GP with cross-covariance function $C$, there exists a \emph{unique} and information-theoretically optimal GGP %with respect to the original GP with covariance $C$
among the class of all GGP$(\calG_\calV)$. % with respect to $\calG_\calV$. 
Proofs of all subsequent results are provided in the supplement.

\begin{theorem}\label{th:exists}
Let $\calG_\calV=(\calV,E_\calV)$ be any given graph, $C = (C_{ij})$ be a $q \times q$ stationary cross-covariance function. Let $F(\omega) = (f_{ij}(\omega))$ be the spectral density matrix corresponding to $C$ at frequency $\omega$. Let $f_{ii}(\cdot)$ be square-integrable for all $i$. Then
\begin{enumerate}[(a)]
    \item There exists a unique $q\times 1$ GGP$(\calG_\calV)$ $w(\cdot)$ with cross-covariance function $M = (M_{ij})$ such that $M_{ij} = C_{ij}$ for $i=j$ and for all $(i,j) \in E_\calV$;
    \item If $\tilde F(\omega)$ denotes the spectral density matrix of $w(\cdot)$ and $\calF$ is the set of spectral density matrices of all possible GGP$(\calG_\calV)$, then % with respect to $\calG_\calV$, then
    \[
     \tilde F(\cdot) = {\arg \min}_{K(\cdot) \in \calF} \int_\omega d_{KL}(F(\omega)\|K(\omega))d\omega\; ,
    \]
    where $d_{KL}(F\|K)=\tr(K^{-1}F) + \log\det(K)$ denotes the Kullback-Leibler divergence between two positive definite matrices $F$ and $K$.
\end{enumerate}
\end{theorem}
Theorem \ref{th:exists} shows that the optimal GGP approximating a GP, given a graph, needs to exactly preserve the marginal distributions of the univariate processes, which is also critical to retain interpretation of the spatial properties of each univariate surface. This optimal GGP also preserves cross-covariances for variable pairs included in $\calG_\calV$. Theorem~\ref{th:exists}, however, is of limited practical value %. For data observed over a (possibly irregular) set of locations, the likelihood is specified by the cross-covariance function. Theorem~\ref{th:exists} 
because it does not present a convenient way to construct cross-covariances. %In the next Section we pursue a more practicable approach we refer to as {\em stitching} to construct the marginal-preserving GGP for use in likelihood-based analysis of irregularly placed spatial data. 
We develop a practicable method of {\em stitching} $q$ univariate random fields (Section~\ref{sec:stitch}) to construct marginal-preserving GGPs for modelling irregular spatial data.

%exactly satisfying conditions (i) - (iii) and is the  (in terms of the integrated reverse Kullback-Leibler divergence on the SDMs) approximation of the 
%whose components will satisfy process-level conditional independence structure as specified by the edges of the graph $\calG_V$. 

%We will first define and derive a \emph{Graphical Gaussian Process} (GGP) from a given cross-covariance function such that the same marginal spatial covariance functions for each variable and the same cross-covariance functions for pairs of variables $(i,j) \in E_\calV$ are retained.  
 %To prove this, we use the following Lemma. 
%\begin{lemma}[Covariance selection \citep{dempster1972covariance}]\label{thm:cov-sel}
%Given $\calG_\calV=(\calV,E_\calV)$ and any positive definite matrix $F=(F_{ij})$ indexed by $\calV \times \calV$, there exists a unique positive definite matrix $\tF=(\tF_{ij})$ such that $\tF_{ij}=F_{ij}$ for $i=j$ or for $(i,j) \in E_{\calV}$, and $(\tF^{-1})_{ij}=0$ for $(i,j) \notin E_\calV$.
%\end{lemma} 
%\begin{proof}
%This is the main result developed in \cite{dempster1972covariance}.
%\end{proof} The above is a seminal result on covariance selection that we use recurrently in this paper. 

%\begin{proof}
%See Supplementary Materials.
%\end{proof}

\subsection{Stitching of Gaussian Processes}\label{sec:stitch}
Given any $\calG_\calV$ and a cross-covariance function $C$, we seek a multivariate GP $w(\cdot)$ %aussian process %henceforth referred to as a {\em graphical Gaussian Process (GGP)} $w(\cdot)=(w_1(\cdot),\ldots,w_q(\cdot))$, 
that
\vspace{-0.1in}
\begin{enumerate}[(i)]
%\noindent (i) 
\item exactly preserves the marginal distributions specified by $C$, i.e., $w_i(\cdot) \sim GP(0,C_{ii}) \,\forall i$;

%\noindent (ii) 
\item is a GGP($\calG_\calV$), i.e., satisfies process-level conditional independence according to $\calG_\calV$; %, i.e., for every $(i,j)$ such that $(i,j) \notin E_\calV$, the processes $w_i(s)$ and $w_j(s)$ are conditionally independent given the remaining processes, and
and

%\noindent (iii) 
\item exactly or approximately retains the  cross-covariances specified by $C$ for pairs of variables included in $\calG_\calV$, i.e., for $(i,j) \in E_\calV$, $\mbox{Cov}(w_i(s),w_j(s')) \approx C_{ij}(s,s')$.
\end{enumerate}

\vskip-4mm \begin{figure}[thb]
    \begin{center}
    \hspace*{-1mm}\includegraphics[scale=0.36,trim={0 0 0 15},clip]{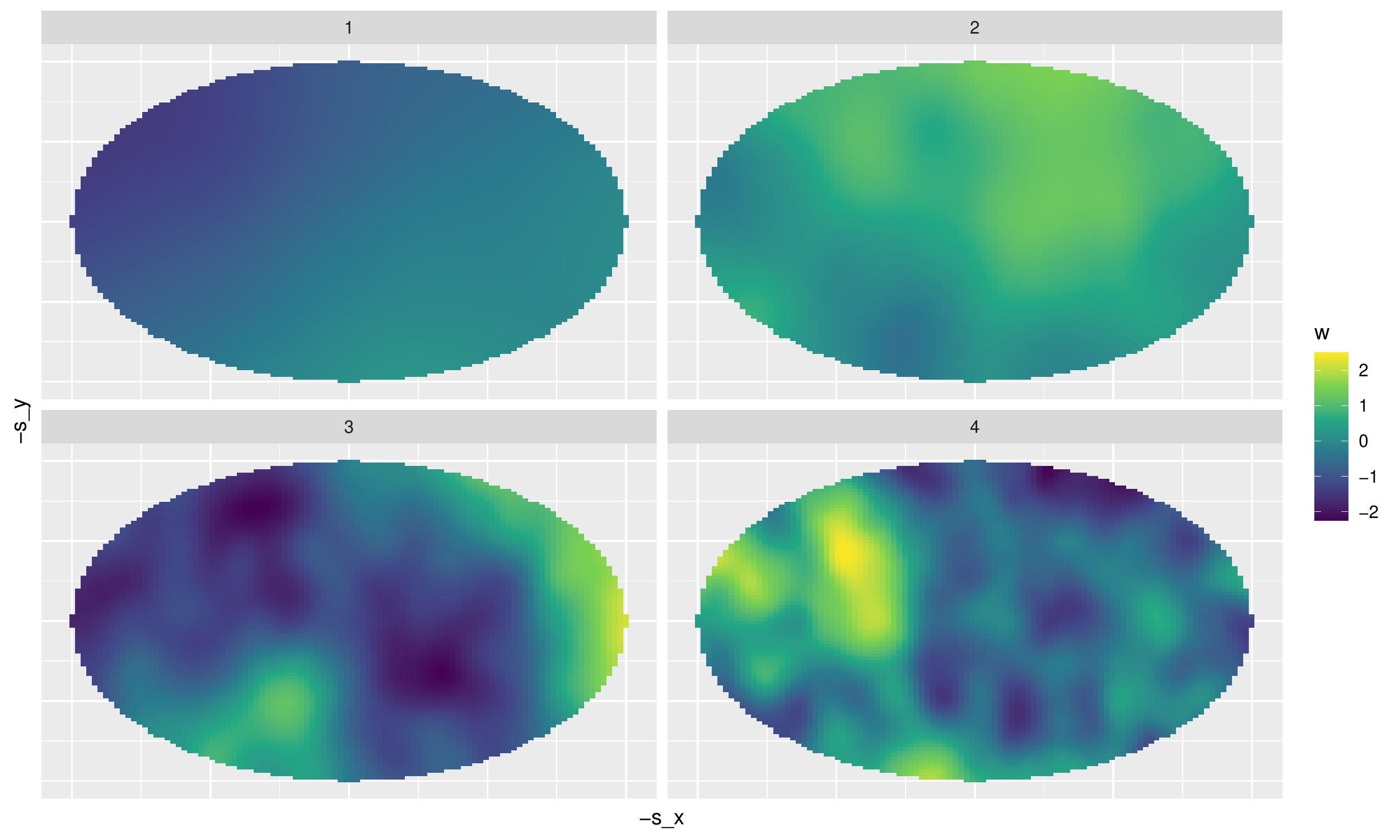}
   \hspace*{-1mm}\includegraphics[scale=0.26,trim={0 0 0 15},clip]{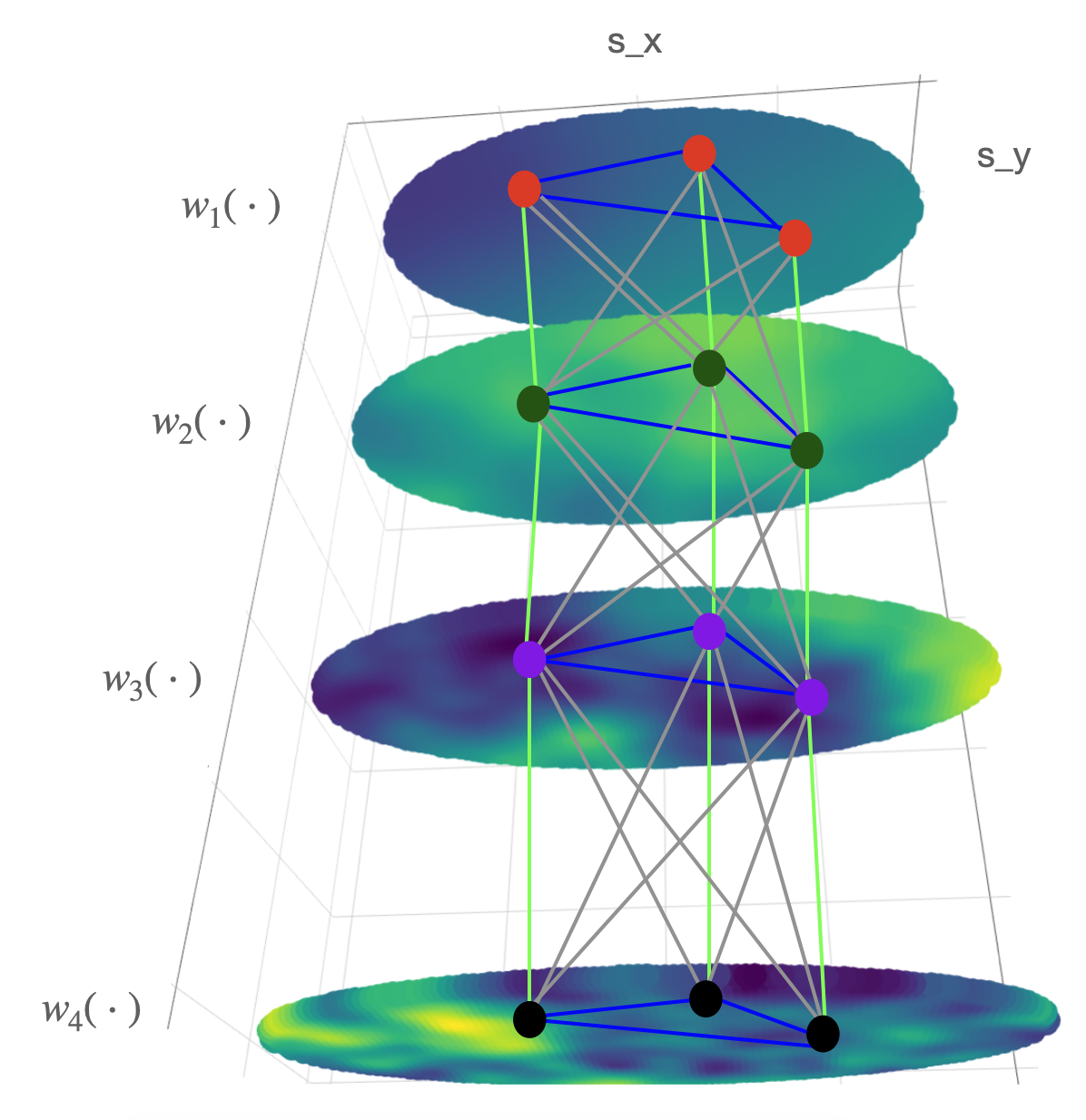}
   \end{center}
        \setlength{\belowcaptionskip}{-10pt}
\setlength{\abovecaptionskip}{-5pt}
\caption{Stitching Gaussian Processes. Left: Realizations of 4 univariate GPs. Right: Realization of a multivariate (4-dimensional) GGP created by stitching together the 4 univariate GPs from the left figure using the strong product graph over the 4 variables and 3 locations.}
    \label{fig:stitch}
\end{figure}

%%We ``stitch'' univariate GPs to build a GGP \black{$w(\cdot)$, satisfying (i)-(iii) above. 
We visually illustrate stitching of univariate GPs to build a GGP $w(\cdot)$, satisfying (i)-(iii) above. Figure~\ref{fig:stitch} (left) shows realizations of 4 univariate Mat\'ern GPs $w_i(\cdot)$, $i=1,\ldots,4$, each with a different smoothness and spatial range. 
Figure~\ref{fig:stitch} (right) shows a multivariate GGP constructed by stitching together the 4 processes using a path-graph as $\calG_\calV$ with $E_\calV=\{(i,i+1): i=1,2,3\}$. We begin our construction on $\calL$, a finite but otherwise arbitrary set of locations in $\calD$ (the 3 locations in Figure~\ref{fig:stitch} (right)). We first ensure that $w(\calL)=(w_1(\calL),\ldots,w_q(\calL))^\T$%, the realization of the process on $\calL$, 
 satisfies conditions (i)-(iii) when the domain is restricted to $\calL$. This is achieved by stitching together the variables at the 3 locations in $\calL$ such that there is a {\em thread} (edge) between two variable-location pairs if and only if there is an edge between the two corresponding variables in $\calV$. We then stitch each of the remaining surfaces independently so that they have the same distribution as the univariate surfaces from the left panel and conforms to the graph at the process-level. This resembles stitching the four surfaces together at the locations $\calL$, while exactly preserving each univariate surface. The graph edges serve as the threads holding the surfaces together.

Turning to the formal development, we first create $w(\calL)$---the realisation of our target process $w(\cdot)$ on $\calL$ that satisfies properties (i)-(iii) on $\calL$. %To satisfy (i), each $w_i(\calL)$ needs to be a realization from a univariate GP with covariance function $C_{ii}(s,s')$, i.e., $\mbox{Cov}(w_{i}(\calL)) = C_{ii}(\calL,\calL)$. For (iii), for any edge $(i,j) \in E_\calV$, the cross-covariances needs to be preserved, i.e., $\mbox{Cov}(w_i(\calL),w_j(\calL)) = C_{ij}(\calL,\calL)$. Finally, for (iii) $w(\calL)$ has to satisfy the conditional independence relations specified by $\calG_\calV$ on $\calL$.
Combining the three requirements, we model
%Given a valid $q\times q$ cross-covariance function $C=(C_{ij})$ and a graph $\calG_\calV$, 
 $w(\calL) \sim N(0, \covm)$, seeking a p.d. matrix $\covm$ such that 
\vspace{-0.1in} 
\begin{enumerate}[(a)]
    \item $M_{ii}(\calL,\calL) = C_{ii}(\calL,\calL)$ for all $i=1,\ldots,q$, to satisfy (i);
    \item $(M(\calL,\calL)^{-1})_{ij} = 0$ for all $(i,j) \notin E_\calV$ to satisfy (ii). 
    \item $M_{ij}(\calL,\calL) = C_{ij}(\calL,\calL)$ for all $(i,j) \in E_\calV$, to satisfy (iii).
\end{enumerate}%(a) $M_{ii}(\calL,\calL) = C_{ii}(\calL,\calL)$ for all $i=1,\ldots,q$, to satisfy (i), and (b) $(M(\calL,\calL)^{-1})_{ij} = 0$ for all $(i,j) \notin E_\calV$ to satisfy (ii),  \footnote{Condition (b) only ensures conditional independence of the process restricted to $\calL$. Process-level conditional independence over the entire domain $\calD$ follows from the subsequent extension in (\ref{eqn:extend}) as proved in Theorem \ref{thm: multgp}.} and  (c) $M_{ij}(\calL,\calL) = C_{ij}(\calL,\calL)$ for all $(i,j) \in E_\calV$, to satisfy (iii).
\vspace{-0.1in}
Existence of such a matrix $\covm$ is a covariance selection problem \citep{dempster1972covariance}. %We first state the main result on covariance selection from \cite{dempster1972covariance} in the following Lemma.
\begin{lemma}[Covariance selection \citep{dempster1972covariance}]\label{thm:cov-sel}
Given a graph $\calG=(\calS,E)$ and any p.d. matrix $F=(F_{rs})$ indexed by $\calS \times \calS$, there exists a unique p.d. matrix $\tF=(\tF_{rs})$ such that $\tF_{rs}=F_{rs}$ for $r=s$ or for $(r,s) \in E_{\calS}$, and $(\tF^{-1})_{rs}=0$ for $(r,s) \notin E_\calS$.
\end{lemma} 
%\begin{proof}
%This is the main result developed in \cite{dempster1972covariance}.
%\end{proof}

To ensure that the covariances and cross-covariances are preserved over $\calL$ for all $i$ and all $(i,j) \in E_\calV$ and the conditional independence among elements of $w(\calL)$ are inherited from $\calG_\calV$, $w(\calL)$ needs to conform to a graph with edges between variable-location pairs as in Figure~\ref{fig:stitch}. Formally, let $\mathcal{G_L}=(\calL,E_\calL)$ be the complete graph on the set of locations $\calL$. The variable-location graph from Figure~\ref{fig:stitch} (right) is the \emph{strong product graph} $\mathcal{G_V} \boxtimes \mathcal{G_L}$. %. the distribution of $w(\calL)$,where $\boxtimes$ denotes the {\em strong product} of  graphs, i.e., 
Here, $\calG_V\boxtimes\calG_L = (\calV \times \calL, E_{\calV \times \calL})$ with $\calV \times \calL = \{(i,l) : i\in \calV,\; l\in \calL\}$ % is the Cartesian product of the two vertex sets
and $E_{\calV\times \calL}$ comprises edges between vertex-pairs $(i,l)$ and $(i',l')$ based upon the following strong-product adjacency rules: (i) $i=i'$ and $(l,l') \in E_\calL$; or (ii) $l=l'$ and $(i,i') \in E_\calV$; or (iii) $(i,i') \in E_\calV$ and $(l,l') \in E_\calL$. %is adjacent to $v'$ in $\calG_V$  and $l$ is adjacent to $l'$ in $\calG_L$.  

Applying Lemma~\ref{thm:cov-sel} with the vertex set $\calS=\calV \times \calL$, positive definite matrix $F=C(\calL,\calL)$ and the graph $\calG_V\boxtimes\calG_L$, ensures the existence and uniqueness of a positive definite matrix $\tF=\covm$ satisfying conditions (a), (b) and (c) above. In practice, $\covm$ can be obtained using an iterative proportional scaling (IPS) algorithm \citep{speed1986gaussian,xu2011improved}. 

Note that Condition (b) only ensures conditional independence of the process restricted to $\calL$. Process-level conditional independence over the entire domain $\calD$ follows from the subsequent extension in (\ref{eqn:extend}) as proved in Theorem \ref{thm: multgp}. Having built the finite-dimensional distribution of $w(\calL)$ from $\calG_\calV \boxtimes \calG_\calL$,  we now suitably extend it to a well-defined multivariate GP $w(\cdot)$ over the domain ${\mathcal D}$, %$w(\calD \setminus \calL)$ 
%so that the resulting process $\{w(s) \given s \in  \calD\}$
which conforms to the conditional dependencies implied by $\calG_\calV$. 
\black{We leverage the following well-known decomposition of a GP $w_i(\cdot)$ as sum of a finite rank {\em predictive process} $w_i^*(\cdot)=E(w_i(\cdot) \given w_i(\calL))$ and an independent {\em residual process} $z_i(\cdot)$ \citep{banerjee2008gaussian, finley2009improving}:
\begin{equation} \label{eqn:extend}
w_i(s) = w_i^*(s) + z_i(s) = C_{ii}(s,\calL)C_{ii}(\calL,\calL)^{-1}w_i(\calL) + z_i(s)\quad \mbox{ for all}\quad s\in \calD\setminus\calL \;,
%\begin{array}{cc}
%   w_i(s) = C_{ii}(s,\calR)C_{ii}(\calL,\calL)^{-1}w_i(\calL) + z_i(s)\;; \\
%   z_i(s) \ind GP(0, C_{ii \given \calL}),  \mbox{ for } %s \in \calD \setminus \calL, 
%   i=1,\ldots,q, \\
%\mbox{ where } C_{ii \given \calL}(s_i, s_j) = C_{ii}(s_i, s_j ) - C_{ii}(s_i, \calL )C^{-1}_{ii}(\calL, R)C_{ii}(\calL,s_j)\;.
%\end{array}
\end{equation}
where each $z_i(\cdot)$ is a zero-centred Gaussian Process, %independent across $i$ and 
independent of $w(\calL)$, with the valid covariance function $C_{ii \given \calL}(s,s') = C_{ii}(s,s') - C_{ii}(s, \calL)C^{-1}_{ii}(\calL, \calL)C_{ii}(\calL,s')$. 

The first part of stitching ensures that $w(\calL)$ conforms to $\calG_\calV$ when restricted to $\calL$. The next result establishes process-level conditional independence for the stitched predictive process. 
\begin{lemma}\label{lem:pp}
The predictive process $w^*(\cdot)=(w^*_1(\cdot),\ldots,w_q^*(\cdot))^{\T}$ is a GGP$(\calG_\calV)$ on $\calD$. 
\end{lemma}
 %is a GGP with respect to $\calG_\calV$ over the entire domain $\calD$. 
%the $i$-th and $j$-th elements of $w^*(\cdot)=(w^*_1(\cdot),\ldots,w_q^*(\cdot))'$ are conditionally independent given all elements of $w(\calL)$ except $w_i(\calL)$ and $w_j(\calL)$. 
We now extend the finite-rank GGP $w^*(\cdot)$ to a full-rank GGP $w(\cdot)$ over the entire domain $\calD$ through (\ref{eqn:extend}). %by adding to the predictive processes $w_i^*(\cdot)$ 
We construct $z_i(\cdot) \sim GP(0,C_{ii|\calL})$ such that $z_i(\cdot) \perp z_j(\cdot)$ for all $i \neq j$, and $z_i(\cdot) \perp w(\calL)$ for all $i$. % that are independent both to $w(\calL)$ and to each other. 
Independence among $z_i(\calL)$ and $w(\calL)$ and the marginal covariance of $z_i(\calL)$ in (\ref{eqn:extend}) ensures that each $w_i(\cdot)$ on $\calD$ is exactly $GP(0,C_{ii})$. However, independence among the $z_i(\cdot)$'s is a neat choice ensuring that the conditional independence relations in $\calG_\calV$ is extended from the finite set $\calL$ to the spatial process over $\calD$. We prove this formally in Theorem~\ref{thm: multgp}.}

\begin{theorem}\label{thm: multgp}
Given a cross-covariance function $C$ %(s,s')=(C_{ij}(s,s'))$ 
and an inter-variable graph $\calG_\calV$, stitching creates a valid multivariate GGP $w(\cdot)$ %with there exists a multivariate GP with a valid 
with a valid (p.d.) cross-covariance function $M$ %(s,s')=(M_{ij}(s,s'))$ 
such that:% if $w(\cdot) \sim GP(0,M)$ then:
\begin{enumerate}[(a)]
	\item $w_i(\cdot) \sim GP(0,C_{ii})$ , i.e., $M_{ii}(s,s')=C_{ii}(s,s')$ for all $s,s'\in \calD$ and for each $i=1,\ldots,q$,  %i.e.,  individual processes $w_i(\cdot)$ on $\calD$ will be $GP(0,C_{ii})$.
\item $w(\cdot)$ is a GGP$(\calG_\calV)$ on $\calD$,%If variables $(i,j) \notin E_\calV$, then the processes $w_{i}(\cdot)$ and  $w_{j}(\cdot)$ will be conditionally independent on $\calD$ given $w_B(\cdot)$, where $w_{B}(\cdot)=\{w_{k}(\cdot); k \in \calV \setminus \{i,j\}\}$ are all the other processes,  
\item if $(i,j) \in E_\calV$, then $M_{ij}(s,s')=C_{ij}(s,s')$ for all $s,s'\in \calL$.
\end{enumerate}
\end{theorem}
%\begin{proof}
%See Supplementary Materials.
%\end{proof}
\black{Stitching produces a multivariate GP $w(\cdot)$ that exactly satisfies the first two conditions sought in Section \ref{sec:exist}.} %The marginal covariance functions for each $w_i(s)$ over $\calD$ is exactly $C_{ii}(,s,s')$ \black{thereby satisfying Condition (i)}. 
%The above completes the specification of a multivariate GP using stitching. The first specification in (\ref{eq:covsel}) ensures $w_i(\calL) \sim N(0,C_{ii}(\calL))$. This along with the component-wise extension in (\ref{eqn:extend}) using variable-specific independent GP $z_i(s)$ with the conditional covariance function $C_{ii \given \calL}(\cdot,\cdot)$ ensures that the marginal covariance functions for $w_i$ is exactly $C_{ii}(\cdot)$ over the entire domain $\calD$. 
%Process-level conditional independence relations \black{(Condition (ii))} are established using $\calG_V \boxtimes \calG_L$ on $\calL$ and then using independent processes $z_i(\cdot)$ for each variable $i$ for the extension to $\calD$ without violating the  conditional independences. 
%The following Theorem formally states and proves the properties of GGP using stitching. 
%We now show that constructing $w(\calL)$ on $\calG_V \boxtimes \calG_L$ followed by a suitable extension to $w(\calD \setminus \calL)$ using (\ref{eqn:extend}) will ensure that the resulting process $w(s)$ will be a GGP with respect to $\calG_V$ in the entire domain $\calD$.
%We provide some additional intuition here. 
%The process $w(s) \sim GP(0,M(\cdot,\cdot))$, constructed as above, is referred to as a $q\times 1$ GGP on $\calD$ conforming to the graph $\calG_V$. 
%Also, if we begin with a stationary $C(s,s')$, then (a) of Theorem~\ref{thm: multgp} implies $M_{ii}(h)=C_{ii}(h)$ for all $h\in \calD$. 
\black{Regarding Condition (iii)}, we point out some differences between the GGP ensured by Theorem~\ref{th:exists} and the one produced by stitching. For pairs of variables $(i,j) \in E_\calV$, the cross-covariance for the former is exactly the same as the given cross-covariance $C_{ij}$ on the entire domain $\calD$, whereas for the latter \black{$M_{ij}(s,s') = C_{ij}(s,s')$ for locations in $\calL$. For a pair $s, s'\notin \calL$ and $i \neq j$ it is straightforward to verify that 
\begin{equation}\label{eq:cross}
M_{ij}(s,s') = C_{ii}(s, \calL) C_{ii}(\calL,\calL)^{-1} M(\calL,\calL)_{ij} C_{jj}(\calL,\calL)^{-1}C_{jj}(\calL, s')\;.
\end{equation}
Stitching, thus, produces a computationally feasible GGP with desired full-rank marginal covariance and process-level conditional independence at the expense of allowing a fixed rank cross-covariance. Choosing $\calL$ to be reasonably dense (well-spaced) in $\calD$, we have $M_{ij}(s,s') \approx C_{ij}(s,s')$ for $(i,j) \in E_\calV$, $s,s' \in \calD \setminus \calL$.  Hence, %for stitching,
condition (iii) is satisfied exactly on $\calL$ and approximately on $\calD \setminus \calL$ for the stitched GP.} 
%On the other hand, unlike Theorem~\ref{th:exists}, stitching does not rely on the given cross-covariance $C$ to be stationary and can be used with asymmetric or even non-stationary $C$ \black{(as discussed in Section \ref{sec:ns}).}
 %Also the constraints like $(b_{ij})$'s being a correlation matrix. themselves often need $O(q^3)$ operations to be verified (adding to the computation) \citep{gneiting2010matern} or \cite{apanasovich2012valid}.  
%In the next section, we discuss the challenges for large $q$ and develop in \ref{eq:joint}, as a) the $nq \times nq$ matrix $\calM(\calL,\calL)$% pose computational challenges, and evaluating the likelihood for the reference set $w(\calL)$. 
% \pink{Move Theorem A1 here (keep the proof in the appendix). Since the original covariance function is denoted by $C$, maybe use $M$ to denote the graphical covariance function. Add a third condition that on $\calL$, if $i,j$  belongs to $E$, then $M_{ij}(s_1,s_2)=C_{ij}(s_1,s_2)$. }

\section{Highly multivariate Graphical Mat\' ern Gaussian processes}\label{sec:high}
\subsection{\black{Incompatibility of multivariate Mat\' ern with graphical models}}\label{sec:mat}
\black{Theorems~\ref{th:exists}~and~\ref{thm: multgp} establish, respectively, the existence of and the construction of a marginal-preserving GGP given any valid cross-covariance $C$ and any inter-variable graph $\calG_\calV$. %While these results hold for any $C$ and $\calG_\calV$, 
We are particularly interested in developing a novel class of {\em multivariate graphical Mat\'ern GPs} that are GGP$(\calG_\calV)$ %s with respect to $\calG_\calV$ 
such that each univariate process is a Mat\'ern GP. This is appealing for inference as we retain the ability to interpret the parameters for each univariate spatial process. We achieve this using stitching, which is necessary as we argue below that no non-trivial parametrisation of the existing multivariate Mat\'ern GP yields a GGP.}% with respect to $\calG_\calV$.} 

The isotropic multivariate Mat\'ern cross-covariance function on a $d$-dimensional domain is $C_{ij}(s,s')=\sigma_{ij}H_{ij}(\|s - s' \|)$, where $H_{ij}(\cdot)=H(\cdot \given \nu_{ij}, \phi_{ij})$, $H$ being the Mat\'ern correlation function \citep{apanasovich2012valid}. If  $\theta_{ij}=\{\sigma_{ij},\nu_{ij},\phi_{ij}\}$, then for a multivariate Mat\'ern GP the $i$th individual variable is a Mat\'ern GP with parameters $\theta_{ii}$. This is attractive because it endows each univariate process with its own variance $\sigma_{ii}$, smoothness $\nu_{ii}$, and spatial decay $\phi_{ii}$. Another nice property is that under this model, $\Sigma = (\sigma_{ij})=\mbox{Cov}(w(s))$ is the covariance matrix for $w(s)$ within each location $s$. The cross-correlation parameters $\nu_{ij}$ and $\phi_{ij}$ for $i \neq j$, are generally hard to interpret, especially since $\nu_{ij}$ does not correspond to the smoothness of any surface. Recent work by \cite{kleiber2017coherence} on the concept of \emph{coherence} %defined by the scaled spectral density matrix of a stationary multivariate process 
has facilitated some interpretation  of these  parameters. 
The {\em parsimonious multivariate Mat\'ern} model of \cite{gneiting2010matern} emerges from this general specification as a special case with $\nu_{ij}=(\nu_{ii}+\nu_{jj})/2$ and  $\phi_{ij}=\phi$.

To ensure a valid multivariate Mat\'ern cross-covariance function, it is sufficient to constrain the intra-site covariance matrix $\Sigma=(\sigma_{ij})$  to be of the form \citep[Theorem 1,][]{apanasovich2012valid}%One of the constraints on the The sufficient constraints on the parameter space is specified by a sufficient constraint (:
\begin{equation}\label{eq:constraints}
\begin{array}{cc}
   % \nu_{ij} =& \frac 12 (\nu_{ii} + \nu_{jj}) + \Delta_A (1 - A_{ij}) \mbox{ where } \Delta_A \geq 0, A=(A_{ij}) \mbox{ for all } i \geq 0, A_{ii}=1 \\
    %\sum_{i,j} c_ic_j\phi_{ij} \leq 0  &\mbox{ is a conditionally non-negative definite matrix } \\
    \sigma_{ij} =& b_{ij} \frac{\Gamma(\frac 12 (\nu_{ii}+\nu_{jj} + d))\Gamma(\nu_{ij})}{\phi_{ij}^{2\Delta_A+\nu_{ii}+\nu_{jj}}\Gamma(\nu_{ij} + \frac d2)} \mbox{ where } \Delta_A \geq 0, \mbox{ and } B=(b_{ij}) > 0, \mbox{ i.e., is p.d.}
\end{array}
\end{equation} This is equivalent to $\Sigma$ being constrained as $\Sigma = (B \odot (\gamma_{ij}))$, where %$D$ is a diagonal matrix with entries $\sqrt{\sigma_{ii}}$. Here 
$\gamma_{ij}$ are constants collecting the terms in (\ref{eq:constraints}) involving only $\nu_{ij}$'s and $\phi_{ij}$'s, and $\odot$
denotes the Hadamard (element-wise) product. Similarly, the spectral density matrix takes the form $F(\omega) = (B \odot (g_{ij}(\omega)))$, where $g_{ij}(\omega)$ are functions involving the parameters $\phi_{ij}$ and $\nu_{ij}$. The matrix $B=(b_{ij})$'s are the $O(q^2)$ parameters (free of $\phi_{ij}$'s or $\nu_{ij}$'s) that are constrained to ensure $B$ is positive-definite. \black{Process-level conditional independences introduce zeros in the inverse of the spectral density matrix for stationary processes (see, e.g., Theorem~2.4 in \cite{dahlhaus2000graphical}). This implies that, for any parametrisation of the multivariate Mat\'ern GP to be a GGP, we need $(F(\omega)^{-1})_{ij}=0$ for every $(i,j) \notin E_\calV$ and almost all $\omega$. From the Hadamard product $F(\omega)= (B \odot (g_{ij}(\omega)))$,} it is clear that zeros in $B^{-1}$ or $\Sigma^{-1}$ do not generally imply zeros in $F^{-1}(\omega)$ for the multivariate Mat\'ern. An exception occurs when each component is posited to have the same smoothness $\nu$ and the same spatial decay parameter $\phi$, whence both $\Sigma$ and $F(\omega)$ become proportional to $B$. In this case, zeros in $B^{-1}$ (specified according to $\calG_\calV$) will correspond to zeros in $\Sigma^{-1}$ and $F^{-1}(\omega)$ yielding a GGP with respect to $\calG_\calV$. However, assuming $\nu_{ij}=\nu$ and $\phi_{ij}=\phi$ for all $(i,j)$ implies that the univariate GPs have the same smoothness and rate of spatial decay, which is restrictive. Beyond this separable model, there is, to the best of our knowledge, no known parameter choice for the multivariate Mat\'ern GPs that will allow it to be a GGP$(\calG_\calV)$. % with respect to a given $\calG_V$.

\subsection{Computational considerations for stitching}\label{sec:comp}
\black{Stitching univariate processes corresponding to a valid multivariate Mat\'ern cross-covariance $C$ and a graph $\calG_\calV$ yields a multivariate graphical Mat\'ern GP such that (i) the univariate processes are exactly Mat\'ern; (ii) the multivariate process conforms to process-level conditional independence relations as specified by $\calG_\calV$; and (iii) the cross-covariances for pairs of variables in $\calG_\calV$ are exactly or approximately Mat\'ern (see Eq.~\ref{eq:cross}).} %The reference set $\calL$ used in stitching is arbitrary and can, but need not, overlap with the set of data locations. 
\black{For each $i=1,2,\ldots,q$ let $D_i$ be the set of $n_i$ locations where the $i$-th variable has been observed.} The joint probability density of $w_i(D_i)$ and $w(\calL)$ is specified by $w(\calL) \sim N(0,M(\calL,\calL))$ and
\begin{equation}\label{eq:joint}
\begin{array}{c}
w_i(D_i) \given w(\calL) \stackrel{ind}{\sim} N(C_{ii}(D_i,\calL)C_{ii}(\calL,\calL)^{-1}w_i(\calL),C_{ii \given \calL}(D_i,D_i))\; \mbox{ for } i=1,\ldots,q\;.
\end{array}
\end{equation}
The \black{covariance matrix} for $\{w_i(D_i):\, i=1,\ldots,q\} \given w(\calL)$ \black{is block-diagonal with variable-specific blocks and is cheap to compute if all of the $n_i$'s are small.} If some $n_i$'s are large, we can use one of the several variants of scalable GPs for very large number of locations \citep{heaton2019case}. For example, a nearest neighbour GP \citep[NNGP, ][]{nngp} yields a sparse approximation of $C_{ii \given \calL}(D_i,D_i)$ with linear complexity, but the joint distribution still preserves the conditional independence implied by $\calG_\calV$. 

When $q$ is large, note that $\{w_i(D_i):\, i=1,\ldots,q\} \given w(\calL)$ in (\ref{eq:joint}) has $q$ conditionally independent factors and is easy to compute in parallel. However, the likelihood for $w(\calL) \sim N(0,M(\calL,\calL))$ presents the bottleneck for this highly multivariate case. In particular, there are two challenges for large $q$. As discussed earlier, the multivariate Mat\'ern $C$ required for stitching needs to constrain $B=(b_{ij})$ to be p.d. on an $O(q^2)$-dimensional parameter space. Searching in such a high-dimensional space is difficult for large $q$ and verifying positive definiteness of $B$ incurs an additional cost of $O(q^3)$ flops. Second, evaluating $w(\calL) \sim N(0,M(\calL,\calL))$ involves matrix operations for the $nq \times nq$ matrix $M(\calL,\calL)$. While the precision matrix, $M(\calL,\calL)^{-1}$, is sparse because of $\calG_V$, its determinant is usually not available in closed form and the calculation can become prohibitive even for small $n$. 

\subsection{Decomposable variable graphs}\label{sec:decomp}
To facilitate scalability in highly multivariate settings, we consider decomposable inter-variable graphs% $\calG_\calV$
. For $\calG_\calV = (\calV, E)$, and a triplet $(A,B,O)$ of disjoint subsets $\calV$, $O$ is said to {\em separate} $A$ from $B$ if every path from $A$ to $B$ passes through $O$. If $\calV= A \cup B \cup O$, and $O$ induces a complete subgraph of $\calV$, then $(A,B,O)$ is said to decompose $\calG_\calV$. The graph $\calG_\calV$ is said to be decomposable if it is complete or if there exists a proper decomposition $(A,B,O)$ into decomposable subgraphs $\calG_{A\cup O}$ and $\calG_{B\cup O}$. \black{Several naturally occurring dependence structures like low-rank dependence or autoregressive dependence correspond to decomposable graphs (see Section \ref{sec:ex}).} More generally, if a graph is non-decomposable, it can be embedded in a larger decomposable graph. 
Hence, assuming decomposability is conspicuous in graphical models \citep[see, e.g.,][]{dobra2003markov, wang2009bayesian} since fitting Bayesian graphical models is cumbersome for non-decomposable graphs \citep{roverato2002hyper, atay2005monte}. %We will show below that 

\begin{table}[t]
        \setlength{\belowcaptionskip}{-5pt}
\setlength{\abovecaptionskip}{-0pt}
\caption{Properties of any $q$-dimensional multivariate Mat\'ern GP of \cite{gneiting2010matern} or \cite{apanasovich2012valid} and a multivariate graphical Mat\'ern GP stitched using a decomposable graph $\calG_V$ with largest clique size $q^*$ (typically $\ll q$), length of perfect ordering $p$, and maximal number of cliques $p^*$ sharing a common vertex.}\label{tab:comp}
\begin{small}
\begin{center}{
\begin{tabular}{|l|l|l|}\hline
Model attributes &  Multivariate Mat\'ern & Multivariate Graphical Mat\'ern  \\ \hline
 Number of parameters & $O(q^2)$ & $O(|E_\calV|+q)$ \\
 Parameter constraints & $O(q^3)$ & $O(p^* (q^{*3}))$ (worst case) \\
 Storage & $O(n^2q^2)$ & $O(pn^2q^{*2})$ (worst case)\\
Time complexity &  $O(n^3q^3)$ &  $O(p n^3 q^{*3})$ (worst case)\\
Conditionally independent processes & No & Yes \\
Univariate components are Mat\'ern GPs & Yes & Yes \\\hline
\end{tabular}}\end{center}\end{small}
\end{table}

For stitching of Mat\'ern GPs using decomposable graphs we can significantly reduce the dimension of the parameter space, storage and computational burden. %An important property of a decomposable graph is that the 
%Cliques of a decomposable graph can be ordered into a {\em perfect sequence} \citep{lauritzen1996graphical}. %\begin{definition}
Let $K_1,\cdots, K_p$ be a sequence of subsets of the vertex set $\calV$ for an undirected graph $\calG_\calV$. Let, $F_m = K_1 \cup \cdots \cup K_m$ and $S_m = F_{m-1} \cap K_m$. The sequence $\{K_m\}$ is said to be {\em perfect} if (i) for every $l > 1$, there is an $m < l$ such that $S_l \subset K_m$; and (ii) the {\em separator} sets $S_m$ are complete for all $m$. 
%\end{definition}
If %our model, we take 
$\mathcal{G}_V$ is %to be 
decomposable, %and $\mathcal{G_L}$ to be the complete graph between $n$ location.  This particular structure helps in
then it has a perfect clique sequence \citep{lauritzen1996graphical} and the joint density of $w(\calL)$ can be factorized as follows. %From now on, we will shorten $\covr$ and $\covm$ as $C$ and $M$ when suitable and denote by 
\begin{corollary}
\label{cor: fact-dens}
If $\mathcal{G_V}$ has a perfect clique sequence $\{K_1,K_2,\cdots,K_p\}$ with separators $\{S_2,\ldots,S_m\}$, %and we denote by $f$ the corresponding densities and by subscripts, the corresponding variable indices for $w$, 
then the GGP likelihood on $\calL$ can be decomposed as
\begin{equation}\label{eq:factor}
    f_M(w(\calL)) = \frac{\Pi_{m=1}^{p} f_C(w_{K_m}( \calL))}{\Pi_{m=2}^{p} f_C(w_{S_m}( \calL))}\;,
\end{equation}
where %$S_m=F_{m-1} \cap K_m$ and $F_{m-1}= K_1 \cup \cdots \cup K_{m-1}$, and 
$f_A$ denotes the density of a GP over $\calL$  with covariance function $A$ for $A \in \{M,C\}$.
\end{corollary}
%\begin{proof}
%See Supplementary Materials.
%\end{proof}
%Corollary \ref{cor: fact-dens} is a consequence of the factorization theorem for decomposable graphical models in \cite{lauritzen1996graphical}, where we apply it for the strong product of the decomposable variable graph $\calG_V$ with the complete location graph $\calG_L$. We now discuss how the result helps with dimensionality and constraints on the parameter space, storage and  computational complexity of stitching. 

Corollary~\ref{cor: fact-dens} helps us manage the dimension and constraints of the parameter space and the computational complexity of stitching. For an arbitrary $\calG_V$, the parameter space for the stitching covariance function $M$ is the same as the parameter space $\{\theta_{ij} | 1 < i,j \leq q\}$ for the original covariance function $C$. %If $C(h)$ is the multivariate Mat\'ern cross-covariance, then the parameters $\{B_{ij}\}$ are constrained to be a $q \times q$ correlation matrix. Compliance with this constraint requires checks on positive definiteness that require $O(q^3)$ flops for every new value of $B_{ij}$'s in an iterative sampling or optimization scheme. 
For a decomposable $\calG_V$, the likelihood (\ref{eq:factor}) and, in turn, the stitched GGP is only specified by the parameters $\{ \theta_{ij} \given (i=j) \mbox{ or } (i,j) \in E_\calV \}$. Therefore, the dimension of the parameter space reduces from $O(q^2)$ to $O(|E_\calV|+q)$ , where $|E_\calV|$ is the number of edges on $\calG_\calV$, which is small for sparse graphs. When using a multivariate Mat\'ern cross-covariance $C$ for stitching, the parameter space for $B$ in the stitched graphical Mat\'ern is the intersection of the parameter spaces of the low-dimensional clique-specific multivariate Mat\'ern covariance functions $C_{K_1}, \ldots,C_{K_p}$. Hence, the parameter space becomes $\{b_{ij} | (i=j) \mbox{ or } (i,j) \in E_\calV \}$ and needs to satisfy the constraint that $B_{K_l}=(b_{ij})_{i,j \in K_l}$ is p.d. for all $l=1,\ldots,p$. %Also, to update $b_{ij}$'s given all other $b_{i'j'}$'s for $(i'j') \neq (i,j)$ in a Bayesian Gibbs sampler or a coordinate descent algorithm, we only need to check positive definiteness for the sub-matrices of $C(h)$ corresponding to the cliques and separators containing the edge $(i,j)$. 
This reduces the computational complexity of parameter constraints from $O(q^3)$ to at most $O(p^*q^{*3})$, where $q^*$ is the largest clique size and $p^*$ is the maximum number of cliques sharing a common vertex. 
%\begin{proof}
%See Supplementary Materials.
%\end{proof}
The precision matrix of $w(\calL)$ satisfies \citep[Lemma~5.5,][]{lauritzen1996graphical}
\begin{equation}\label{eq:m-decomp}
M(\calL,\calL)^{-1} = \sum_{m=1}^{p} [{C}_{[K_m \boxtimes \mathcal{G_L}]}^{-1}]^{\calV \times \calL} - \sum_{m=2}^{p} [{C}_{[S_m \boxtimes \mathcal{G_L}]}^{-1}] ^{\calV \times \calL}\;, %= \sum_{m=1}^{p} [{M}_{[K_m \boxtimes \mathcal{G_L}]}^{-1}] ^\calV - \sum_{m=2}^{p} [{M}_{[S_m \boxtimes \mathcal{G_L}]}^{-1}] ^\calV\;
\end{equation}
where, for any symmetric matrix $A=(a_{ij})$ with rows and columns indexed by $\calU \subset \calV \times \calL$, $A^{\calV \times \calL}$ denotes a $|{\calV \times \calL}| \times |{\calV \times \calL}|$ matrix such that $(A^{\calV \times \calL})_{ij} = a_{ij}$ if $(i,j) \in \calU$, and $(A^{\calV \times \calL})_{ij} = 0$ elsewhere. %the principal submatrix of $A^{\calV \times \calL}$ corresponding to the indices in $\calU$ is $A$, and $A^\calV$ is $0$ elsewhere. 
From (\ref{eq:factor}) and (\ref{eq:m-decomp}) we see that the stitching likelihood evaluation avoids the large matrix $\covm$ and all matrix operations are limited to the sub-matrices of $\covm$ corresponding to the cliques $K_m \boxtimes \mathcal{G_L}$ and separators $S_m \boxtimes \mathcal{G_L}$.  % thus requiring to invert submatrices of the dimension of $K_m \times \mathcal{G_L}$ and $S_m \times \mathcal{G_L}$. 
The entire process requires at most $O(pn^3q^{*3})$ flops and $O(pn^2q^{*2})$ storage, where $p$ is the length of the perfect ordering.  %If $n$ is large we can use NNGP \citep{nngp} or any other popular big-GP approach to make the computation and storage linear in $n$. Since large $n$ is not the focus of this paper, we do not discuss this further. Thus, for decomposable graphs $\calG_V$, stitching, apart from ensuring a valid GGP with process-level conditional independence and marginal GPs with Mat\'ern kernels, also ensures significant reduction in the dimension and complexity of the parameter space, storage and computation time. 
Table~\ref{tab:comp} summarizes these gains from stitching with decomposable graphs.

 \black{The computational efficiency of stitching is clear from the above. In addition, the following result shows that the GGP likelihood from stitching yields unbiased estimating equations for all parameters included in the GGP (all marginal and cross-covariance parameters for any pairs of variables included in $\calG_\calV$) under model misspecification when the data is generated from a multivariate Mat\'ern GP, but is modelled as a graphical Mat\'ern GP with a decomposable $\calG_\calV$. 
%clique or separator specific parameters even in a mis-specified setting when the true process is  (Lemma \ref{lemma: ub-ee}). See Section \ref{appn: proofs} for the proof.
\begin{proposition}\label{lemma: ub-ee}
Let $w(\cdot) \sim GP(0,C(\cdot,\cdot))$, where $C$ is a valid $q \times q$ multivariate Mat\'ern cross-covariance function with parameters $\{\theta_{ij}: 1 \leq i,j \leq q\}$, and $f_M(w(\calL))$ denotes the multivariate graphical Mat\'ern GP likelihood (\ref{eq:factor}) from stitching using a decomposable graph $\calG_\calV$. Then $E(\partial \log f_M(w(\calL))/ \partial \theta_{ij})=0$ for any $i=j$ or $(i,j) \in E_\calV$. %  the true process is a multivariate Mat\'ern GP, maximum likelihood estimating equations using a multivariate graphical Mat\'ern GP with a decomposable graph $\calG_\calV$ are unbiased for cross-covariance parameters corresponding to all the cliques and separators of the graph $\calG_\calV$.
\end{proposition}}

\subsection{Chromatic Gibbs sampler}\label{sec:chrom}
\begin{figure}[h!]
    \begin{center}
    \includegraphics[scale=0.25,trim={0 10 10 15},clip]{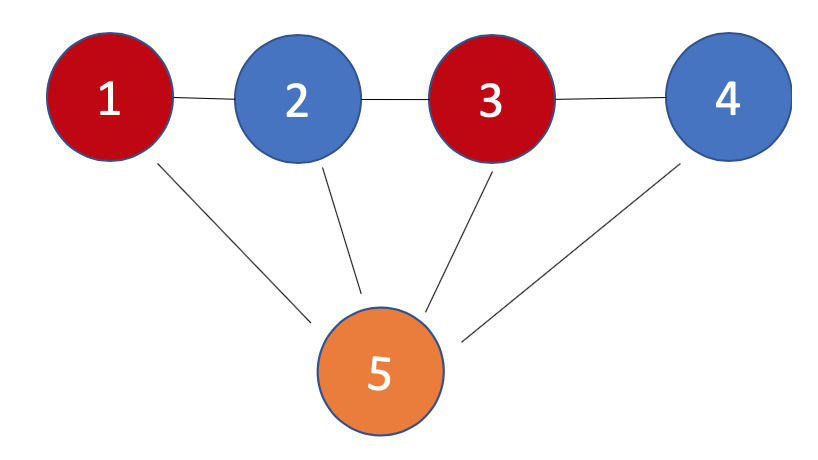}
    \hspace*{2.4cm}\includegraphics[scale=0.2,trim={0 15 0 15},clip]{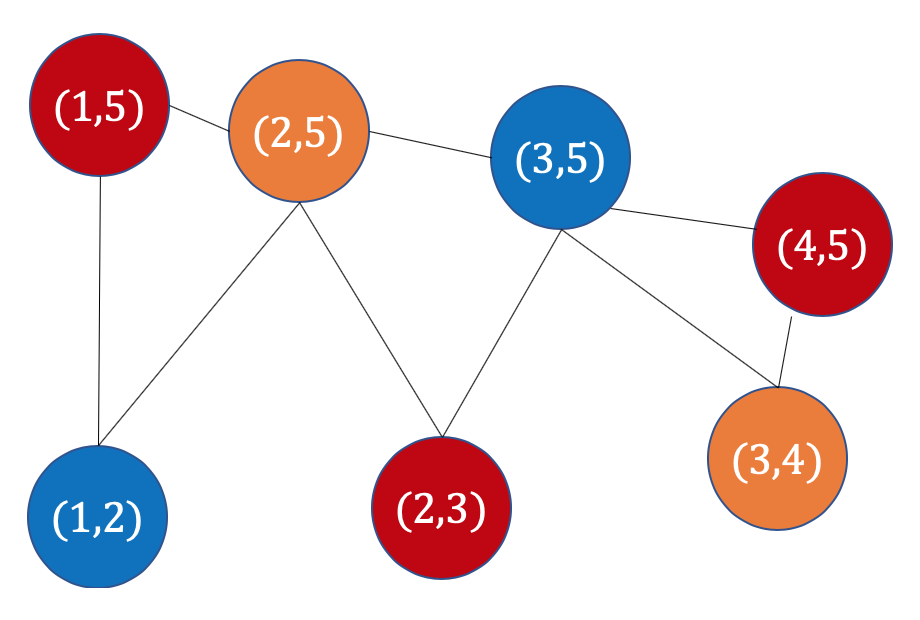}
    \end{center}
	%\subfloat[Network of 8 plant traits \cite{traittrait}]{
	%\hskip -1mm\subfloat[]{\includegraphics[scale=0.15, trim={26 26 26 26},clip]{Figures/gem_graph_model_700x820.png}\label{fig:stitchgem}}
	%\hskip -1mm\subfloat[]{\includegraphics[scale=0.15, trim={26 26 26 26},clip]{Figures/path_gem_model_700x820.png}\label{fig:stitchpathgem}}
        \setlength{\belowcaptionskip}{-10pt}
\setlength{\abovecaptionskip}{2pt}\caption{Chromatic sampling for GGP with a gem graph  between $5$ variables: Left: Gem graph and colouring used for chromatic sampling of the variable-specific parameters. Right: Colouring of the corresponding edge graph $\calG_E(\calG_V)$ used for chromatic sampling of the cross-covariance parameters $b_{ij}$'s.}% In chromatic sampling, we can use this coloring to sample nodes belonging to same color in parallel bringing down the complexity by significant amount.} %    multivariaFour graphs are (a) full multivariate GP, (b) stitched GP with path graph between variables, (c) stitched GP using Diamond graph between variables, (d) stitched GP using $1$ nearest-neighbor location graph (path graph) between locations and diamond graph between variables.}
    \label{fig: chrom-w}
    
%Turning to predictions, spatial modelling using GGP delivers inference on the response process $y(\cdot)$ and the latent process $w(\cdot)$ over a set of unknown locations. In Bayesian settings, we evaluate (or sample from) the posterior predictive distribution $p(y(\cdot),w(\cdot)\given y)$ by using the sampled $\{\beta,\theta, w\}\sim p(\beta,\theta,w\given y)$, where $\theta$ is the collection of all unknown process parameters, and then sampling $\{y(\cdot),w(\cdot)\}\sim p(y(\cdot), w(\cdot)\given \beta, w,\theta))$. With the GGP specification on $w$, this conditional predictive distribution is Gaussian and predictive inference is implemented as in usual multivariate spatial settings \citep[see, e.g., Chapter~8 in][]{ban14}. 
\end{figure}

\black{With a valid process specification for $w(\cdot)$, we cast (\ref{eqn:mgp}) into a hierarchical model over the $n$ observed locations in ${\calS}$ and sample from the posterior distribution derived from
\begin{equation}\label{eqn:mgp_hierarchical}
p(\beta, \tau, \theta) \times N(w(\calS)\given 0, C_{\theta}(\calS,\calS))\times \prod_{j=1}^n N(y(s_j)\given X(s_j)\beta + w(s_j), D_{\tau})\;,
\end{equation}
where $X(s_j) = \mbox{diag}(x_1(s_j)^{\T}, x_2(s_j)^{\T},\ldots, x_q(s_j)^{\T})$ is $q\times (\sum_{i=1}^q p_i)$, $\beta = (\beta_1^{\T}, \beta_{2}^{\T}\ldots,\beta_q^{\T})^{\T}$, $w(\calS) = (w(s_1)^{\T}, w(s_{2})^{\T}\ldots,w(s_n)^{\T})^{\T}$, $D_{\tau} = \mbox{diag}(\tau^2_1, \tau^2_2,\ldots,\tau^2_q)$, $\theta$ is the set of parameters in the cross-covariance function and $p(\beta, \tau, \theta)$ is a prior distribution on model parameters. Besides the computational benefits described in Table \ref{tab:comp}}, stitched GGP models are \black{also} amenable to parallel computing. In a Bayesian implementation of a stitched GGP model (described in Section \ref{sec: bayes-model} of the Supplement), we can exploit the graph $\calG_V$ and deploy a chromatic Gibbs sampler \citep{gonzalez2011parallel} to simultaneously update batches of random variables in parallel. Let $\eta_i$ be the vector grouping  variable-specific parameters (regression coefficients, spatial parameters, noise variance and latent spatial random effects). Under a graph colouring of $\calG_V$, $\eta_i$ and $\eta_{i'}$ can be updated simultaneously if $i$ and $i'$ share the same colour, as illustrated in Figure \ref{fig: chrom-w} (left).  %The graph colorings for $\calG_V$ and $\calG_E$ are demonstrated in Figure \ref{fig: chrom-w} for an example graph. \vskip-3mm

This brings down the number of sequential steps in sampling of the $\eta_i$'s from $q$ to the chromatic number $\chi(\calG_V)$. \black{We can also employ a chromatic sampling scheme for the $b_{ij}$'s, but using a different graph. We exploit the fact that the parameters $b_{ij}$ and $b_{i'j'}$ belongs to the same factor in (\ref{eq:factor}) for a pair of edges $(i,j)$ and $(i',j')$ in $E_\calV$ if and only if the variables $i,j,i',j'$ belongs to the same clique.} Thus, if $\calG_E(\calG_V)=(E_\calV,E^*)$ denotes this graph on the set of edges $E_\calV$, i.e., there is an edge $((i,j),(i',j'))$ in this new graph $\calG_E(\calG_V)$ if $\{i,i',j,j'\}$ are in some clique $K$ of $\calG_V$, then we can batch the updates of $b_{ij}$'s based on the colouring of the graph $\calG_E(\calG_V)$ (Figure \ref{fig: chrom-w} (right)). The number of such sequential batch updates will be the chromatic number $\chi(\calG_E(\calG_V))$, a potentially drastic reduction from $|E_\calV|$ sequential updates for $b_{ij}$.

\section{Extensions}\label{sec:ex}

\subsection{\black{Factor models}}\label{sec:lmc}
\black{The construction of GGP and its implementation described in Sections~\ref{sec:meth}~and~\ref{sec:high} assumes a known graphical model. Here, we describe different avenues for choosing or estimating the graph and offer extensions of GGP to model different spatial and spatiotemporal structures.}

\black{
In many multivariate spatial models, the inter-variable graphical model arises naturally and is decomposable. A large subset of multivariate spatial models are  process-level factor models (emerge from more general linear models of coregionalization (LMC)), where each of the $q$ observed univariate processes are a weighted sum of $r \leq q$ latent univariate factor processes with the weights being component-specific \citep{schmidt2003bayesian,gelfand2004nonstationary,wackernagel2013multivariate}. In %its most 
general% form
, a linear model of coregionalization can be expressed as
\begin{equation}\label{eq:lmc}
    w_i(s) = \sum_{j=1}^r a_{ij}(s) f_j(s) + \xi_i(s)\;,
\end{equation}
where each $f_j(\cdot)$ is a latent factor process such that $f(\cdot)=(f_1(\cdot),\ldots,f_r(\cdot))^\T$ is a multivariate GP, $a_{ij}(\cdot)$'s are component-specific weight functions and $\xi(\cdot)$ are independent processes representing the idiosyncratic spatial variation in $w_i(\cdot)$ not explained by the latent factors. 
%The factors $f(\cdot)=(f_1(\cdot),\ldots,f_r(\cdot))^\T$ follow a multivariate GP. %In practice, they are often assumed to be independent. The weights $a_{ij}$ can be fixed or spatially varying. The processes $\xi(\cdot)$ are often not included, i.e., $\xi(s) \equiv 0$ as the component-specific variation is already considered in the response model (\ref{eqn:mgp}) via the white noise processes $\eps_i(\cdot)$. We choose $r \ll q$ in (\ref{eq:lmc}) when modelling processes driven by a small number of latent variables. % (e.g., speciated particulate matter (PM) attributable to a handful of emission sources). 
If $q$ is large, choosing $r \ll q$ in (\ref{eq:lmc}) also facilitates dimension reduction \citep{Lopes2008, ren2013hierarchical, taylor2019spatial, zhangbanerjee2021}. %, whereas choosing $r$ to be equal to or close to $q$ allows modeling complex dependencies among the variables. 
%We have now added a new Section comparing LMC with GGP. In particular, we have established aa theoretical result that all LMC are GGP with a fixed graphical model on the joint set of observed and latent (factor) processes. %However, unlike our marginal-preserving GGP, LMC does not retain the ability to interpret spatial properties of each process separately, with every process now sharing the smoothness of the roughest latent factor process. 
 %   We have generalized the implementation of GGP to allow a graphical model jointly on the observed and latent processes and have demonstrated its success to model data generated from LMC.  %LMC are a mainstay of multivariate spatial modeling and our result shows that they implicitly use a known graphical model.Many applications of multivariate spatial models where the graphical model arises naturally. 
%In applications such as modeling speciated particulate matter, we can often encounter situations such as one or two compounds influencing the rest of the particle peaks in the spectrum. 
%Linear model of coregionalization (LMC) \citep{goulard1992linear, schmidt2003bayesian, wackernagel2013multivariate} can be considered as a candidate model for this scenario where the observed Gaussian processes are assumed to result from small number of latent processes. We summarize this finding in Lemma~\ref{lemma: ggp-lmc} and provide the proof in Section~\ref{appn: proofs}. 
We next show that any linear model of coregionalization can be formulated as a GGP with a decomposable graph on the elements of $w(\cdot)$ and $f(\cdot)$.

\begin{proposition}\label{lemma: ggp-lmc}
Consider the linear model of coregionalization (\ref{eq:lmc}) where $f(\cdot)$ is an $r \times 1$ multivariate GP with a complete graph between component processes, and $\xi_i(\cdot)$'s are independent univariate GPs. Then $(w_1,w_2,\ldots,w_q,f_1,\ldots,f_r)^\T$ is a GGP on vertices $\{1,\dots,q+r\}$ and a decomposable graph $\{(i,j)|i \in 1,\ldots,(q+r),j \in (q+1),\ldots,(q+r), j \neq i\}$. 
\end{proposition}

Proposition \ref{lemma: ggp-lmc} dictates that the assumption of multivariate dependence induced through factor processes can be translated into a decomposable graph between the observed and factor processes. Hence, GGPs can be used as a richer alternative to the linear model of coregionalization. While the linear model of coregionalization enforces all processes $w_i(\cdot)$ to have the smoothness of the roughest $f_j(\cdot)$ \citep{genton2015cross}% %due to the linearity in (\ref{eq:lmc})
, 
the GGP enables us to model and interpret the spatial smoothness of each component process (e.g., with the graphical Mat\'ern GP). 
The complete graph between the $r$ component processes of $f(\cdot)$ can be assumed without loss of generality as even for a sparse graph between latent factors (e.g., when the factors are independent processes), they will generally be conditionally dependent given the observed processes $w(\cdot)$, thereby yielding the same joint graph. Due to $r \ll q$, this joint graph of observed and latent processes will still be sparse even after considering all possible edges between latent processes.
  %a full rank between observed and latent processes are shown in . The embedded graphical structure allows us to apply the Graphical Gaussian Process framework to model these types of data. \vskip-6mm 
  Figure~\ref{fig: graph-lmc} illustrates two examples of the decomposable graphs arising from linear model of coregionalization.

\begin{figure}[htb]
\begin{center}
\hspace{-3cm}\subfloat[$2$ observed (red) and $2$ latent (blue) processes]{\hspace*{2cm}\includegraphics[scale=0.5,trim={0cm 0cm 0cm 0cm},clip]{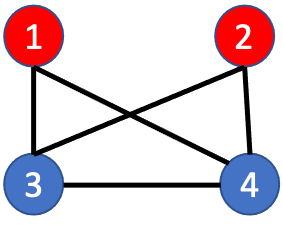}\label{fig: graph-lmc-1}}\hspace{1.5cm}
\subfloat[$5$ observed (red) and $1$ latent (blue) processes]{\hspace*{2cm}\includegraphics[scale=0.28,trim={0cm 1.2cm 0cm 0cm},clip]{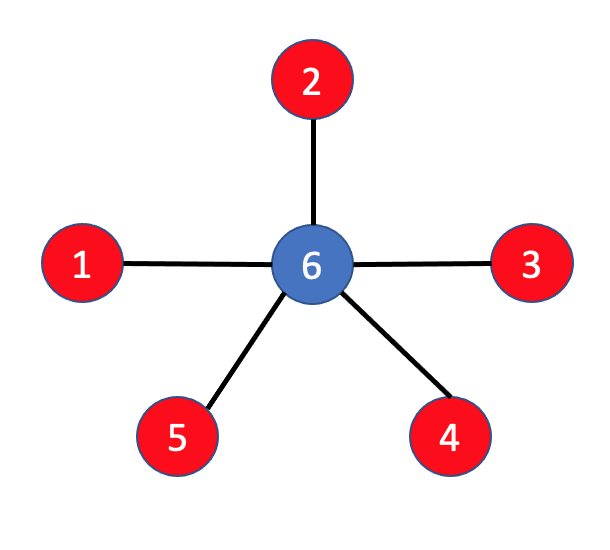}\label{fig: graph-lmc-2}}
\end{center}
    \setlength{\belowcaptionskip}{-11pt}
\setlength{\abovecaptionskip}{-2pt}\caption{Decomposable graphs for (a) a full rank and (b) a low-rank linear model of coregionalization.}
    \label{fig: graph-lmc}
\end{figure}

%For analysing datasets that arise from a small set of common sources (e.g., multivariate pollutant levels), a 

An alternative approach to linear models of coregionalization builds multivariate spatial processes by sequentially modelling a set of univariate GPs conditional from some ordering of the $q$ variables \citep{cressie2016multivariate}. A sparse partial ordering can facilitate dimension reduction for large $q$. This approach does not attempt to preserve marginals or introduce process-level conditional independence. However, a partial ordering yields a directed acyclic graph (DAG), which, when moralised, produces a decomposable undirected graph that can be used in our stitched GGPs.}

%Apart from LMC, another major strand of literature on multivariate spatial models consider sequentially constructing distributions for each variable conditional on the previous ones \citep{jin2005generalized,cressie2016multivariate}. This idea relies on knowledge of a known true ordering of the outcome variables, and different orderings lead to different models. The set of all possible orderings is large and assumption of known ordering is akin to assumption of a known graphical model. In fact, a (given) partial ordering among the variables leads to a directed acyclic graph which can be moralized to obtain an undirected graphical model. Hence, any data modeled using such a conditional approach with a DAG can also be modeled using a GGP on the moralized graph.}

\subsection{Non-separable spatial time-series modelling}\label{sec:time} GGPs are natural candidates for non-separable (in space-time), non-stationary (in time) modelling of univariate or multivariate spatial time-series. Consider a univariate spatial time-series modelled as a GP $\{w(s,t)\}$ for $s \in \calD$ evolving over a discrete set of time points $t \in \calT=\{1,2,\ldots,T\}$. We envision this as a $T\times 1$ GP $w(s) = (w_1(s),\ldots,w_T(s))^{\T}$, where $w_t(s) = w(s,t)$.  %\textbf{[SB notes:  Should we not develop this as $qT\times 1$ vector $w(s)$ so that a multivariate time-series can be accommodated? Or do you prefer to have a univariate time-series, which is how you have developed it here?]} %[DD comments: I think, q=T here, so, I didn't understand this note fully. By multivarite time series, was Sudiptoda referring to multiple variable for each location and time?]}.\pink{ AD: I have kept this as a univariate spatial time-series recast as multivariate spatial data, as our data application falls in this category. As you point out correctly we can more generally do this for a multivariate time-series. We have added that extension below.}
Temporal evolution of processes is often encapsulated using a directed acyclic graph (DAG), which, when moralized, produces an undirected graph $\calG_\calT$ over $\calT$. We can then recast the spatial time-series model as a $T\times 1$ GGP with respect to $\calG_\calT$. A multivariate Mat\'ern used for stitching will produce a GGP with each $w_t(\cdot)$ being a Mat\'ern GP with parameters $\theta_{tt}$. Time-specific process variances and spatial parameters enrich the model without imposing stationarity of the spatial process over time and space-time separability \citep{gneiting2002nonseparable}. 

Any autoregressive (AR) structure over time corresponds to a decomposable moralized graph $\calG_\calT$. For example, the $AR(1)$ model corresponds to a path graph with edges $\{(t,t+1)\given t=1,\ldots,T-1\}$, $q^*=2$ and $p^*=2$. An $AR(2)$ is specified by the DAG $t-2 \to t$ and $t-1 \to t$ for all $t \in \{3,\ldots,T\}$ (Figure~\ref{fig:ardag} in the Supplement), which, %for $T=6$ time-points.
when moralized, yields the sparse decomposable graph %graphical model 
$\calG_\calT$ (with $q^*=3$) in Figure~\ref{fig:ar} of the Supplement. %This will be a  with $q^*=3$. Hence, even for large $T$, 
Hence, Corollary~\ref{cor: fact-dens} accrues computational gains for GGP models for autoregressive spatial time-series. An added benefit of using the GGP is that the auto-regression parameters need not be universal, but can be time-specific, thus relaxing another restrictive stationarity condition. 

GGP allows the marginal variances and autocorrelations of the processes to vary over time and be estimated in an unstructured manner. However, more structured temporal models for stochastic volatility can be easily accommodated by a GGP if forecasting the process at a future time-point is of interest. This can be achieved by adding a model for the time-specific variances like the  log-AR(1) model as considered in \cite{jacquier1993priors}. Bayesian estimation of these model parameters has been discussed in \cite{jacquier2002bayesian} and can be seamlessly incorporated into our Bayesian framework for estimation of GGP parameters.

% \pink{Could you elaborate this a little bit more and expand on the frequentist estimation part (which you have mentioned briefly in the next para). Also, I forget why  we  need the pd checks. If we are using Apanasovich's parametrization within each clique, aren't the whole thing guaranteed to be pd as long as within each clique the $b_{ij}$ parameters are pd?}
% from $O(q^3)$ to $O(q^{*3})$ where $q^*$ is the size of the largest clique. Similarly, computation burden of stitching will reduce to $\calO(n^3q^{*3})$ FLOPs. This is a drastic reduction as $q^*$ is often substantially smaller than $q$. For example, a path graph of length $q$ for any $q$ will only have $q^*=2$. 
% \textbf{Put the Bayesian section before Simulations.} 
Multivariate spatial time-series can also be modelled using GGP. We envision $q$ variables recorded at $T$ time-points resulting in $qT$ variables. We now specify $\calG_{\calV\times\calT}$ on the variable-time set. %This can be constructed in a separable (in variable-time) manner, using the strong product graph $\calG_\calV \boxtimes \calG_\calT$. If both $\calG$ and $\calT$ are decomposable, then so is their strong product. If variable-time separability is deemed inappropriate, one can directly specify the graph $\calG_{\calV \times \calT}$. 
Common specifications for multivariate time-series like graphical vector autoregressive (VAR) structures \citep{dahlhaus2003causality} will yield decomposable $\calG_{\calV\times\calT}$. For example, consider the non-separable graphical-VAR of order 1 with $q=2$ and specified by the DAG $(1,t-1) \to (1,t)$, $(1,t-1) \to (2,t)$, and $(2,t-1) \to (2,t)$ (Figure \ref{fig:vardag} of the Supplement). This yields the decomposable $\calG_{\calV\times\calT}$ in Figure \ref{fig:var} of the Supplement, also with $q^*=3$. 

\subsection{\black{Graph estimation}}\label{subsec: unknown-graph}
%Let $y_i= (y_i(s_{i1}),y_i(s_{i2}),\ldots,y_i(s_{in_i}))^{\T}$ be the $n_i\times 1$ vector of measurements for the $i$-th response or outcome over the set of $n_i$ locations in $\calD$. Let $X_i=(x_i(s_{i1}),x_i(s_{i2}),\cdots,x_i(s_{in_i}))^{\T}$ be the known $n_i\times p_i$ matrix of predictors on the set $\calS_i= \{s_{i1}, \cdots, s_{in_i}\}$. We specify the spatial linear model as $y_i = X_i\beta_i + w_i + \eps_i$,
%\end{equation}\label{eqn: bayes-model} 
%where $\beta_i$ is the $p_i\times 1$ vector of regression coefficients, $\eps_i$ is the $n_i\times 1$ vector of normally distributed random independent errors with marginal common variance $\tau^2_i$, and $w_i$ is defined analogously to $y_i$ for the latent spatial process corresponding to the $i$-th outcome. The distribution of each $w_i$ is derived from the specification of $w(s)$ as the $q\times 1$ multivariate graphical Mat\'ern GP with respect to a decomposable $\calG_\calV$. We have talked about estimating the parameters of the above model conditioned on a graph $\calG_\calV$ through a Gibbs sampler implementation in Section \ref{sec: bayes-model}. 

\black{Sections~\ref{sec:lmc}~and~\ref{sec:time} present %multivariate and spatio-temporal 
settings where the decomposable graph for a GGP arises naturally. For gridded spatial data, one can use a spatial graphical lasso to estimate the graph from the sparse inverse spectral density matrix \citep{jung2015graphical}, and plug-in the estimated graph in subsequent estimation of GGP likelihood parameters. For irregularly located spatial data, we now extend our framework in (\ref{eqn:mgp_hierarchical}) to % accommodate an unknown decomposable graph. We conduct inference using MCMC to 
infer about the graphical model itself along with the GGP parameters %. We adapt 
by adapting an MCMC sampler for decomposable graphs \citep{green2013sampling}.
%We first introduce the  ourselves with the definition of a junction tree. 

The {\em junction graph} $G$ of a decomposable $\calG_\calV$ is a complete graph with the cliques of $\calG_\calV$ as its nodes. Every edge in the junction graph is represented as a link, which is the intersection of the two cliques, and can be empty. A {\em spanning tree} of a graph is a subgraph comprising all the vertices of the original graph and is a tree (acyclic graph). Suppose a spanning tree $J$ of the junction graph of $G$ satisfies the following property: for any two cliques $C$ and $D$ of the graph, every node in the unique path between $C$ and $D$ in the tree contains $C \cap D$. Then $J$ is called the {\em junction tree} for the graph $\calG_\calV$ \citep[see Figure 2 of][ for an illustration]{thomas2009enumerating}. %An example of a junction tree is shown in Figure 2 of . 
%Some key properties of junction trees are - (a) a 
A junction tree exists for $\calG_\calV$ if and only if $\calG_\calV$ is decomposable. Also, a decomposable graph can have many junction trees but each junction tree represents a unique decomposable graph. This allows us to transform a prior on decomposable graphs to a prior on the junction trees. If $\mu(\calG_\calV(J))$ is the number of junction trees for the decomposable graph $\calG_\calV$ corresponding to $J$, then a prior $\pi$ on decomposable graphs gives rise to a prior $\tilde{\pi}$ on the junction trees as $\tilde{\pi}(J) = {\pi(\calG_\calV(J))}/{\mu(\calG_\calV(J))}$. In our application, we assume $\pi$ to be uniform over all decomposable graphs with a pre-specified maximum clique size, i.e., $\tilde{\pi}(J) \propto {1}/{\mu(\calG_\calV(J))}$.

With junction trees as a representative state variable for the graph, % faciliates %. This Jump between junction trees as tis 
%computationally faster jumps but incurs the tradeoff of exploring less-connected state space, i.e., the decomposable graphs that could have been reached in one step can now require more moves along the junction tree route. 
the jumps are governed by constrained addition or deletion of single/multiple edges so that the resulting tree is also a junction tree for some decomposable graph. Each graph corresponds to a different GGP model using a specific subset of the cross-covariance parameters. To embed sampling this graph within the Gibbs sampler in Section \ref{sec: bayes-model}, jumps between graphs need to be coupled with introduction or deletion of cross-covariance parameters depending on addition or deletion of edges. We use the reversible jump MCMC (rjMCMC) algorithm of \cite{barker2013bayesian} to carry out the sampling of the graph and cross-covariance parameters and lay out the details in Section \ref{sec:rjmcmc}. %This prompts us to resort to the second step, i.e. 
}

\subsection{Asymmetric covariance functions}\label{sec:ns}
Our examples of stitching have primarily involved the isotropic (symmetric) multivariate Mat\'ern cross-covariances. Symmetry implies $C_{ij}(s,s')=C_{ij}(s',s)$ for all $i,j,s,s'$ and is not a necessary condition for validity of a cross-covariance function. An asymmetric cross-covariance function %s have been discussed in 
\citep{apanasovich2010cross,li2011approach} %and \cite{li2011approach}. An asymmetric cross-covariance 
$C^{a}$ can be specified in-terms of a symmetric cross-covariance $C$ as 
$C^{a}_{ij}(s,s')=C^a_{ij}(s-s')=C_{ij}(s - s' + (a_i - a_j))$, where $a_i$, $i=1,\ldots,q$ are distinct variable specific parameters. Stitching works with any valid cross-covariance function, and if $C^a$ is used for stitching, then the resulting graphical cross-covariance $M^a$ will also be asymmetric, satisfying   
$M^{a}_{ij}(s,s')=C^a_{ij}(s,s')$ for all $(i,j) \in E_\calV$, and $s,s' \in \calL$.

\subsection{Response model}\label{subsec:bgmr}
We outline a Gibbs sampler in Section~\ref{sec: bayes-model} of the Supplement for the multivariate spatial linear model in (\ref{eqn:mgp}), where the latent $q\times 1$ process $w(s)$ is modelled as a GGP. If $|\calL|=n$, then the algorithm needs to sample $\sim O(nq)$ latent spatial random effects $w(\calL)$ at each iteration. % of the model ($w$).  thus resulting $O(nq)$ many parameters being sampled in the Gibbs sampler in \eqref{eqn: gibbs-sam-latent}. Here we present an alternate GGP model directly on the response process that effectuates a low-dimensional MCMC. %The high-dimensionality of the Markov chain can make its convergence slow. 

\begin{figure}[h]
    \begin{center}
        \subfloat[GGP on the latent process. ]{\hspace*{2cm}\includegraphics[scale=0.4,trim={17cm 9cm 1cm 2cm},clip]{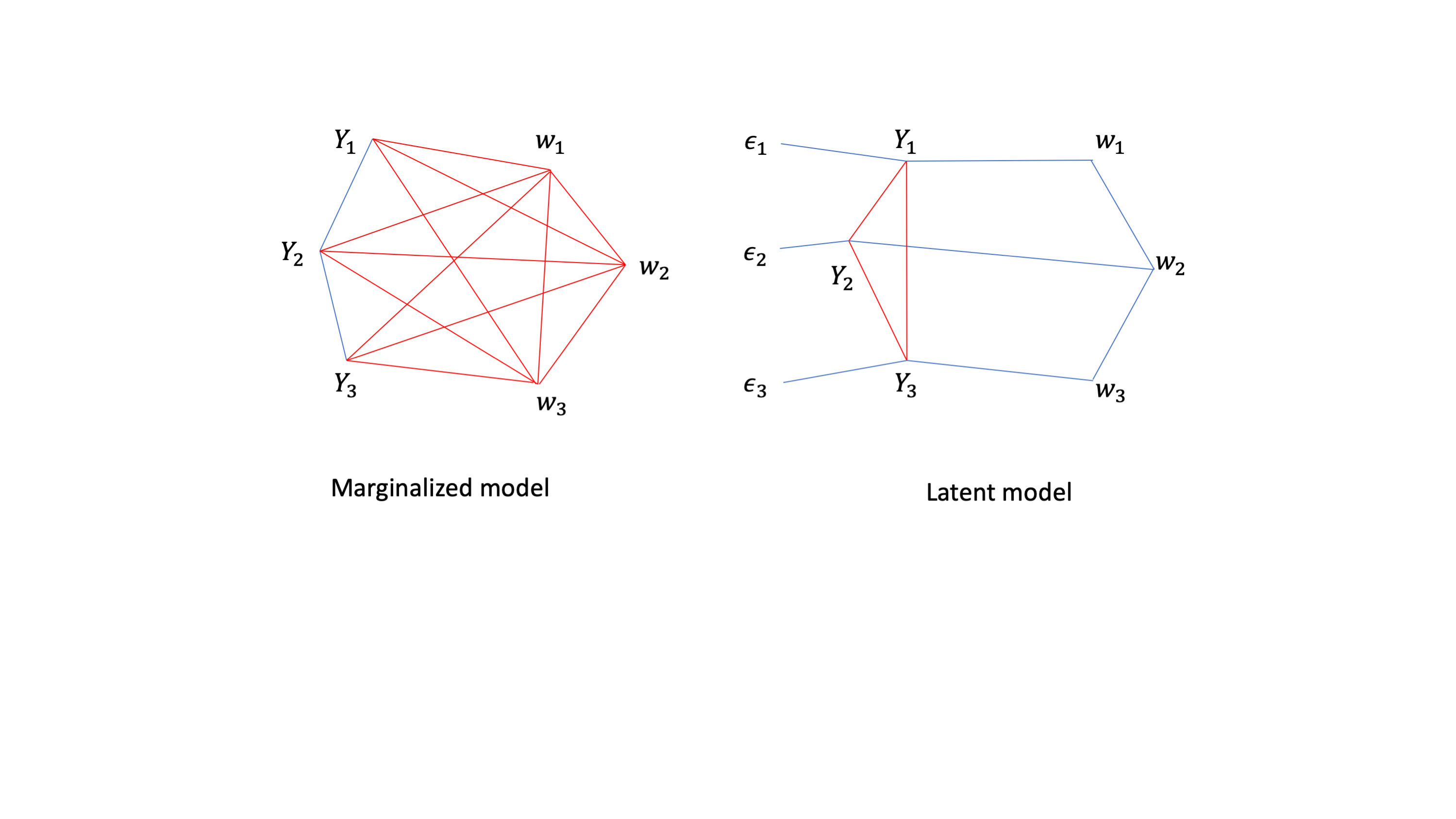}}\hspace{0cm}
        \subfloat[GGP on the response process. ]{\hspace*{1cm}\includegraphics[scale=0.4,trim={6cm 9cm 17cm 2cm},clip]{Figures/BGMR-L.png}}
        \end{center}
    \setlength{\belowcaptionskip}{-10pt}
\setlength{\abovecaptionskip}{-0pt}    
    \caption{Comparison of induced graphs for $3$ processes (obeying a path graph) from marginalized model and latent model. Blue edges indicate the dependencies modelled and red edges denote the marginal dependencies induced from the model construction.}
    \label{fig: marg-unmarg}
\end{figure}

A popular method for estimating spatial process parameters in (\ref{eqn:mgp}) is to integrate out the spatial random effects $w(\calL)$ and directly use the marginalized (or collapsed) likelihood for the response process $y(\cdot)=(y_1(\cdot),\ldots,y_q(\cdot))^{\T}$, which is also a multivariate GP. However, $w(\cdot)$ modelled as a GGP does not ensure that the marginalized $y(\cdot)$ will be a GGP. %However, if $w(\cdot)$ is a GGP, then marginally $y(\cdot)$ is not a GGP. 
We demonstrate this in Figure~\ref{fig: marg-unmarg}(a) with a path graph $\calG_\calV$ between $3$ latent processes $w_1(\cdot)$, $w_2(\cdot)$ and $w_3(\cdot)$. The response processes $y_i(\cdot) = w_i(\cdot) + \epsilon_i(\cdot)$ have complete graphs. This is because  $\mbox{Cov}(y)=\mbox{Cov}(w) + \mbox{Cov}(\eps)$, and the zeros in $\mbox{Cov}(w)^{-1}$ do not correspond to zeros in $\mbox{Cov}(y)^{-1}$. Hence, modelling the latent spatial process as a GGP and subsequent marginalization is inconvenient because the marginalized likelihood for $y$ will not factorize like (\ref{eq:factor}).

\vskip -0.5mm 

Instead, we can directly create a GGP for the response process by stitching the marginal cross-covariance function $\mbox{Cov}(y(s),y(s+h)) = C(h) + D(h)$ using $\calG_\calV$, where $D(h)=\mbox{diag}(\taus_1, \ldots,\taus_q) I(h=0)$ is the diagonal white-noise covariance function. With a Mat\'ern cross-covariance $C$,  the resulting GGP model for $y(\cdot)$ endows each univariate GP $y_i(\cdot)$ with mean $x_i(\cdot)^{\T}\beta_i$ and \black{retaining the marginal} covariance function $C_{ii}(h) + \tau_i^2I(h=0)$ \black{(i.e., Mat\'ern plus a nugget)}. %, where $C_{ii}(h)$ is Mat\'ern. 
The cross-covariance between $y_i(\cdot)$ and $y_j(\cdot)$ is also Mat\'ern for $(i,j) \in E_\calV$ and locations in $\calL$. For $(i,j) \notin \calG_V$, the response processes $y_i(\cdot)$ and $y_j(\cdot)$ will be conditionally independent. We %denote the covariance function of this GGP by $M^*$ and 
outline the Gibbs sampler for this {\em response GGP} in Section~\ref{sec:gibbsmarg} of the Supplement. 

The response model drastically reduces the dimensionality of the sampler from $O(nq+|E_\calV|)$ for the latent model to $O(q+|E_\calV|)$. What we gain in terms of convergence of the chain is traded off in interpretation of the latent process. As we see in Figure~\ref{fig: marg-unmarg}(b), using a graphical model on the response process leads to a complete graph among the latent process. If, however, conditional independence on the latent processes is not absolutely necessary, then the marginalized GGP model is a pragmatic alternative for modelling highly multivariate spatial data. 

\section{Simulations}\label{sec:sims}

\subsection{Known graph}\label{sec:sims-known}
% \textbf{The frequentist simulation section has too many figures and too few (none?) comparisons with either the parsimonious mat'ern of Gneiting or the full multivariate Mat\'ern of Apanasovich. Present maybe two cases: one a small(-ish) gem graph, and one the path graph with 50/100 variables. Compare the parameter estimates with these two methods.}
We conducted multiple simulation experiments to compare three models: (a) PM: Parsimonious Multivariate Mat\'ern of \cite{gneiting2010matern};  (b) MM: Multivariate Mat\'ern of \cite{apanasovich2012valid} with $\nu_{ij}=\nu_{ii}=\nu_{jj} = \frac{1}{2}$, and $\Delta_A=0$  and $\phi^2_{ij}=(\phi^2_{ii}+\phi^2_{jj})/2$; and %described in Section \ref{sec: bayes-model} of the Supplement; 
(c) GM: Graphical Mat\'ern (GGP on the latent process, stitched using multivariate Mat\'ern model (b)). % Bayesian Graphical Mat\'ern on the latent process; (d) BGMR: Bayesian  Graphical Mat\'ern on the response process (Sections \ref{subsec:bgmr}~and~\ref{sec:gibbsmarg}); (e) GM: Graphical Mat\'ern using maximum likelihood estimation (with a co-ordinate descent algorithm outlined in Section~\ref{sec:freq} of the Supplement). All GGP were stitched using the MM model of (b). % with marginal and cross-smoothness  parameters $\nu_{ij}$ fixed to be $0.5$ for model fitting. % all the models.
\begin{table}[t]
\setlength{\abovecaptionskip}{-0pt}
\caption{Different simulation scenarios considered for the comparison between methods.}
\resizebox{\textwidth}{!}{\begin{tabular}{|cccccccc|}\hline
Set & $q$ & Graph $\calG_\calV$ & $B$                                                                                           & Nugget & Locations                                                                                                                      & Data model & Fitted models        \\ \hline
1A   & 5   & Gem (Figure \ref{fig: chrom-w}(a))   & Random                                                                               & No     & Same location for all variables                                                                                                & GM                    & GM, MM, PM\\%, BGMR    \\ 
1B   & 5   & Gem (Figure \ref{fig: chrom-w}(a))   & Random                                                                               & No     & Same location for all variables                                                                                                & MM                    & GM, MM, PM\\%, BGMR    \\ 
2A  & 15  & Path  & $b_{i-1,i}=\rho_i$                                                                             & Yes    &Partial overlap in  locations for variables %\\ 50 different locations hold out for prediction
%\end{tabular} 
& GM                    &  GM, PM\\%, BGML, BGMR \\ 
2B  & 15  & Path  & $b_{i-1,i}=\rho_i$                                                                             & Yes    &Partial overlap in  locations for variables %\\ 50 different locations hold out for prediction
%\end{tabular} 
& MM                    &  GM, PM\\%, BGML, BGMR \\ 
%2B  & 15  & Path  & AR (1) structure                                                                                & Yes    & ,,                                                                                                                             & MM                    &  PM, BGML, BGMR \\ \hline
3A  & 100 & Path  & $b_{i-1,i}=\rho_i$ & Yes    & Partial overlap in  locations for variables                                                                                                                             & GM                    & GM \\%BGML, BGMR\\%, \black{SpDynLm}         \\ 
3B  & 100 & Path  & $b_{i-1,i}=\rho_i$ & Yes    & Partial overlap in  locations for variables                                                                                                                           & MM                    & GM %BGML, BGMR
\\\hline%, \black{SpDynLm}          \\ 
\end{tabular}}
\label{tab: sim-scen}
\end{table}

We consider the 6 settings in Table \ref{tab: sim-scen}. In Sets 1A, 2A, and 3A, we generate data from GM. Set 1A has $q=5$ and uses a gem graph (Figure \ref{fig: chrom-w} (a)).   %We assumed no spatial misalignment, i.e, all variables observed at each location in $\calL$. Hence, we used GM for estimation. 
%As $q$ is small, for this set, we could implement both PM and MM and compare their performance with GM. %Set~1A and 1B did not use a nugget, hence BGML and BGMR are the same for this set. 
For Set~2A, we considered $q=15$ %spatially misaligned 
outcomes and used a path graph, while % $\calG_\calV$ to generate data from GM. %, and included the nugget processes to generate the responses. For Set~2B, we changed the data generating process to be multivariate Mat\'ern. The models fitted are PM, BGML and BGMR. 
Set~3A considers the highly multivariate case with $q=100$ outcomes and  %. The data is generated using GM in 3A with 
 a path graph. % among the variables. %, and using MM in 3B. 
\black{Sets~1B--3B are same as Sets~1A--3A, respectively, except that we generate data from MM. Thus the scenarios 1A--3A correspond to correctly specified settings for the GGP, while scenarios 1B--3B serve as misspecified examples where data is generated from MM.} For all scenarios, we generated data on  $n=250$ locations uniformly chosen over a grid. % within the unit square. 
We simulated 1 covariate $x_j(s_i)$ for each variable $j$, generated independently from a $N(0,4)$ distribution and the true regression coefficients $\beta_j$ from \emph{Unif(-2,2)} for $j=1,2,\ldots,q$.
\begin{figure}[t]
    \begin{center}
     \hspace*{-1mm}\subfloat[Set 1B]{\includegraphics[scale=0.2]{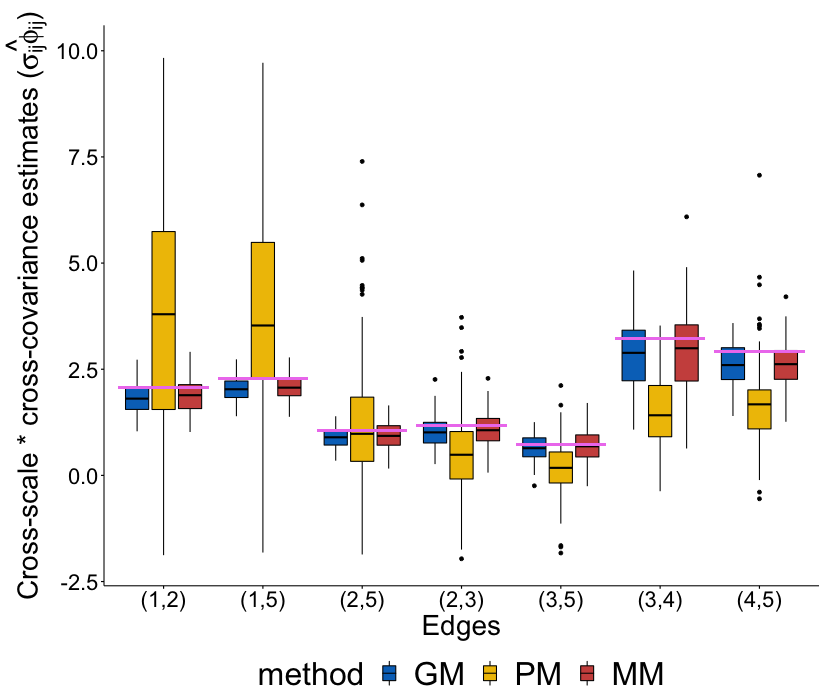}\label{fig:corset1B}}
    \hspace*{-1mm}\subfloat[Set 2B]{\includegraphics[scale=0.2]{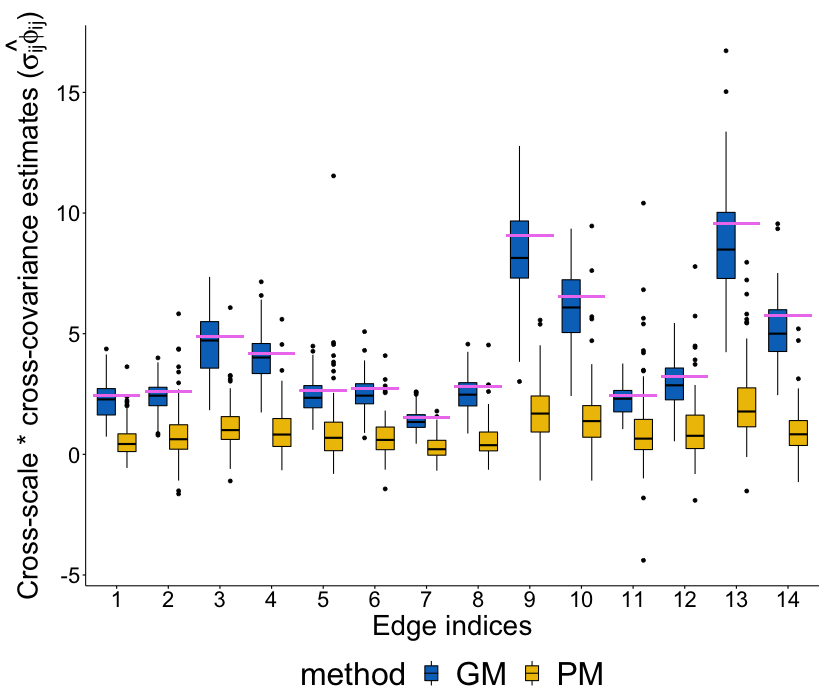}\label{fig:corset2B}}\\
     \hspace*{-1mm}\subfloat[Set 3B]{\includegraphics[scale=0.2]{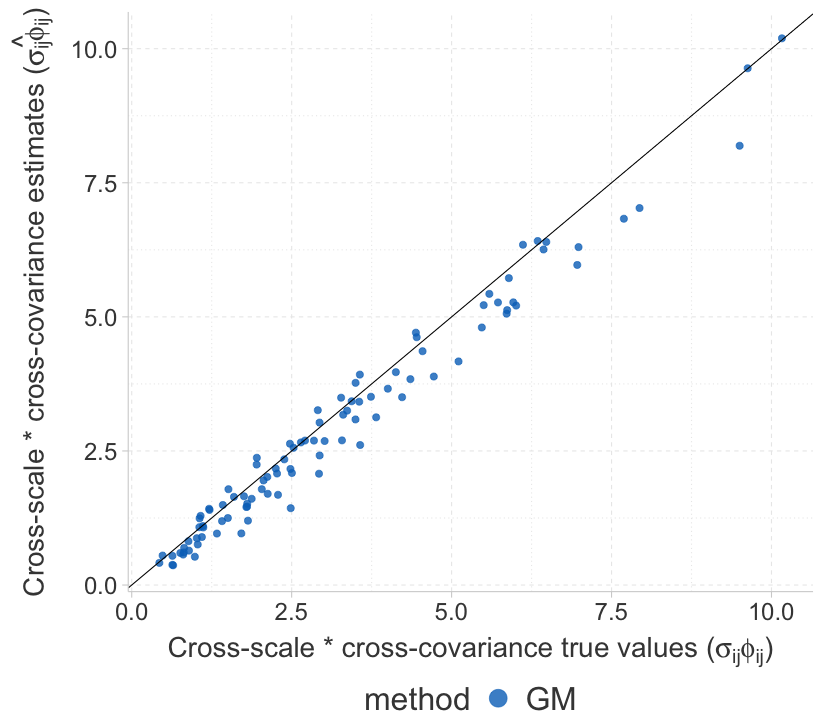}\label{fig:corset3gm}}
         \hspace*{-1mm}\subfloat[Set 1B]{\includegraphics[scale=0.2,trim={0 0 0 0},clip]{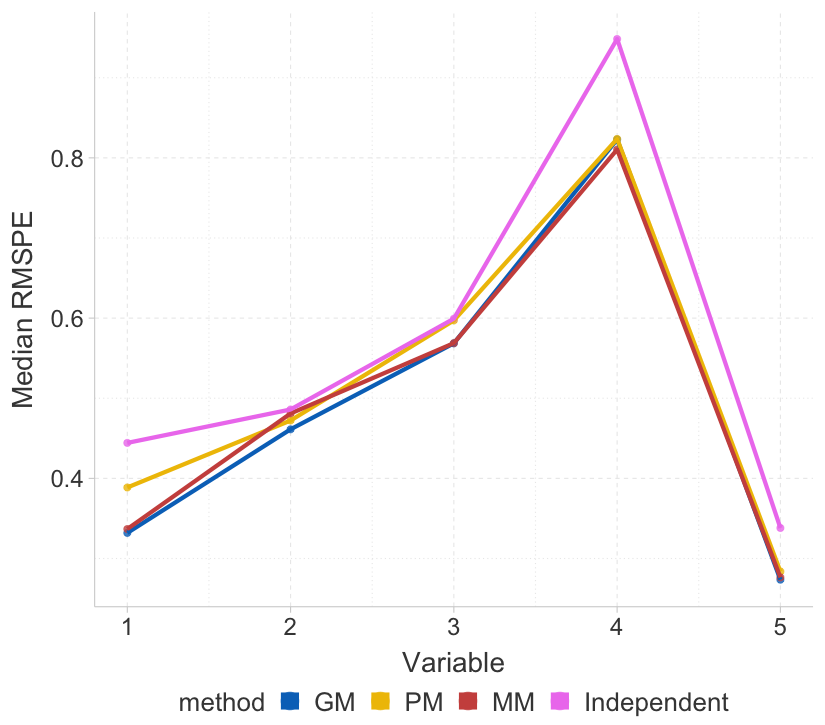}\label{fig:pred1b}}
\setlength{\belowcaptionskip}{-15pt}
\setlength{\abovecaptionskip}{-0pt}
\caption{Performance of graphical Mat\'ern under misspecification:  (a), (b) and (c): Estimates of the cross-covariance parameters $\sigma_{ij}\phi_{ij}=\Gamma(1/2)b_{ij}$, $(i,j) \in E_\calV$ for the sets 1B, 2B and 3B respectively. The pink lines in Figures (a) and (b) indicate true parameter values.  (d): Median RMSPE for GM, MM, PM and Independent GP model for Set~1B.} %violet line to indicate truth. Three plots (a) Product of marginal scale and variance parameter estimates, (b) product of cross-scale and cross-covariance parameter estimates}
    \label{fig:cross}
    \end{center}
\end{figure}
The $\phi_{ii}$ and $\sigma_{ii}$ were equispaced numbers in $(1,5)$, while the $b_{ij}$'s where chosen as in Table \ref{tab: sim-scen}.
For all of the candidate models, each component of the $q$-variate process is a Mat\'ern GP. Following the recommendation outlined in \cite{apanasovich2012valid}, the marginal parameters $\theta_{ii}$ for the univariate Mat\'ern processes were estimated apriori using only the data for the $i$-th variable. The \texttt{BRISC} R-package \citep{saha2018brisc} was used for estimation. %Subsequently, these estimates were plugged into (\ref{eq:factor}) and the co-ordinate descent algorithm was used to obtain the MLE of the cross-covariance parameters $b_{ij}$.
%To demonstrate and compare our methods with existing ones, we consider simulation scenarios considering different kind of underlying graphs among variables as laid out in Table \ref{tab: sim-scen}. 

To compare estimation performance, we primarily focus on the cross-covariance parameters $b_{ij}$, $(i,j) \in E_\calV$, as they specify the cross-covariances in stitching. Specifically, we compare the estimates of $\sigma_{ij}\phi_{ij}=\Gamma(1/2)b_{ij}$, which are the $b_{ij}$'s rescaled to be at the same scale as the marginal microergodic parameters $\sigma_{ii}\phi_{ii}$.  \black{Model evaluations under the correctly specified settings of 1A--3A are provided in Supplementary Figure~\ref{fig:supcross}, which reveals that the GGP accurately estimates cross-covariance parameters for all the edges in the graph for all 3 scenarios. Figures~\ref{fig:cross} (a), (b), and (c), evaluate the estimates of GM for the misspecified settings 1B, 2B and 3B, respectively.} 
 %instead of the individual parameters. 
For Set~1B we see that MM, and GM produce reasonable estimates of the true cross-covariance parameters included, whereas the estimates from PM are biased and more variable. For Set~2B the estimates of PM are once again biased, while GM is more accurate. % estimates much closer to the truth. 

For the highly multivariate settings in Sets~3A and 3B, neither PM nor MM can be implemented because $B$ involves $4,950$ parameters and likelihood evaluation requires inverting a $25,000 \times 25,000$ matrix in each iteration. \black{Hence, we only compare the estimates from GGP to the truth. Figure \ref{fig:corset3mm} shows that the GGP performs well in the highly multivariate setting with misspecification (3B) with GM once again accurately estimating all the $b_{ij}$'s for $(i,j) \in E_\calV$. These simulations under misspecification confirm the accuracy of GGP in  estimating $b_{ij}$ for the MM for pairs $(i,j)$ included in the graph and aligns with the conclusion from Proposition~\ref{lemma: ub-ee}. % used for stitching. We observed that GGP offers accurate estimates for these parameters even when the data is generated from MM. } 
%This is true even for the misspecified case of Set~3B, where both BGML and BGMR accurately estimated all the cross covariances for variable pairs belonging to the graphical model. 

We also evaluate the impact of misspecification on the predictive performance. Figure~\ref{fig:pred1b} plots the root mean square predictive error (RMSPE) based on hold-out data for Set~1B. In addition to the models listed in Table~\ref{tab: sim-scen} we also consider a model where each component GP is an independent Mat\'ern GP serving as a reference for the impact of not modelling dependence. We find GM performs competitively with MM (the correctly specified model) yielding nearly identical RMSPEs for all the 5 variables. PM yields higher RMSPE for variables $1$ and $3$, while the independent model is, unsurprisingly, the least accurate. 
 Additional analyses and discussions are in the Supplementary materials (Section~\ref{sec:suppsim}). These include comparison of marginal parameter estimates (Section~\ref{sec:marg}), impact of excluding edges on estimation of cross-correlation functions (Section~\ref{sec:crossest}), comparison of GGP with dynamic linear models for spatial time series (Section~\ref{sec:simdlm}), comparison of GGP with linear model of coregionalization  (Section~\ref{sec:simlmc}), and comparison among different variants of the GM model (Section~\ref{sec:variants}).}

\subsection{\black{Unknown graph}}\label{sec:simgraph}
\begin{figure}[h]
    \begin{center}
        \subfloat[Posterior edge selection probabilities for Set 2A.]%  (Data generated from GM, True graph is path graph among $15$ variables).]
        {\includegraphics[scale=0.2,trim={0cm 0cm 0cm 1.2cm},clip]{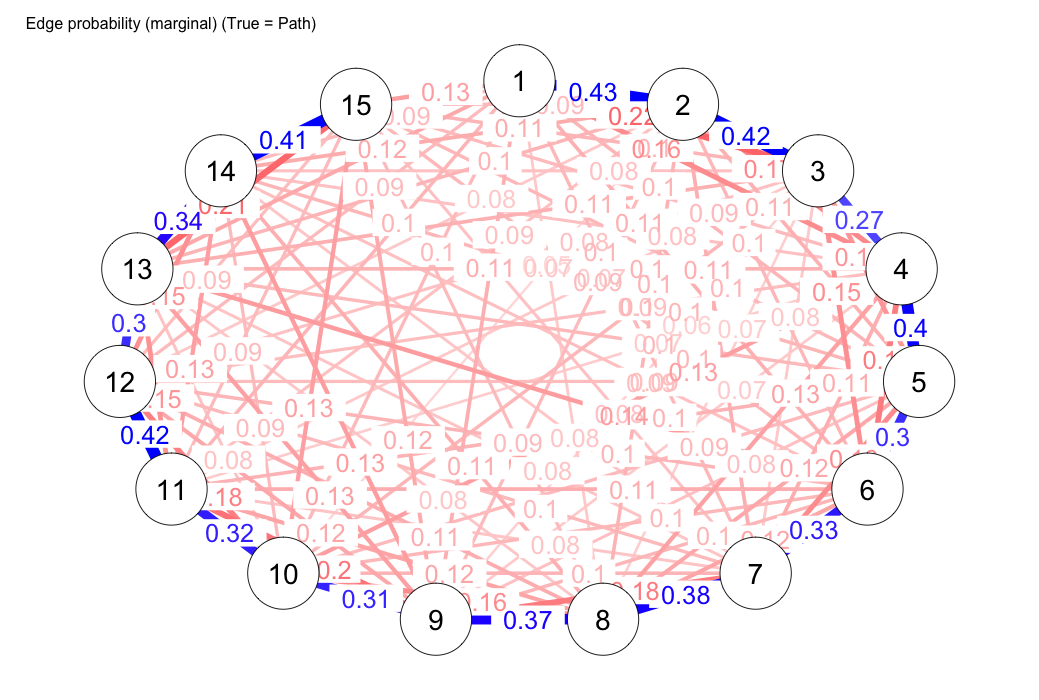}}\hspace{0 cm}
            \hspace*{-1mm}\subfloat[Cross-covariance parameter estimates for Set 2A while estimating the unknown graph]{\includegraphics[scale=0.2,trim={0 0 0 0},clip]{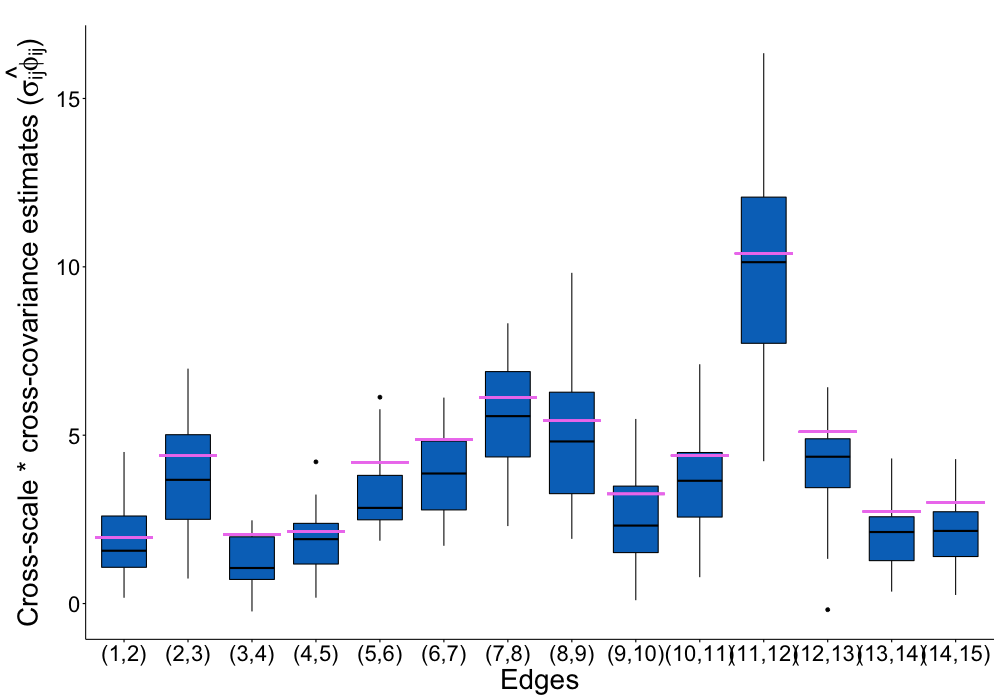}}
        \end{center}
        \setlength{\belowcaptionskip}{-6pt}
        \setlength{\abovecaptionskip}{-2pt}
    \caption{Performance of GGP with unknown graph for Set 2A: (a): Marginal edge probabilities estimated from the reversible jump MCMC sampler. Blue edges denote the true edges and red denotes the non-existent edges. Edges are weighted proportional to the estimated posterior selection probabilities. (b) GM estimates of cross-correlation parameters ($b_{ij}$) corresponding to true edges when the graph is unknown, with horizontal pink lines indicating the true values.}\label{fig: est-graph}
    %\label{fig: est-graph}
\end{figure}

\black{We also evaluated our model when the graph is unknown and is sampled using the reversible jump MCMC  sampler described in Section~\ref{subsec: unknown-graph}. We consider simulation scenarios in Sets~1A~and~2A from Table~\ref{tab: sim-scen}, where the true multivariate process is a graphical Mat\'ern. We assess the accuracy of inferring about the graphical model and the estimates of the cross-covariance parameters. We visualise the estimated edge probabilities for Set 2A in Figure~\ref{fig: est-graph}(a). The blue edges correspond to the true edges, while red ones correspond to false edges. The width of the edges are proportional to the posterior probability of selecting that edge. We see that most of the false edges have narrow width indicating their low selection probability. We report the top 20 probable edges estimated by our model in Table~\ref{tab:edge-prob} of the Supplement and observe that our approach ranks all the $14$ true edges higher than any of the false edges in terms of marginal probability. Figure~\ref{fig: est-graph}(b) shows that the cross-covariance parameters corresponding to true edges are also estimated correctly. The results for Set~1A are similar and presented in Figure~\ref{fig: est-graph1A}.}

\vskip-7mm

\section{Spatial modelling of PM$_{2.5}$ time-series}\label{sec:data}
We demonstrate an application of GGP for non-stationary (in time) and non-separable (in space-time) modelling of spatial time-series (Section~\ref{sec:time}). We model daily levels of PM$_{2.5}$ measured at monitoring stations %is an important public health problem. From United States Environmental Protection Agency website, we have downloaded daily PM 2.5 data collected at monitoring stations 
across 11 states of the north-eastern US and Washington DC for a three month period from February, 01, 2020, until April, 30th, 2020. The data is publicly available from the website of the United States Environmental Protection Agency (EPA). 
%We also gathered PM$_{2.5}$ levels for the same dates (except February 29) for the year 2019 to treat it as a baseline covariate for our analysis. 

We selected $n=99$ stations with at least two months of measured data for both $2020$ and $2019$. Meteorological variables such as temperature, barometric pressure, wind-speed and relative humidity are known to affect PM$_{2.5}$ levels. Since all of the pollutant monitoring stations do not measure all these covariates, we collected the data from NCEP North American Regional Reanalysis (NARR) database, and merged it with the available weather data from EPA to impute daily values of these covariates at pollutant monitoring locations using multilevel B-spline smoothing. Also to adjust for baseline PM$_{2.5}$ levels,  for each station and day in 2020, we included a $7$-day moving average of the PM$_{2.5}$ data for that station  centered around the same day of $2019$  as a baseline covariate %, and was included as a covariate.} %for baseline information. % available for 2019.
We adjust for weekly periodicity of PM$_{2.5}$ levels by subtracting day-of-the-week specific means from raw PM$_{2.5}$ values% and use these adjusted values as outcomes
. Following Section~\ref{sec:time}, we view the spatial time-series at $n=99$ locations and $T=89$ days %of information on PM2.5 levels and the aforementioned covariates. We can treat this as a 
as a highly multivariate ($89$-dimensional) spatial data set. %envisioning the time-points as individual variables. %Akin to the simulation scenarios in Sets~3A~and~3B, 
Neither the parsimonious Mat\'ern nor the multivariate Mat\'ern were implementable as they involve \black{$89^2/2 \approx 4000$} cross-covariance parameters and $9000 \times 9000$ matrix computations \black{($99 \times 89 \approx 9000$)} in each iteration. 

\begin{figure}[!htb]
    \centering
    \hspace*{-1mm}\subfloat[Prediction performance for fortnightly analysis]{\includegraphics[scale=0.2,trim={0 0 0 35}, clip]{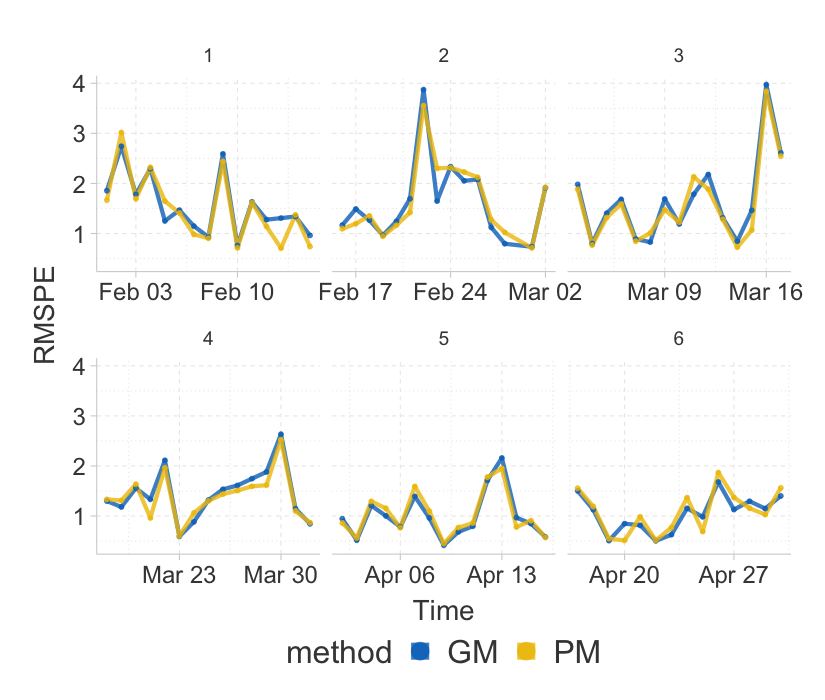}\label{fig:predfortnight}}
    \hspace*{-1mm}\subfloat[Prediction performance for full analysis]{\includegraphics[scale=0.2]{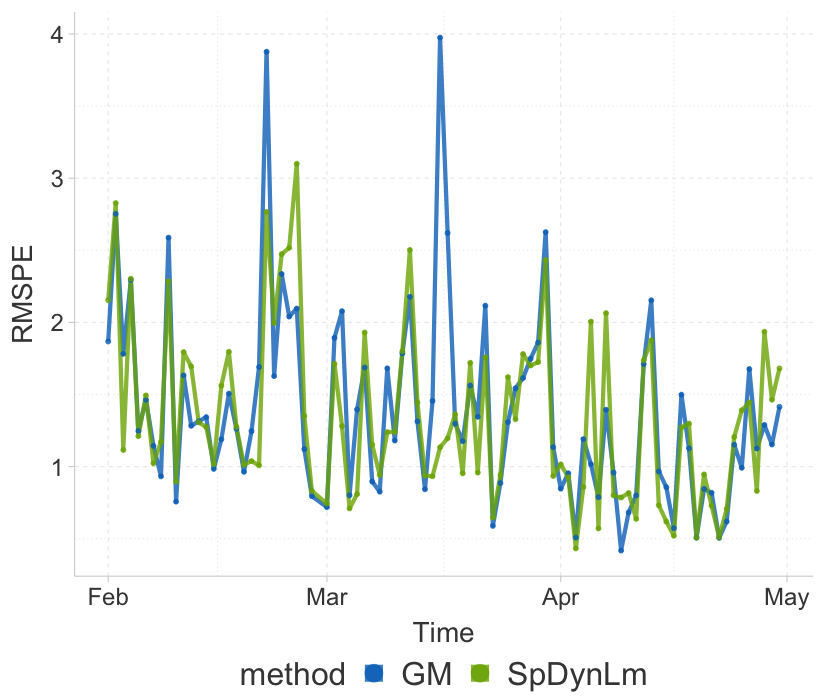}\label{fig:pred-data}}\\
    \hspace*{-1mm}\subfloat[Log-variance estimates for the full analysis]{
\includegraphics[scale=0.2]{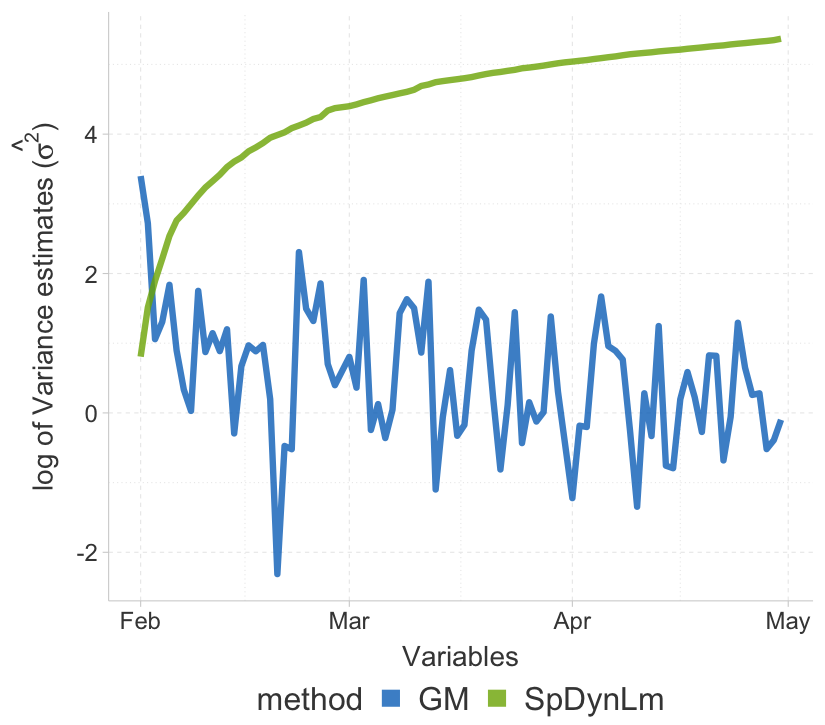}\label{fig:log-sigma-set3-data}}
    \hspace*{-1mm}\subfloat[Estimates of time-specific cross-correlations]{\includegraphics[scale=0.2]{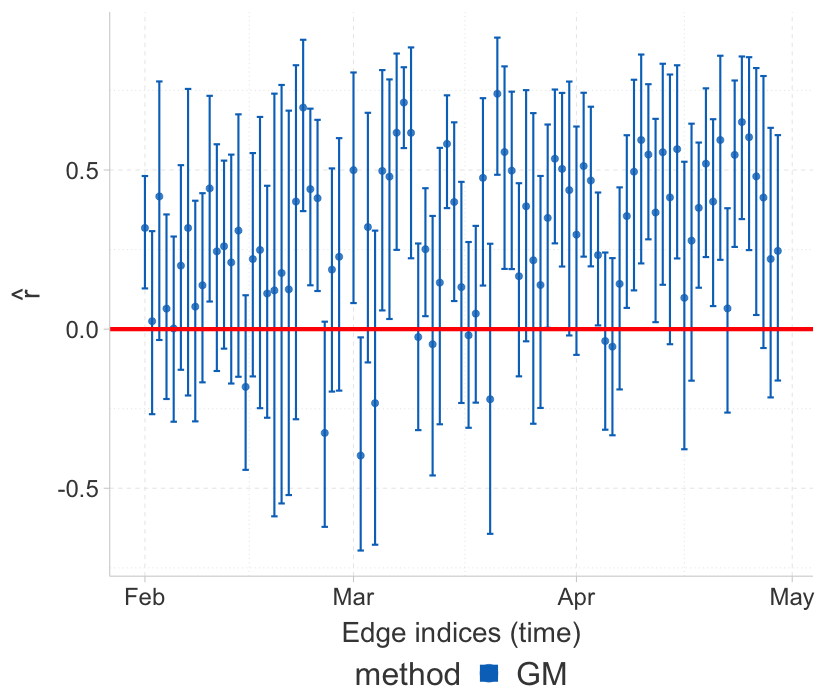}\label{fig:cor-data-ci}}\\
      %  \includegraphics[scale=0.4, angle=0, trim={0 0 0 0},clip]{Figures/Data_analysis/Feb_apr20/Correlation3_set3_ci.png}
	%\caption{Estimates of cross-correlation parameters (inter-day) for BGML and BGMR with confidence intervals} 
    %\label{fig:cor-data-ci}
	%\caption{(a) Product of cross-scale and covariance parameters and (b) RMPSE of test locations for modeling covariate-adjusted PM2.5 over northeastern US locations continuously modelled between February 1 and April 30, 2020}
    %\label{fig: data-set3}
%\end{figure} 
%
%\begin{figure}
%    \centering
    \hspace*{1mm}\subfloat[Average residual surfaces ]{\includegraphics[scale=0.4, angle=0, trim={0 30 0 40},clip]{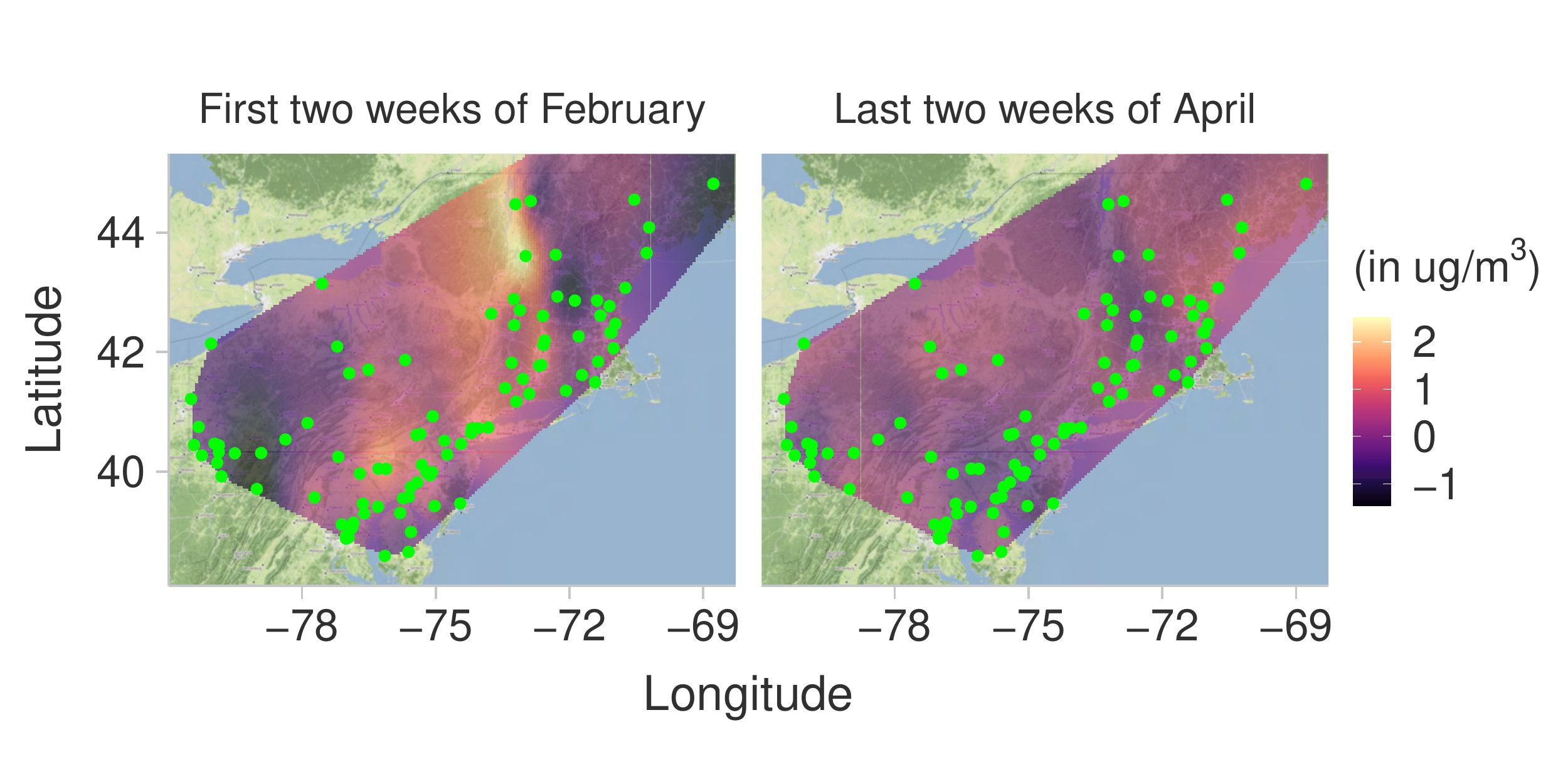}\label{fig:data-heatmap}}
    \setlength{\belowcaptionskip}{-15pt}
\setlength{\abovecaptionskip}{1pt}
\caption{PM$_{2.5}$ analysis: (a) Daily RMSPE for the 6 fortnightly analyses, (b) Daily RMSPE for the full analyses, (c) Estimates of the time-specific process variances, (d) Estimates and credible intervals of the cross-correlation parameters $r_{t,t-1}$ (corresponding to the cross-covariances $b_{t,t-1}$), (e) Estimates of the residual spatial processes from GM (after adjusting for covariates and baseline) for first two weeks of February and last two weeks of April.}\label{fig:pm}
\end{figure} 

We used a graphical Mat\'ern GP with an $AR(1)$ graph based upon exploratory analysis that revealed  autocorrelation among pollutant processes on consecutive days after adjusting for covariates. The marginal parameters for day $t$ were $\sigma_{tt}, \phi_{tt}$ and $\taus_t$. The autoregressive cross-covariance between days $t-1$ and $t$ is $b_{t-1,t}$. % which was parametrized via the corresponding cross-correlation parameter $r_{t,t-1}$. 
Hence, GGP offers the flexibility to model non-separability across space and time, time-varying marginal spatial parameters and autoregressive coefficients. %We implemented both the latent (BGML) and the response (BGMR) models.  

We first present a subgroup analysis breaking $89$ days worth of data into $6$ fortnights. Data for each fortnight is only $14$ or $15$ dimensional and, hence, we are able to analyse each chunk separately using the parsimonious Mat\'ern (PM). %We compare the daily RMSPE based on hold-out data among the three methods (BGML, BGMR, PM). 
Figure~\ref{fig:predfortnight} presents hold-out RMSPE and reveals that GM and PM produce very similar predictive performance when analysing each fortnight of data separately. We analyse the full dataset using the GGP model (GM) as other multivariate Mat\'ern GPs like PM are precluded by the highly multivariate setting. The GGP model involves only $88$ cross-covariance parameters. Since the largest clique size in an AR(1) graph is $2$, the largest matrix we deal with for the data at $99$ stations is only $198 \times 198$. \black{We also consider spatiotemporal models that can model non-stationary and non-separable relationships in the data. \cite{gneiting2002nonseparable} developed general classes of non-separable spatiotemporal models. However, these models assume a stationary temporal process. More importantly, likelihood for this model will involve a dense $9000 \times 9000$ matrix over the set of all space-time pairs and is generally impracticable for modelling long spatial time-series. %Instead, we compare our results with dynamic linear models (DLM) that, like GGP, can parsimoniously model the temporal evolution using an $AR(1)$ structure and allows both non-separability and time-specific parameters. %SpDynLm \citep{finley2013spbayes} similar to the simulation Set~3A and 3B. 

For the full analysis, we compare GGP with a spatial dynamic linear model \citep{stroud2001, gel05} that, like GGP, can parsimoniously model the temporal evolution using an $AR(1)$ structure and allows both non-separability and time-specific parameters. We use the \texttt{SpDynLm} function currently offered in the \texttt{spBayes} package (see Section~\ref{sec:simdlm} of the Supplement for details). Predictive performance is similar for both models with respect to both point predictions (Figure~\ref{fig:pred-data}) and interval predictions (Figure \ref{fig:pred-data-ci}). Figure \ref{fig:log-sigma-set3-data} plots variance estimates (in the log-scale) over time of the latent processes. The implementation in \texttt{spBayes} uses the customary random-walk prior to model for the AR(1) evolution. This enforces these marginal variances to be monotonically increasing resulting in unrealistically large variance estimates for later time-points. The estimates from GGP show substantial variation across time with generally a decreasing trend going from February to April. The estimates and credible intervals for the auto-correlation parameters $r_{t,t-1}$ (normalized $b_{t,t-1}$) from GGP are presented in %Figure~\ref{fig:phisigpm} and auto-correlation parameters in 
Figure~\ref{fig:cor-data-ci}. There is large variation in these estimates across time with many spikes indicating high positive autocorrelation. Quantitatively, $95\%$ Bayesian credible intervals for $40$ out of the $88$ ($45\%$) $r_{t,t-1}$ estimates from GM  %($28$ out of $88$ or $(32 \%)$ from BGMR) 
exclude $0$ providing strong evidence in favour of non-stationary auto-correlation across time. \texttt{SpDynLm} does not have an analogous auto-correlation parameter, and, hence, cannot be compared in this regard.} %BGML, BGMR and SpDynLm show comparable predictive performance with the coverage being slightly better for SpDynLm.  

The estimated average residual spatial surface, $y_\calP(s) = (1/|\calP|) \sum_{t \in \calP} (y_t(s) - x_t(s)^{\T}\hat\beta_t)$, is depicted in Figure~\ref{fig:data-heatmap} for two choices of the time-period $\calP$--the first two weeks of February, 2020 (left), and the last two weeks of April, 2020 (right). These two periods represent the beginning and end of the time period for our study and also correspond to before and during lock-downs imposed in the north-eastern US due to COVID-19.  
%In this way, we compare the average residual spatial surface from the first two weeks of February with last two weeks of April. 
We observe a slight decrease in the magnitude of the residual process from February \black{(median across locations: $0.181$)} to April \black{(median across locations: $0.164$) (Figure \ref{fig:pred-resid})} suggesting a decrease in the PM$_{2.5}$ levels during this period even after accounting for the meteorological covariates \black{and the previous year's level as a baseline.  The  residuals for April also showed much lesser variability compared to that in February, suggesting a decrease in the latent process variance over time. This agrees with the estimates of $\sigma_{tt}$ from GGP (Figure \ref{fig:log-sigma-set3-data}) and contradicts the strongly increasing variance estimates from SpDynLm %. Thus, while the DLM is competitive in terms of predictive performance, it does not offer meaningful quantitative insights on the latent process due to the rigidity of parameter constraints %. unnecessary constraints on the process variance and lack of ability to model time-varying auto-correlation 
(see Section \ref{sec:simdlm} for a broader discussion).}

\section{Discussion}\label{sec: discussion}
% We have addressed modelling highly-multivariate spatial data using GGP with cross-covariance functions that exploit graphical models to ensure process-level conditional independence among the variables, while preserving attractive interpretation of the marginal covariance functions (e.g., the Mat\'ern family) for each univariate process. \black{The existence of such processes and optimality in approximating non-graphical multivariate GPs are formally established.} A pragmatic variant using ``stitching'' is developed.  %We achieve this construction by ``stitching" univariate GPs with a graphical model to construct a new class of multivariate GGP with process-level conditional independence among variables as implied by the graphical model. 
% The highly multivariate setting, where we seek joint modelling of a very large number of spatially dependent variables, is scaled by stitching with decomposable graphs and Mat\'ern GPs. %This development has focused upon the highly multivariate setting that require joint modelling of a very large number of spatially dependent variables. 

This high-dimensional problem we address here accounts for large number of variables and is distinctly different from the burgeoning literature on high-dimensional problems referring to the massive number of spatial locations. %Nevertheless, there are some obvious connections between these two problems in lieu of the recent interest in DAG-based GPs for modelling the latter situation \cite{nngp,pbf2020}.  %,katzfuss2021,pbf2020}.
A future direction will be to simultaneously address the problem of big $n$ and big $q$ by extending stitching to nearest neighbor location graphs with sparse variable graphs. Relaxing the assumption of linear covariate effects $x_i^\T \beta_i$ in (\ref{eqn:mgp}) can also be pursued as discussed recently by \cite{saha2021random}% offered an avenue for this by melding machine learning approaches within the mixed-effects model framework, facilitating scalability via use of sparse GP precision matrices
. A multivariate analogue of this would benefit from the sparse precision matrices available from stitching (\ref{eq:m-decomp}). Finally, the idea of stitching can be transported to the discrete spatial (areal) setting to create multivariate analogs of the interpretable Directed Acyclic Graph Auto-regressive (DAGAR) models \citep{datta2019spatial}%.% Unlike the recent multivariate extensions of DAGAR \citep{gao2021hierarchical}
, 
where
stitching would preserve the univariate marginals being exactly DAGAR distributions.  % that preserve process level. % used in can be addressed by modifying our stitching algorithm on the following lines...\textbf{Abhi, please add (at most) 2-3 sentences what is needed to scale in terms of $n$ and identify this as a future area of research.}  

\subsection{ACKNOWLEDGEMENT}
A. Datta gratefully acknowledges financial support from the National Science Foundation Division of Mathematical Sciences grant DMS-1915803. S. Banerjee gratefully acknowledges support from  NSF grants, DMS-1916349 and DMS-2113778, and from NIH grants NIEHS-R01ES030210 and NIEHS-R01ES027027. This work was completed through a fellowship supported by a Joint Graduate Training Program between the Department of Biostatistics at the Johns Hopkins Bloomberg School of Public Health and the Intramural Research Program of the National Institute of Mental Health. The authors are grateful to the Editor, Associate Editor and anonymous reviewers for their feedback which helped to improve the manuscript.

\bibliographystyle{biometrika}
\bibliography{ref}

\renewcommand\thesection{S\arabic{section}}
\renewcommand\theequation{S\arabic{equation}}
\renewcommand\thefigure{S\arabic{figure}}
\renewcommand\thetable{S\arabic{table}}
\setcounter{section}{0}
\setcounter{equation}{0}
\setcounter{figure}{0}

\pagebreak

\begin{center}
\Large{	Supplementary Materials for ``Graphical Gaussian Process Models for Highly Multivariate Spatial Data"}
\end{center}

\section{Proofs}\label{appn: proofs}
\begin{proof}[of Theorem \ref{th:exists}] \black{Part (a).} 
	For the original GP, $F(\omega)=\{f_{ij}(\omega)\}$ is a valid spectral density matrix. Therefore, following Cramer's Theorem \citep{cramer1940theory,parra2017spectral}, $F(\omega)$ is positive definite (p.d.) for (almost) every frequency $\omega$. Using Lemma~\ref{thm:cov-sel} we derive a unique $\tF(\omega)=(\tf_{ij}(\omega))$, which is also positive definite and satisfies $\tF(\omega)_{ij}=F(\omega)_{ij}=f_{ij}(\omega)$ for $i=j$ or $(i,j) \in E_{\calV}$, and $\tF(\omega)^{-1}_{ij}=0$ for $(i,j) \notin E_{\calV}$. The square-integrability assumption of $f_{ii}(\omega)$ is sufficient to ensure that $\int |\tf_{ij}(\omega)|d\omega < \infty$ using the Cauchy-Schwarz inequality. Thus, we have a spectral density matrix $\tF(\omega)$, which is positive definite for (almost) all $\omega$,  $\tf_{ii}(\omega) = f_{ii}(\omega) > 0$ for all $i$, $\omega$, and  $\int |\tf_{ij}(\omega)|d\omega < \infty$ for all $i,j$. By Cramer's theorem, there exists a GP $w(\cdot)$ with spectral density matrix $\tF(\omega)$ and some cross-covariance function $M$. As by construction $\tf_{ij}(\omega)=f_{ij}(\omega)$ for $i=j$ or $(i,j) \in E_\calV$, we have $M_{ij}=C_{ij}$ for $i=j$ or $(i,j) \in E_\calV$. Since $\tF^{-1}(\omega)_{ij}=0$ for $(i,j) \notin E_\calV$ and almost all $\omega$, using the result of \cite{dahlhaus2000graphical}, $w(s)$ has process-level conditional independence on $\calD$ as specified by $\calG_\calV$, completing the proof.\\ 
	
	\black{Part (b). Let $K(\omega) \in \calF$. By definition, $K(\omega)$ corresponds to the spectral density matrix of a GGP with respect to $\calG_\calV$. By  Theorem~2.4 in \cite{dahlhaus2000graphical}), $(K(\omega)^{-1})_{ij}=0$ for all $(i,j) \notin E_\calV$ and almost all $\omega$. Let $S(K)$ denote the collection of $\omega$ for which this happens. From the construction of $\tilde F$ in part (a), for each $\omega \in S(K)$ we thus have $(K(\omega)^{-1})_{ij}=(\tilde F(\omega)^{-1})_{ij}=0$ for all $(i,j) \notin E_\calV$ and $\tilde F(\omega)_{ij}=F(\omega)_{ij}$ for all $(i,j) \in E_\calV$. Using property (c) of \cite{dempster1972covariance}, we have
		$$ \tr(K(w)^{-1}F(\omega)) + \log \det (K(\omega)) \geq \tr(\tilde F(w)^{-1}F(\omega)) + \log \det (\tilde F(\omega))\; \forall\; \omega \in S(K).$$
		As $S(K)^c$ has measure zero, integrating this over $S(K)$ produces the inequality in part (b). 
	}\\
\end{proof}

\begin{proof}[of Lemma \ref{lem:pp}]
		Consider any $(i,j) \notin E_\calV$, and $s,s' \in \calD$. Let $B=\calV \setminus \{i,j\}$ and $\calA$ denote the $\sigma$-algebra $\sigma(\{w^*_k(s) \given s \in \calD, k \in B\})$. As $w^*_k(s)=C_{kk}(s,\calL)C_{kk}(\calL,\calL)^{-1}w_k(\calL)$ is a deterministic function of $w_k(\calL)$ for all $k,s$, we have $\calA \subseteq \sigma(w_B(\calL))$. On the other hand, as $w_k(s)=w_k^*(s)$ for all $s \in \calL$ (predictive process agrees with the original process at the knot locations) we have $\sigma(w_B(\calL)) \subseteq \sigma(\calA)$. Hence, $\sigma(w_B(\calL)) = \sigma(\calA)$. Now we have
		
		\begin{align}
			& Cov(w_i^*(s),w_j^*(s') \given \calA) = Cov(w_i^*(s),w_j^*(s') \given \sigma(w_B(\calL))\\
			&\quad = C_{ii}(s,\calL)C_{ii}(\calL,\calL)^{-1} Cov(w_i(\calL),w_j(\calL)\given \sigma(w_B(\calL))C_{jj}(\calL,\calL)^{-1}C_{jj}(\calL,s') \\
			&\qquad = 0.
		\end{align}
		
		The last equality follows directly from the construction of stitching for $w(\calL)$.\\
\end{proof}

\begin{proof}[of Theorem \ref{thm: multgp}] For two arbitrary locations $s_1, s_2 \in  \calD$, % \setminus \calL$, 
	we can calculate the covariance function from our construction as follows: 
	\begin{equation}
		\begin{array}{cl}
			M_{ij}(s_1,s_2) & = Cov \left(C_{ii}(s_1,\calL)C_{ii}(\calL,\calL)^{-1}w_i(\calL) + z_i(s_1),
			\right.\\
			&\qquad \left. C_{jj}(s_2,\calL)C_{jj}(\calL,\calL)^{-1}w_j(\calL) + z_j(s_2) \right) \\
			& = C_{ii}(s_1,\calL) C_{ii}(\calL,\calL)^{-1} Cov(w_i(\calL),w_j(\calL))C_{jj}(\calL,\calL)^{-1}C_{jj}(\calL,s_2)) + \\
			&\qquad \mathbb{I}(i=j) C_{ii \given \calL}(s_1,s_2)\\
			& = \mathbb{I}(i = j) [ C_{ii}(s_1,\calL) C_{ii}(\calL,\calL)^{-1} C_{ii}(\calL,s_2)) + C_{ii \given \calL}(s_1,s_2)] + \\
			&\qquad \mathbb{I}(i \ne j)  C_{ii}(\bs_1,\calL) C_{ii}(\calL,\calL)^{-1} M_{ij}(\calL,\calL)C_{jj}(\calL,\calL)^{-1}C_{jj}(\calL,s_2)\\
			& = \mathbb{I}(i = j) C_{ii}(s_1,s_2) + \\
			& \qquad \mathbb{I}(i \ne j)  C_{ii}(s_1,\calL) C_{ii}(\calL,\calL)^{-1}M_{ij}(\calL,\calL)C_{jj}(\calL,\calL)^{-1}C_{jj}(\calL,s_2))
			%    &  C_{ii}(s_1,s_2) - C_{ii}(s_1,\calL) C_{ii}(\calL,\calL)^{-1}C_{ii}(\calL,s_2))]\\
			%    & = \mathbb{I}(i \ne j)  C_{ii}(s_1,\calL) C_{ii}(\calL,\calL)^{-1} M_{ij}(\calL,\calL)C_{jj}(\calL,\calL)^{-1}C_{jj}(\calL,s_2)) + \mathbb{I}(i = j) C_{ii}(s_1,s_2)
		\end{array} \label{eqn: cross-cov-graphmatern}
	\end{equation}
	The second equality follows from the independence of $z_i$ and $z_j$ for $i \neq j$, the third equality uses $M_{ii}(\calL\calL) = C_{ii}(\calL,\calL)$ and the fourth uses the form of the conditional covariance function $C_{ii \given \calL}$ from (\ref{eqn:extend}). It is now immediate, that $w_i$ has the covariance function $C_{ii}$ on the entire domain $\calD$, proving Part~(a). 
	%So, clearly, from \eqref{eqn: cross-cov-graphmatern}, we can infer that, individual processes $\{w_i(.)\}$ defined on the domain $\calD$ will be univariate processes respecting marginal covariance function $C_{ii}$ and 
	
	%Hence we have proved part (1) and (2) of the theorem. 
	
	To prove part~(b), without loss of generality we only consider $q=3$ processes $w_1(s), w_2(s), w_3(s)$ which is constructed via stitching,  %according to \eqref{eq:refset} and \eqref{eqn:extend} 
	with the assumption that $(1,3) \notin E_\calV$. First, we will show that, for any two locations $s_1,s_2 \in \calD$, $w_1(s_1)$ is conditionally independent of $w_3(s_2)$ given $w_2(\calL)$, which we denote as $w_1(s_1) \independent w_3(s_2) \given w_2(\calL)$. 
	
	As $(1,3) \notin E_\calV$, the sets $\{1 \times \calL\} = \{(1,s) \given s \in \calL\}$ and $\{3 \times \calL\}$ are separated by $\{2 \times \calL\}$ in the graph $\calG_V \boxtimes \calG_L$. Hence, using the global Markov property of Gaussian graphical models, we have $w_1(\calL) \independent w_3(\calL) \given w_2(\calL)$. %If $\bs_1, \bs_2 \in \calL$, then this implies that $w_1(\bs_1) \independent w_3(\bs_2) \given w_2(\calL)$. 
	
	%Next we consider the case, if
	For any $s_1,s_2 \in \calD$ we have, similar to (\ref{eqn: cross-cov-graphmatern}),% \setminus \calL$. 
	\begin{equation*}
		\begin{array}{cl}
			Cov(w_1(s_1)w_3(s_2)\given w_2(\calL)) \\
			= C_{11}(s_1,\calL)C_{11}(\calL,\calL)^{-1}Cov\left(w_1(\calL),w_3(\calL)\given w_2(\calL)\right)C_{33}(\calL,\calL)^{-1}C_{33}(\calL,s_2)= 0.
		\end{array}
	\end{equation*}
	%Using the results from \eqref{eq: thm2-2} and \eqref{eq: thm2-3}, we have shown that for any two arbitrary locations $\bs_1, \bs_2$, $Cov(w_1(s_1),w_3(s_2)\given w_2(R))= 0$ and 
	Hence, $w_1(s_1) \independent w_3(s_2) \given w_2(\calL)$ for any $s_1,s_2 \in \calD$. Now  
	\begin{equation}
		\begin{array}{cl}
			Cov(w_1(s_1),\,w_3(s_2)\given\sigma\left(\{w_2(s) \given s \in \calD\}\right)) \\
			= Cov(w_1(s_1),\;w_3(s_2)\given\sigma\left(w_2(\calL),\{ z_2(s)  \given s \in \calD\}\right)) \\
			= Cov(w_1(s_1),\;w_3(s_2)\given\sigma\left(w_2(\calL)\right)) = 0.
		\end{array}\label{eq:processindep}
	\end{equation}
	The second inequality follows due to the agreement of the two conditioning $\sigma$-algebras (similar to the argument in the proof of Lemma ~\ref{lem:pp}). The third inequality follows from the fact that for any three random variables $X,Y$ and $Z$ such that $X$ and $Y$ are independent of $Z$, $E(X|Y,Z) = E(X|Y)$. Equation~(\ref{eq:processindep}) establishes process level conditional independence for $w_1(\cdot)$ and $w_3(\cdot)$ given $w_2(\cdot)$, thereby proving part (b).

	Finally, if $(i,j) \in E_\calV$, and $(s_1,s_2)\in \calL$, in (\ref{eqn: cross-cov-graphmatern}), we will have  $M_{ij}(s_1,s_2)=C_{ij}(s_1,s_2)$ directly from the construction of $\covm$. %To see this, without loss of generality, let $s_1$ and $s_2$ be the first two members of $\calL$. Then $C_{ii}(s_1,\calL)C_{ii}(\calL,\calL)^{-1}$ is going to be $e_1$, a vector with $1$ as its first co-ordinate and $0$ elsewhere. Similarly, $C_{jj}(s_2,\calL)C_{jj}(\calL,\calL)^{-1} = e_2$, and we will have $M_{ij}(s_1,s_2)=C_{ij}(s_1,s_2)$. 
	This proves part (c). \\
	%Now, we can prove the general case of part (2) by following similarly as above, but now treating $w_1,w_3$ as $w_i, w_j$ respectively and replacing $w_2(\calL)$ (a vector) by $w_B(\calL)$ ($B \times R$ matrix). We also need to use the fact that $z_k$'s are independent processes for $k \in B$.
\end{proof}

\begin{proof}[of Corollary \ref{cor: fact-dens}]
	%The conditional independence of multivariate Gaussian distribution is ensured through zeroes in precision matrices. Our construction embeds the Global Markov property for $w(\calL)$ in this way. Since the multivariate normal density is a positive and continuous density, the global markov property implies the factorization of density (p. 35 of \cite{lauritzen1996graphical}). 
	Recall from the construction of $\covm$ that the Gaussian random vector $w(\calL)$ %=\{w_i(s_j); j=1,\cdots,n; i=1,\cdots, q\}$ 
	satisfies the graphical model $\cal{G}= \mathcal{G_V} \boxtimes \mathcal{G_L}$, where $\cal{G_L}$ is the complete graph between $n$ locations. The strong product graph $\calG$ is decomposable and $K_m \boxtimes \mathcal{G_L}; m =1,\cdots,p$ form a perfect sequence for $\calG$ with $S_m \boxtimes \mathcal{G_L};m=2,\cdots, p$ being the separators. 
	% By perfect sequence properties, $S_2 = K_1 \cap K_2$ and $\{S_2 \boxtimes  \mathcal{G_L}\}$ separates $\{\{K_1\boxtimes \mathcal{G_L}\} \setminus \{S_2 \boxtimes  \mathcal{G_L}\}\}$ from $\{\{K_2 \boxtimes \mathcal{G_L}\}  \cup \cdots \cup  \{K_m \boxtimes \mathcal{G_L}\} \setminus \{S_2 \boxtimes  \mathcal{G_L}\}\}$.
	Thus, using results (3.17) and (5.44) from \cite{lauritzen1996graphical}, we are able to factorize $w(\calL)$ as (\ref{eq:factor}). 
	% \begin{equation}\label{eq:factor1}
	%     f_M(w(\calL)) = \frac{f_C(w_{K_1 \boxtimes \mathcal{G_L}}) f_C(w_{\{K_2 \boxtimes \mathcal{G_L}\}  \cup \cdots \cup  \{K_m \boxtimes \mathcal{G_L}\}})}{ f_C(w_{S_2 \boxtimes \mathcal{G_L}})}\;,
	% \end{equation}
	% Repeating this step on $\{K_2 \boxtimes \mathcal{G_L}\}  \cup \cdots \cup  \{K_m \boxtimes \mathcal{G_L}\}$ and so on, we will arrive at (\ref{eq:factor}). 
\end{proof}

\black{\begin{proof}[of Proposition \ref{lemma: ub-ee}]
		
		Suppose, we observe a Multivariate Mat\'ern process. Under the assumption of Graphical Gaussian Processes, the resulting maximum likelihood estimating equations for parameters $\theta_{ij}$ belonging to cliques or separators ($i=j$ or $(i,j) \in E_\calV$) are given by - 
		\begin{equation}
			\begin{split}
				& \frac{\partial \log (f_M(w(\calL))}{\partial{\theta_{ij}}} = 0 \\
				\implies & \frac{\partial}{\partial{\theta_{ij}}} \left(\sum_{K}  \log(f_C(w_K(\calL)) - \sum_{S}  \log(f_C(w_S(\calL))\right) = 0 \\
				\implies & \frac{\partial}{\partial{\theta_{ij}}} \left(\sum_{K \ni (i,j)}  \log(f_C(w_K(\calL)) - \sum_{S \ni (i,j)}  \log(f_C(w_S(\calL))\right) = 0\;,
			\end{split}
		\end{equation}
		where for a subset $a$,  $\log(f_C(w_a(\calL)) =  -\frac{1}{2} w_a(\calL)^T C_a(\theta)^{-1} w_a(\calL) - \log|\det(C_a(\theta)|$. 
		
		Below, we will show that for every subset (clique or separator), the maximum likelihood estimating equation is unbiased. 
		\begin{equation}\label{eq: est-mm}
			\begin{split}
				& E_w\left[\frac{\partial}{\partial{\theta_{ij}}}(-\frac{1}{2} w_a(\calL)^{\T} C_a(\theta)^{-1} w_a(\calL) - \log|\det(C_a(\theta))|)\right]\\
				&\quad = E_w\left[-\frac{1}{2} \tr\left(C_a(\theta)^{-1} w_a(\calL) w_a(\calL)^{\T} C_a(\theta)^{-1} \frac{\partial C_a(\theta)}{\partial \theta_{ij}}\right) + \frac{1}{2} \tr\left(C_a(\theta)^{-1} \frac{\partial C_a(\theta)}{\partial \theta_{ij}}\right)\right] \\
				&\quad = -\frac{1}{2} \tr\left(C_a(\theta)^{-1} E_w\left[w_a(\calL) w_a(\calL)^{\T}\right] C_a(\theta)^{-1} \frac{\partial C_a(\theta)}{\partial \theta_{ij}}\right) + \frac{1}{2} \tr\left(C_a(\theta)^{-1} \frac{\partial C_a(\theta)}{\partial \theta_{ij}}\right) \\
				&\quad = -\frac{1}{2} \tr\left(C_a(\theta)^{-1}\frac{\partial C_a(\theta)}{\partial \theta_{ij}}\right) + \frac{1}{2} \tr\left(C_a(\theta)^{-1} \frac{\partial C_a(\theta)}{\partial \theta_{ij}}\right) = 0\;.
			\end{split}
		\end{equation}
		
		Since the Graphical Gaussian process likelihood is made up of the sums and differences of individual clique and separator likelihoods, the above result ensures that under true parameter values $\theta_{ij}$ of the Multivariate Mat\'ern, we obtain the following which concludes our proof - 
		\begin{equation}
			\begin{split}
				E_w\left[\frac{\partial}{\partial\theta_{ij}} \left(\sum_{K \ni (i,j)}  \log(f_C(w_K(\calL)) - \sum_{S \ni (i,j)}  \log(f_C(w_S(\calL))\right)\right] = 0
			\end{split}\label{eq: unbiased-ee}
		\end{equation}

	\end{proof}
}

\black{
	\begin{proof}[of Proposition \ref{lemma: ggp-lmc}]
		We only need to prove $\mbox{Cov}(w_i(s),w_j(s') \given \sigma(\{f_j(s) \given j \in 1,\ldots,r, s \in\calD\})=0$ for all $i \neq j$ and $s,s' \in \calD$. From Equation (\ref{eq:lmc}), we have $w_i(s)=a_i(s)^{\T}f(s) + \xi_i(s)$ where $a_i(s)=(a_{i1}(s),\ldots,a_{ir}(s))^{\T}$. 
		%$w_i(.), w_j(.)$ which have resulted from a set of latent processes $w_l(.)$. We assumer there exists matrix $A_i$, $A_j$ such that $w_i(s)=A_i w_l(s)$ and $w_i(s)=A_j w_l(s)$ for any $s \in \calD$. Now, for any $s_1,s_2 \in \calD$, 
		\begin{equation*}
			\begin{split}
				&\mbox{Cov}(w_i(s),\,w_j(s')\given\sigma\left(\{f_j(s) \given j=1,\ldots,r, s \in \calD\}\right)) \\
				&\quad = \mbox{Cov}(a_i(s)^{\T} f(s) + \xi_i(s) ,\;a_j(s)^{\T}f(s') + \xi_j(s') \given\sigma\left(\{f_j(s) \given j=1,\ldots,r, s \in \calD\}\right)) \\
				&\quad = \mbox{Cov}(a_i(s)^{\T} f(s),\;a_j(s)^{\T}f(s')\given\sigma\left(\{f_j(s) \given j=1,\ldots,r, s \in \calD\}\right)) + \mbox{Cov}(\xi_i(s) , \xi_j(s'))\\ 
				&\quad = 0 + 0 = 0
			\end{split}
		\end{equation*}
		% \begin{equation*}
		% \begin{array}{cl}
		% Cov(w_i(s),\,w_j(s')\given\sigma\left(\{f_j(s) \given j=1,\ldots,r, s \in \calD\}\right)) \\
		% = Cov(a_i(s)' f(s) + \xi_i(s) ,\;a_j(s)'f(s') + \xi_j(s') \given\sigma\left(\{f_j(s) \given j=1,\ldots,r, s \in \calD\}\right)) \\
		% = Cov(a_i(s)' f(s),\;a_j(s)'f(s')\given\sigma\left(\{f_j(s) \given j=1,\ldots,r, s \in \calD\}\right)) + Cov(\xi_i(s) , \xi_j(s'))\\ = 0 + 0 =0 %\given\sigma\left(\{f_j(s) \given j=1,\ldots,r, s \in \calD\}\right)) \\
		% \end{array}
		% \end{equation*}
		because $a_i(s)^{\T}f(s)$'s are deterministic functions of the conditioning $\sigma$-algebra, and $\xi_i$'s are independent of each other and of the factor processes. 
		%\begin{equation*}
		%\begin{array}{cl}
		%Cov(w_i(s_1),\,w_j(s_2)\given\sigma\left(\{f_j(s) \given j=1,\ldots,r, s \in \calD\}\right)) \\
		%= Cov(A_i w_l(s_1) ,\;A_j w_l(s_2)\given\sigma\left(w_l(s) \given s \in \calD\}\right)) \\
		%= E(A_i w_l(s_1) w_l(s_2)^T A_j^T \given\sigma\left(w_l(s) \given s \in \calD\}\right)) - \\
		%E(A_i w_l(s_1) \given\sigma\left(w_l(s) \given s \in \calD\}\right)) E(w_l(s_2)^T A_j^T \given\sigma\left(w_l(s) \given s \in \calD\}\right)) \\
		%= A_i w_l(s_1) w_l(s_2)^T A_j^T - A_i w_l(s_1) w_l(s_2)^T A_j^T  =0 
		%\end{array}\label{eq:lmcindep}
		%\end{equation*}
		
		Thus we have proved that any pair of observed processes are conditionally independent given the latent processes. When translated into a Graphical Gaussian processes framework, we will observe no edges between the observed nodes and each observed node will be connected to all the factor (latent) nodes. Additionally, we assume  all possible connections (a complete graph) between the factor nodes in their marginal distribution. This gives us a complete graph between the vertices $\{q+j | j \in 1, \ldots, r\}$. Therefore, the graphs on the joint set of observed and factor processes will be decomposable with the perfect ordering of cliques $K_1,\ldots,K_{q}$ where $K_i=\{i\} \cup \{q+i | i \in 1,\ldots,r\}$. % and $K_i=\{i\}$ 
\end{proof}}

\section{Implementation} \label{sec: est}
\subsection{Gibbs sampler for GGP model for the latent processes}\label{sec: bayes-model}
Let $y_i= (y_i(s_{i1}),y_i(s_{i2}),\ldots,y_i(s_{in_i}))^{\T}$ be the $n_i\times 1$ vector of measurements for the $i$-th response or outcome over the set of $n_i$ locations in $\calD$. Let $X_i=(x_i(s_{i1}),x_i(s_{i2}),\cdots,x_i(s_{in_i}))^{\T}$ be the known $n_i\times p_i$ matrix of predictors on the set $\calS_i= \{s_{i1}, \cdots, s_{in_i}\}$. %and let us define $\scrS$ to be the union of the locations on which we observe different variables 
We specify the spatial linear model as $y_i = X_i\beta_i + w_i + \eps_i$,
%\end{equation}\label{eqn: bayes-model}
where $\beta_i$ is the $p_i\times 1$ vector of regression coefficients, $\eps_i$ is the $n_i\times 1$ vector of normally distributed random independent errors with marginal common variance $\tau^2_i$, and $w_i$ is defined analogously to $y_i$ for the latent spatial process corresponding to the $i$-th outcome. The distribution of each $w_i$ is derived from the specification of $w(s)$ as the $q\times 1$ multivariate graphical Mat\'ern GP with respect to a decomposable $\calG_\calV$. Let $\{\phi_{ii}, \sigma_{ii}, \tau^2_{i} | i = 1.\ldots,q\}$ denote the marginal parameters for each component Mat\'ern process %covariance kernels for each  of
$w_i(\cdot)$. 

We elucidate the sampler using a GGP constructed by stitching the simple multivariate Mat\'ern \citep[][]{apanasovich2012valid}, where $\nu_{ij}=(\nu_{ii}+\nu_{jj})/2$, $\Delta_A=0$ in (\ref{eq:constraints}) and $\phi_{ij}^2=(\phi^2_{ii}+\phi_{jj}^2)/2$. Hence, the only additional cross-correlation parameters are $\{b_{ij} | (i,j) \in E_\calV\}$. Any of the other multivariate Mat\'ern specifications in \cite{apanasovich2012valid} that involve more parameters to specify $\nu_{ij}$'s and $\phi_{ij}$'s can be implemented in a similar manner. 
%with joint cross-covariance matrix $M_\scrS$, which is defined as in \eqref{eqn: matern-apas-crosscov}
We consider partial overlap between the variable-specific location sets and take $\calL = \cup_i \calS_i$ as the reference set for stitching. If there is total lack of overlap between the data locations for each variable, we can simply take $\calL$ to be a set of locations sufficiently well distributed in the domain and the Gibbs sampler can be designed analogously. %and we impose conditional independence structure on $M_\scrS$ as defined in $\ref{lemma: precision}$ for a graph $\mathcal{G}$. 

%From now on, we will refer to $\phi$ as one representative element of the set of marginal scale parameters, similarly $\sigma$ as the set of marginal variance parameters and $r$ the set of cross-correlation terms in $Q_V$ in \eqref{eqn: matern-apas-crosscov}. Then, the cross-covariance matrix in \eqref{eqn: matern-apas-crosscov} is defined through $\phi,\sigma,r$. 

%Now, we will show that, a Bayesian approach to model \eqref{eqn: bayes-model} will help us to simultaneously account for varying sets of locations, predict outcome on missing locations treating this as a missing data problem, quantify uncertainty in a proper way and draw inference for marginal parameters based on the full likelihood. 

%\subsection{Latent model} \label{subsec: bgml}

%The latent model can be written out as follows with the priors specified in Equation \eqref{eqn: prior-latent-2}. 
%
%\begin{equation}
%\begin{split}
%  (Y_i|X_i,\beta, w_i, \phi, \sigma, r, \tau_i^2) & \sim N(X_i'\beta_i + w_i, \tau_i^2 I), i=1,2, \cdots, q \\ 
%  w_\calL | \phi, \sigma, r & \sim N(0, M_\calL)
%\end{split}\label{eqn: prior-latent-1}
%\end{equation}

% \textbf{I did not see any mention of how the graph coloring (did you say it was called colored Gibbs sampler?) helps to identify vertices which can be updated in parallel. I thought that explanation was nice and might be worth including along with a small figure demonstrating it.}

%We now outline the steps of the Gibbs sampler. 
Conjugate priors are available for $\beta_i \stackrel{ind}{\sim} N(\mu_i,V_i)$ and $\tau^2_i \stackrel{ind}{\sim} IG(a_i,b_i)$, where $IG$ is the Inverse-Gamma distribution. There are no conjugate priors for the process parameters. %, and we denote their priors as $p(\sigma_{ii})$, $p(b_{ij})$, etc. %\nu_i)$ and $p(b_{ij})$ respectively. %ing techniques to sample from the given posterior density. 
For ease of notation, the collection $M_{a,b}$ the submatrix of $M$ indexed by sets $a$ and $b$, $M_a = M_{a,a}$, and $M_{a|b} = M_{a} - M_{a,b}M_b^{-1}M_{b,a}$. Similarly, we denote $w(a)$ to be the vector stacking $w_i(s)$ for all $(i,s) \in a$. We denote cliques by $K$ and separators by $S$ in the perfect ordering of the graph $\calG_V$. 

The full-conditional distributions for the Gibbs updates of the  parameters  are as follows. 
%\pink{Some of the equations are leaving lots of spaces before and after. Can you fix this?}
\begin{equation*}
	\begin{split}
		p(\beta_i\given \cdot) & \sim N((X_i^{\T}X_i + V_i^{-1})^{-1}(\mu_i + X_i^{\T}(y_i-w_i)), \tau_i^2(X_i^{\T}X_i+ V_i^{-1})^{-1})\;;\\
		p(\tau_i^2\given\cdot) & \sim IG(a+\frac{n_i}{2}, b + \frac{(y_i-X_i^{\T}\beta_i-w_i)^{\T}(y_i-X_i^{\T}\beta_i-w_i)}{2})\;;\\
		p(\sigma_{ii},\phi_{ii},\nu_{ii}\given \cdot) & \propto \frac{\prod_{K \ni i} \frac{1}{|M_{K \times \calL}|^\frac 12} \exp\left(- \frac 12  w(K \times \calL)^{\T} M_{K \times \calL}^{-1} w(K \times \calL)\right)}{\prod_{S \ni i} \frac{1}{|M_{S \times \calL}|^\frac 12} \exp\left(- \frac 12 w(S \times \calL)^{\T} M_{S \times \calL}^{-1} w(S \times \calL)\right)} \times p(\sigma_{ii}) p(\phi_{ii}) p(\nu_{ii})\;; \\
		p(b_{ij}\given \cdot) & \propto \frac{\prod_{K \ni (i,j)} \frac{I(B_{K} > 0 )}{|M_{K\times \calL}|^\frac 12} \exp\left(- \frac 12  w(K \times \calL)^{\T} M_{K \times \calL}^{-1} w(K \times \calL)\right)}{\prod_{S \ni (i,j)} \frac{1}{|M_{S \times \calL}|^\frac 12} \exp\left(- \frac 12 w(S \times \calL)^{\T} M_{S \times \calL}^{-1} w(S \times \calL)\right)} \times p(b_{ij}) \\ 
		&\mbox{ for } (i.j) \in E_\calV\; .
	\end{split}\label{eqn: gibbs-sam-latent-2}
\end{equation*}
To update the latent random effects $w$, let $\calL=\{s_1,\ldots,s_n\}$ and $o_i=\mbox{diag}(I(s_1 \in \calS_i), \ldots, I(s_n \in \calS_i))$ denote the vector of missing observations for the $i$-th outcome. With $X_{i}(\calL) = (x_i(s_1),\ldots,x_i(s_n))^\T$, $y_i(\calL)$ and $w_i(\calL)$ defined similarly, we obtain
\begin{equation*}
	\begin{split}
		p(w_i(\calL)\given \cdot) & \sim N\left(\calM_i^{-1}\mu_i, \calM_i^{-1}\right)\;, \mbox{ where } \\
		\calM_i & =\frac{1}{\tau_i^2}\mbox{diag}(o_i) + \sum_{K \ni i}M_{\{i\} \times \calL|(K\setminus \{i\}) \times \calL}^{-1} - \sum_{S \ni i}M^{-1}_{\{i\} \times \calL|(S\setminus \{i\}) \times \calL}\;, \\
		\mu_i &= \frac{(y_i(\calL) - x_i(\calL)^{\T}\beta_i)\odot o_i}{\tau_i^2} + \\
		& \qquad \sum_{K \ni i}  T_{i}(K) w({(K\setminus \{i\}) \times \calL}) - \sum_{S \ni i} T_i(S) w({(S\setminus \{i\}) \times \calL})\;, \\
		T_{i}(A) &= M_{\{i\} \times \calL|(A\setminus \{i\}) \times \calL}^{-1}  M_{\{i\} \times \calL,(A\setminus \{i\}) \times \calL}M_{(A\setminus \{i\}) \times \calL}^{-1}, \mbox{ for } A \in \{K,S\}.\\
		% U_i(S) &= M_{\{i\} \times \calL|(S\setminus \{i\}) \times \calL}^{-1}  M_{(S\setminus \{i\}) \times \calL,\{i\} \times \calL}M_{(S\setminus \{i\}) \times \calL}^{-1}.
	\end{split}\label{eqn: gibbs-sam-latent}
\end{equation*}

%We denote $\scrT_i= \calL \setminus \scrS_i$. 
% for a clique $K$ in variable graph $\mathcal{G_V}$, the set - $K \times \calL = K \times \scrS$, $S \times \calL= S \times \scrS$,
%$i_{K}=\{K \setminus i\} \times \scrS$, $i \times \scrS_i = i_{\scrS}$ and $i \times \scrT_i= i_{\scrT}$, $i_{K} \cup i_{\scrT}= i_{K\scrT}$ and $i_{K} \cup i_{\scrS}= i_{K \times \calL}$ . Also, for a clique $K$ in containing a variable $i$, let's denote  $M_{i_{{\scrS}.K}} = M_{i_{\scrS}} - M_{i_{\scrS}, i_{K\scrT}}M_{i_{K\scrT}}^{-1}M_{i_{K\scrT},i_{\scrS}}$ and $M_{i_{{\scrT}.K}} = M_{i_{\scrT}} - M_{i_{\scrT}, i_{K \times \calL}}M_{i_{K \times \calL}}^{-1}M_{i_{K \times \calL},i_{\scrT}}$. 

The Gibbs sampler evinces the multifaceted computational gains. The constraints on the parameters $b_{ij}$ no longer require checking the positive-definiteness of $B$, which would require $O(q^3)$ flops for each check. Instead, due to decomposability it is enough to check for positive definiteness of the (at most $q^*$ dimensional) sub-matrices $B_K$ of $B$ corresponding to the cliques of $\calG_\calV$. The largest matrix inversion across all these updates is of the order $nq^* \times nq^*$, corresponding to the largest clique. The largest matrix that needs storing is also of dimension $nq^* \times nq^*$. These result in appreciable reduction of computations from any multivariate Mat\'ern model that involves $nq \times nq$ matrices and positive-definiteness checks for $q \times q$ matrices at every iteration. 

Finally, for generating predictive distributions, note that, as a part of the Gibbs sampler, we are simultaneously imputing $w_i$ at the locations $\calL \setminus \calS_i$. Subsequently, we only need to sample $y_i(\calL \setminus \calS_i) \given \cdot \sim N(X_i(\calL \setminus \calS_i)'\beta_i + w_i(\calL \setminus \calS_i), \tau_i^2 I)$. 

% \textbf{Need to compare the Bayesian model with Apanasovich and parsimonious Mat\'erns in terms of predictions on a hold-out set. Do this for a small graph and for a very large graph.}

\subsection{Gibbs sampler for GGP model for the response processes}\label{sec:gibbsmarg}
Let ${y(\calL)}= (y_1(\calL),\cdots,y_q(\calL))^{\T}$, $X(\calL)= \mbox{bdiag}(X_1(\calL),\ldots,X_q(\calL))$, and $\beta= (\beta_1^\T, \cdots, \beta_q^\T)^\T$. %, and $\mathbb{\tau}= (\tau^2_1, \ldots, \tau^2_q)$. 
We will consider the joint likelihood 
\begin{equation}
	\begin{split}
		\mathbf{y(\calL)}\given X(\calL),{\beta}, \{\phi_{ii}, \sigma_{ii}, \tau^2_i\}_{\{i=1,\ldots,q\}}, \{b_{ij}\}_{\{(i,j) \in E_\calV\}} \sim N\left(X(\calL){\beta}, M^*_{V \times \calL} \right)
	\end{split}\label{eqn: marginal-dist}
\end{equation} 
and impute the missing data $y_i(\calL \setminus \calS_i)$ in the sampler. Let $\calT_i=\{i\} \times (\calL \setminus \calS_i)$,  $U_i(A)= (A \times \calL) \setminus  \calT_i$ for $A \in \{K,S\}$ and $\beta(A)$ be the vector stacking up $\beta_j$ for $j \in A$. %, and $\calU_i(S)= (S \times \calL) \setminus  \calT_i$. 
Also, for any $U \subseteq \calV \times \calL$, let $\tilde X(U)=bdiag(\{ X_j(U \cap (\{j\} \times \calL)) | j \ni U \cap (\{j\} \times \calL) \neq \{\}\}$. %Then, $U_i(A)$ can be written as $\cup_{j \in A} V_{ij}(A)$ where $V_{ij}(A)=U_i(A) \cap (\{j\} \times \calL)$. Defining $\tilde X(U_i(A))=bdiag(\{X_j(V_{ij}(A)): j \in A\})$ and
We have the following updates:
% where each $\tau^2_i$ is repeated $n$ times in the vector. Instead of assuming the conditional independence structure on $M$, in this model, here we assume the structure on $M^* = M + \mathbb{\tau}I$. We layout the Gibbs sampler for this model in \eqref{eqn: gibbs-sam-marginal} in Section \ref{appn: bgmr}. 
%The outcome in missing locations $Y(\scrT_i)$ can be treated as parameters and simulated as similar to  $w(\scrT_i)$ in $\eqref{eqn: gibbs-sam-latent}$ - 
\begin{equation*}
	\begin{split}
		y_i(\calT_i) \given \cdot & \sim N(X_i{(\calT_i)}\beta_i + \\
		& H_i^{-1}\left(\sum_{K \ni i} M_{\calT_i | U_i(K)}^{*-1} M_{\calT_i,\; U_i(K)}^{*-1} M_{U_i(K)}^{*-1} (y(U_i(K))-\tilde X(U_i(K))\beta(K)) \right.  - \\
		& \left. \sum_{S \ni i}M_{\calT_i | U_i(S)}^{*-1} M_{\calT_i,\; U_i(S)}^{*-1} M_{U_i(S)}^{*-1} (y(U_i(S))-\tilde X(U_i(S))\beta(S)) \right),H_i^{-1})\\
		\mbox{ where } & H_i = \sum_{K \ni i}M_{\calT_i | U_i(K)}^{*-1} - \sum_{S \ni i}M_{\calT_i | U_i(S)}^{*-1}
	\end{split}\label{eqn:bgmr-missing}
\end{equation*}
Once again the updates require inversion or storage of matrices of size at most $nq^* \times nq^*$. The updates for the other parameters are similar to that in the sampler of Section ~\ref{sec: bayes-model} of the Supplement with the cross-covariance $M^*$ replacing $M$. The only exception is $\tau^2_i$, which no longer has conjugate full conditionals and are also now updated using Metropolis random walk steps within the Gibbs sampler akin to the other spatial parameters. 

%The response model reduces the dimensionality of the sampler from $O(nq)$ to $O(q + \sum_{i=1}^q (|\calL| - |\calS_i|))$. If the variable specific location sets have substantial overlap then $|\calL| - |\calS_i|$'s are small and the dimension reduction of the parameter space is considerable. % leading to faster MCMC convergence. %, and the marginalized model will be a pragmatic alternative to the latent model. %many parameters and the convergence of chains will be faster. 
%However, what we gain in terms of convergence of the chain gets compromised in interpretation of the latent process. As we see in Figure \ref{fig: marg-unmarg}(b), using a graphical model on the response process leads to a complete graph among the latent process. If, however, inference on the latent processes is not the primary objective, then the marginalized model is a viable alternative for modelling highly multivariate spatial data. 

\subsection{Reversible jump MCMC algorithm}\label{sec:rjmcmc}

We use the reversible jump MCMC (rjMCMC) algorithm of \cite{barker2013bayesian} to carry out the multimodel inference by sampling of the graph and estimating the cross-covariance parameters specific to the graph. We embed the graph sampling described in Section \ref{subsec: unknown-graph} within the Gibbs sampler in Section \ref{sec: bayes-model}. Jumps between graphs need to be coupled with introduction or deletion of cross-covariance parameters depending on addition or deletion of edges. %We use the reversible jump MCMC (rjMCMC) algorithm of \cite{barker2013bayesian} for Bayesian multimodel inference that samples the unknown graph. 
In order to facilitate this, we need a bijection between the parameter sets of the GGP models corresponding to two different graphs. This is achieved by creating a universal parameter (palette) $\psi$ from which all model-specific (graph-specific) parameters can be computed. For example, if we assume the $k$-th graph $\calG_{\calV_k}= (\calV, E_k)$ has $\theta_k$ as the cross-covariance parameter vector, then we need to define an invertible mapping $g_k$ such that $g_k(\psi) = c(\theta_k, u_k)$, where $u_k$'s are irrelevant to graph $k$.  In our case, we define $\psi$ to be the concatenated vector of length $\frac{q(q-1)}{2}$ containing all pairwise cross-covariance parameters, i.e. %it has . %We fix an order apriori for the indices corresponding to pairwise terms, 
$\psi= (\psi_{12}, \psi_{13}, \cdots, \psi_{23},\cdots, \psi_{(q-1),q})$.  We define $g_k(\psi) = \psi^{(k)} = [\{\psi_{ij}^{(k)}: (i,j) \in E_k\},\{\psi_{ij}^{(k)}: (i,j) \notin E_k\}]$ to be the permuted vector of $\psi$.

Using the above setup, we now devise our two-step sampling strategy for the graphs. From the current junction tree $J$, we propose a move to a new junction tree $J'$ by adding or deleting edges. Following \cite{green2013sampling} we calculate the proposal probabilities as $\kappa(J,J')$. %We calculate the posterior likelihood $p(y|\psi, J, .) \tilde{\pi}(J)$ and $p(y|\psi, J', .) \tilde{\pi}(J')$. Combining these two steps, 
The acceptance probability of the new junction tree $J'$ is $\alpha(J,J') = \min\left(1,\frac{p(y|\psi, J', .) \tilde{\pi}(J') \kappa(J,J')}{p(y|\psi, J, \cdot) \tilde{\pi}(J) \kappa(J',J)}\right)$. %During the MCMC sampling, we also randomize the current junction tree by choosing one among its equivalent peers periodically after a certain steps. This step is recommended by \cite{green2013sampling}) to make the complete state space more accessible.

Exploiting the factorisation (\ref{eq:factor})  of stitched GGP likelihoods for decomposable graphs, we can simplify computations in $\alpha(J,J')$. Let $K_J,\; S_J$ be the set of cliques and separators for $J$. % and $K_{J'}, S_{J'}$ be the set of cliques and separators for $J'$. Also, Let
Let $K^{+}(J,J'), K^{-}(J,J')$ and $S^{+}(J,J'), S^{-}(J,J')$ denote, respectively, the cliques and separators added and deleted by the proposed move to $J'$. %and  be the  added and deleted by the proposed moves respectively. 
The ratio for a proposed move from $J$ to $J'$ is % i as follows - 
\begin{equation*}
	\begin{split}
		p(J \longrightarrow J' \given \psi, \cdot) & = \frac{\prod_{K \in K^{+}(J,J')} \frac{I(B_{K} > 0 )}{|M_{K\times \calL}|^\frac 12} \exp\left(- \frac 12  w_K(\calL)^{\T} M_{K \times \calL}^{-1} w_K(\calL)\right) 
		}{\prod_{K \in K^{-}(J,J')} \frac{I(B_{K} > 0 )}{|M_{K\times \calL}|^\frac 12} \exp\left(- \frac 12  w_K(\calL)^{\T} M_{K \times \calL}^{-1} w_K(\calL)\right)} \times \\
		&\frac{\prod_{S \in S^{-}(J,J')} \frac{1}{|M_{S \times \calL}|^\frac 12} \exp\left(-\frac 12 w_S(\calL)^{\T} M_{S \times \calL}^{-1} w_S(\calL)\right) \kappa(J,J') \mu(\calG_\calV(J))}{\prod_{S \in S^{+}(J,J')} \frac{1}{|M_{S \times \calL}|^\frac 12} \exp\left(-\frac 12 w_S(\calL)^{\T} M_{S \times \calL}^{-1} w_S(\calL)\right) \kappa(J',J) \mu(\calG_\calV(J'))}.\\
	\end{split}\label{eqn: gibbs-sam-latent-3}
\end{equation*}
The terms corresponding to the cliques $K^{-}(J,J')$ and separators $S^-(J,J')$ have already been computed from the existing tree $J$. %For the GGP likelihood, 
We only need to evaluate the likelihood factors corresponding to new cliques $K^{+}(J,J')$ and separators $S^{+}(J,J')$ added in the proposed tree $J'$. This makes the jumps between junction trees computationally efficient for the GGP likelihood. Subsequent to moving to a new tree, we modify the Gibbs' sampler (Section~\ref{sec: bayes-model}) to sample the cross-correlation parameters as below. %The priors for logit-transformed $\psi^{(k)}_{ij}$'s are assumed to be normal in line with the prior assumption for $b_{ij}$'s in Section \ref{sec: bayes-model}.
\begin{equation*}
	\begin{split}
		p(\psi^{(k)}_{ij}; (i,j) \in E_k \given J, \cdot) & \propto \frac{\prod_{K_J \ni (i,j)} \frac{I(B_{K} > 0 )}{|M_{K\times \calL}|^\frac 12} \exp\left(- \frac 12  w_K(\calL)^{\T} M_{K \times \calL}^{-1} w_K(\calL)\right)}{\prod_{S_J \ni (i,j)} \frac{1}{|M_{S \times \calL}|^\frac 12} \exp\left(- \frac 12 w_S(\calL)^{\T} M_{S \times \calL}^{-1} w_S(\calL)\right)} \times p(\psi^{(k)}_{ij}) \\
		p(\psi^{(k)}_{ij}; (i,j) \notin E_k  \given J, \cdot) & \propto p(\psi^{(k)}_{ij}).\\
	\end{split}\label{eqn: gibbs-sam-latent-4}
\end{equation*}

\subsection{Co-ordinate descent}\label{sec:freq}
%We lay out efficient estimation strategies for multivariate spatial modeling using Mat\'ern GGP in both frequentist and Bayesian paradigms. From Lemma~\ref{thm: fact-dens}, %it is clear that %Lemma 5.5 in \cite{lauritzen1996graphical} and the definition of multivariate Mat\'ern cross-covariance in (\ref{eq:constraints}) above, 
%we can infer the following: (i) the covariance matrix of the  Mat\'ern GGP for a decomposable $\calG_V$ can be defined only through the $\theta_{ij}$s corresponding to each cliques for the perfect sequence of the variable graph, and (ii) $\theta_{ij}$'s only feature in the  factor densities of (\ref{eq:factor}) corresponding to cliques and separators containing the edge $(i,j)$. %This helps us devise an efficient estimation procedure for classical inference. 

To conduct estimation and prediction using GGP in a frequentist setting, we outline a co-ordinate descent algorithm for maximum likelihood estimation (assuming a known graph). We illustrate the implementation for the case %with no covariates and 
where each of the $q$ variables are measured at $\calL$. The case of spatial misalignment can be handled by an EM algorithm to impute the missing responses for each variable. 
%in this marginal estimation. (b) 
For the frequentist setup, we use the GGP model for the response.  From Corollary~\ref{cor: fact-dens}, the joint  likelihood can be factored into sub-likelihoods corresponding to specific cliques and separators. Let $\theta^{(t)}$ denote the values of the spatial parameters $\theta$ at the $t$-th iteration, and $M^*_\calL=M^*_\calL(\theta)$ denote the GGP covariance matrix of $y(\calV \times \calL)$ from stitching. Let $\theta_{ii}=\{\sigs_{ii},\phi_{ii},\nu_{ii}\}$, $\theta_{-i}=\theta \setminus \theta_{ii}$, $\theta_{-ij}=\theta \setminus \{b_{ij}\}$. %For any subset $A \subset \calV$, let $X(K)$ denote the block diagonal design matrix with blocks $X_i$, $i \in A$. 
Letting %$y(\calL):=y(\calV \times \calL)$, 
$\tilde X(\calL) := \tilde X(\calV \times \calL)$ we immediately have the following updates of the parameters: 
\begin{equation*}
	\begin{split}
		\beta^{(t+1)} &= \left(\tilde X(\calL)^\T M^{*-1}_\calL(\theta^{(t)}) \tilde X(\calL)\right)\tilde X(\calL)^\T M^{*-1}_\calL(\theta^{(t)})y(\calL),\,\\ %\mbox{  where } X=diag(X_1,\ldots,X_q),\\
		\theta_{ii}^{(t+1)} &= \arg\min_{\theta_{ii}} \Big[ \sum_{K \ni i} l_K(\theta_{ii}) - \sum_{S \ni i} l_S(\theta_{ii})\Big], \mbox{ where for any }  A \subset \calV,\\
		l_A(\theta_{ii}) &=  \log(|M^*_{A \times \calL}(\theta_{ii},\theta_{-i}^{(t)})|) + \\
		\qquad & (y(A \times \calL) - \tilde X(A \times \calL)\beta(A))^\T M_{A \times \calL}^{-*1}(\theta_{ii},\theta_{-i}^{(t)}) (y(A \times \calL) - \tilde X(A \times \calL)\beta(A)) , \\
		b^{(t+1)}_{ij} &= \arg\min_{b_{ij}} \Big[ \sum_{K \ni (i,j)} \left( \tilde\ell_K(b_{ij}) - \log(I(B_{K} > 0)) \right)  - \sum_{S \ni (i,j)}  \tilde \ell_S(b_{ij}) \Big] , \mbox{ for } (i,j) \in E_\calV,\\ %\mbox{ where for any }  A \subset \calV, 
		\mbox{ where }& \tilde\ell_A(b_{ij}) =  \log(|M^*_{A \times \calL}(b_{ij},\theta_{-ij}^{(t)})|) + \\
		\qquad & (y(A \times \calL) - \tilde X(A \times \calL)\beta(A))^\T M_{A \times \calL}^{-*1}(b_{ij},\theta_{-ij}^{(t)}) (y(A \times \calL) - \tilde X(A \times \calL)\beta(A)).
	\end{split}
\end{equation*}
The update of $\beta$ involves the large $nq \times nq$ matrix $M^{*-1}_\calL$. However, from (\ref{eq:m-decomp}), $M^{*-1}_\calL$ can be expressed as sum of sparse matrices, each requiring at-most $O(n^3q^{*3})$ storage and computation arising from inverting matrices of the form $C_{K \boxtimes \calL} + D_{K \boxtimes \calL}$. For updates of the spatial parameters $\theta_{ii}$ and $b_{ij}$, coordinate descent moves along the respective parameter and optimizing the negative log-likelihood which is expressed in terms of the corresponding negative log-likelihoods of the cliques and separators  containing that parameter. This process is iterated until convergence. % in the full likelihood. 
Each iteration of the co-ordinate descent has the same complexity of parameter dimension, same computation and storage costs and parameter constraint check as each iteration of the Gibbs sampler, and hence is comparably scalable.

\section{Additional data analyses results}\label{sec:suppsim}

\subsection{Estimation of marginal parameters}\label{sec:marg}

Comparison of the estimates of the marginal (variable-specific) parameters $\theta_{ii}=(\sigma_{ii},\phi_{ii})'$ is of lesser importance because stitching ensures that each univariate process is Mat\'ern GP, similar to the competing multivariate Mat\'ern model. %For the cross-covariance parameters, similarly, we plot estimates $\sigma_{ij}\phi_{ij}$. We present the plots of the cross-correlation parameters $b_{ij}$, $i,j \in \calV$ in the main text (Figure \ref{fig: cross})  The estimates of the marginal parameters are in Figure \ref{fig:marg} of the Supplement.  
The estimates of the marginal microergodic parameters $\sigma_{ii}\phi_{ii}$ are plotted in Figure~\ref{fig:marg} of the Supplement \black{and reveal similar trends to Figure \ref{fig:cross}, with MM and GM accurately estimating the parameters while PM producing poor estimates due to parameter constraints imposed by its simplifying assumptions.} Also, the estimates of the regression coefficients $\beta_j$ were accurate for all models, and are not presented.

\begin{figure}[]
	\centering
	\hspace*{-1mm}\subfloat[1A]{\includegraphics[scale=0.23]{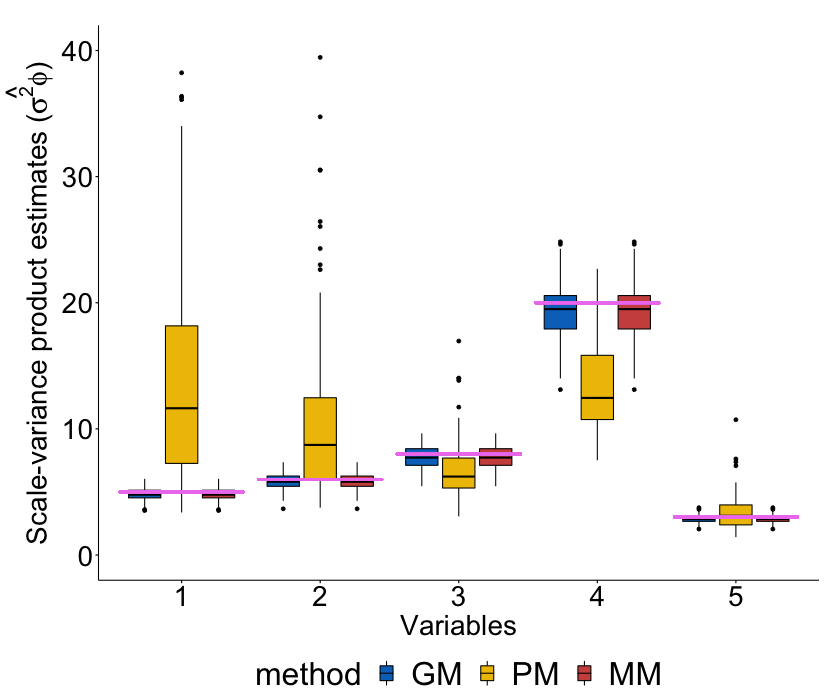}\label{fig:phisigset1A}}
	\hspace*{-1mm}\subfloat[1B]{\includegraphics[scale=0.23]{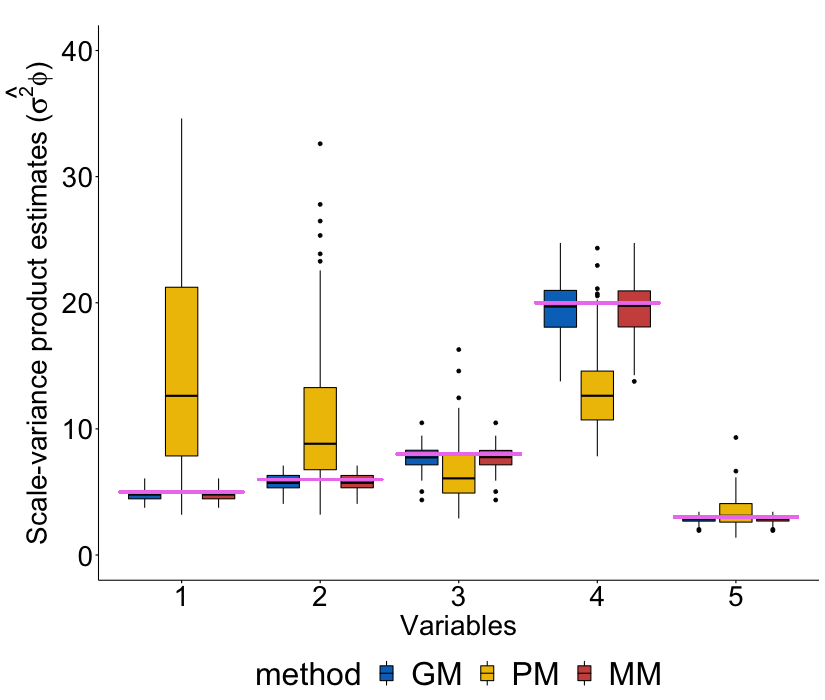}\label{fig:phisigset1B}}\\
	\hspace*{-1mm}\subfloat[2A]{\includegraphics[scale=0.23]{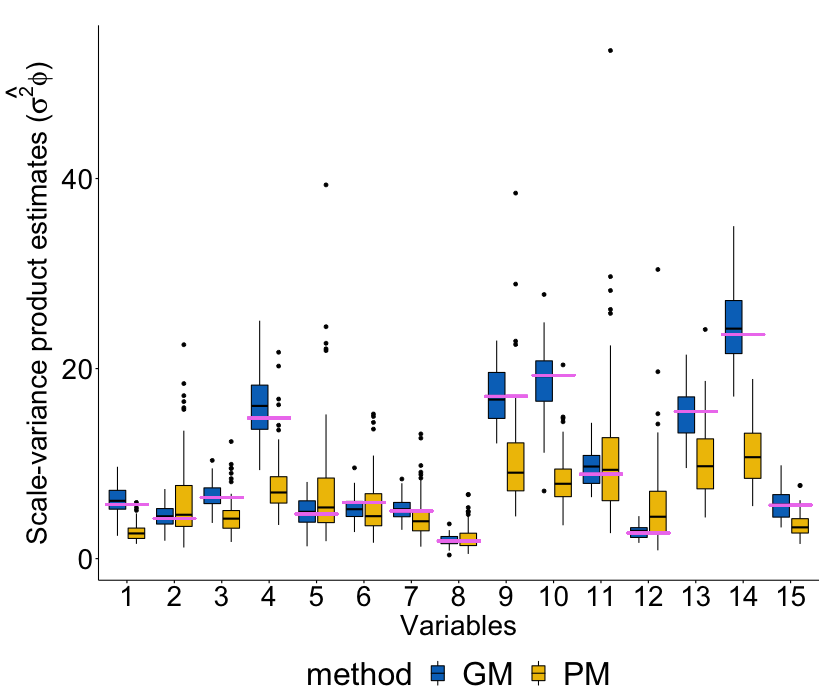}\label{fig:phisigset2A}}
	\hspace*{-1mm}\subfloat[2B]{\includegraphics[scale=0.23]{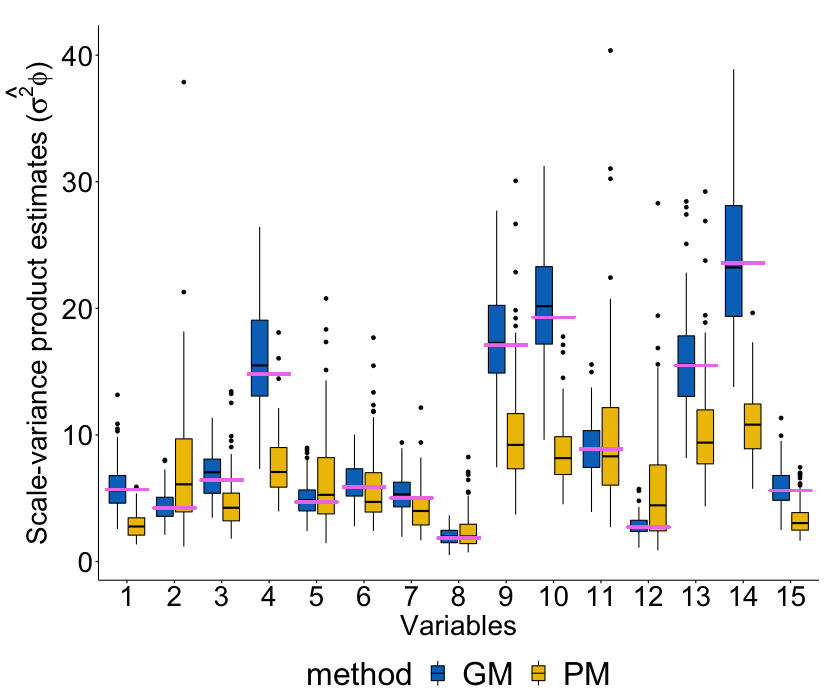}\label{fig:phisigset2B}}\\
	\hspace*{-1mm}\subfloat[3A]{\includegraphics[scale=0.23]{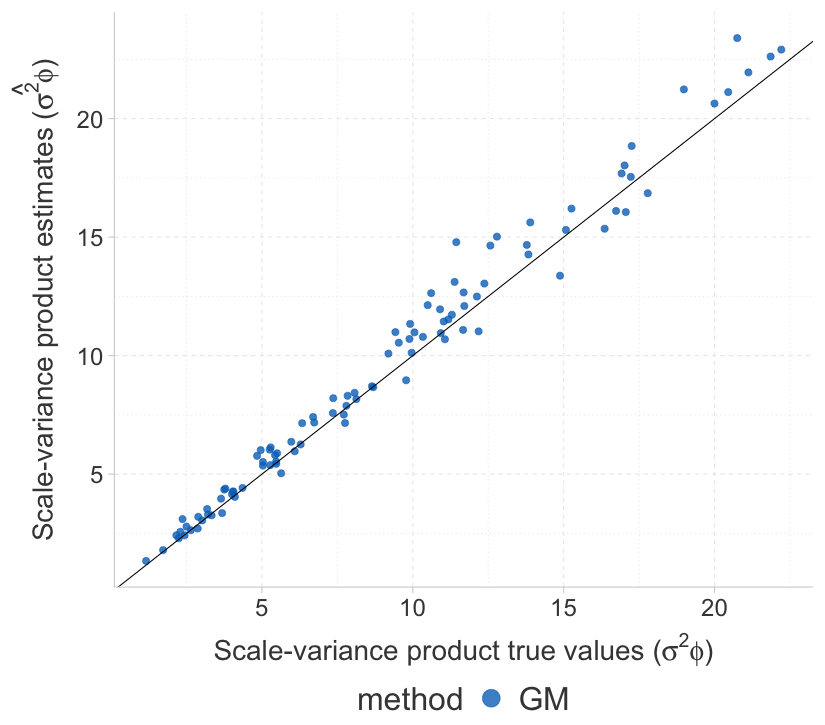}\label{fig:phisigset3mm}}
	\hspace*{-1mm}\subfloat[3B]{\includegraphics[scale=0.23]{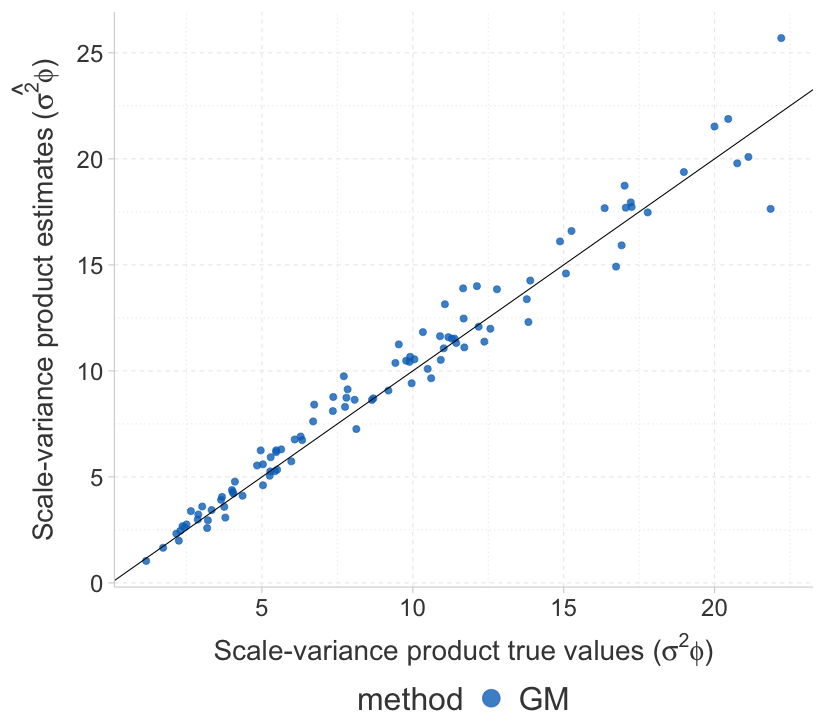}\label{fig:phisigset3gm}}
	%\hspace*{-1mm}\subfloat[]{\includegraphics[scale=0.25]{Figures/Set 2/GM/phisigma_set2.png}\label{fig:phisigset2gm}}
	%\hspace*{-1mm}\subfloat[]{\includegraphics[scale=0.25]{Figures/Set 2/GM/Correlation2_set2.png}\label{fig:corset2gm}}
	%\hskip -1mm\subfloat[]{\includegraphics[scale=0.25]{Figures/Set 2/MM/Prediction_set2.png}\label{fig:predset2mm}}
	%\hskip -1mm\subfloat[]{\includegraphics[scale=0.25]{Figures/Set 2/GM/Prediction_set2.png}\label{fig:predset2gm}}
	\caption{Estimates of the marginal parameters $\sigma_{ii} \phi_{ii}$, $i \in \calV$, for the 6 simulation settings. The horizontal pink lines in Figures (a) and (b) indicate the true parameter values.} %The scale-covariance product estimates coincide for BGML and BGMR.} 
	\label{fig:marg}
\end{figure}

\subsection{\black{Estimates of cross-correlation function under mis-specification}}\label{sec:crossest}
%The above comparisons focused on estimation of the cross-correlation parameters $b_{ij}$ that are included in the graphical model $\calG_\calV$ used for stitching. We observed that GGP offers accurate estimates for these parameters even when the data is generated from MM. This is expected as Proposition \ref{lemma: ub-ee} demonstrates that the GGP likelihood yields unbiased estimating equations for these parameters. 
\black{We also assess the impact of GGP not excluding parameters $b_{ij}$ for all $(i,j) \notin E_\calV$ on the estimates of the cross-correlation functions for these variable pairs. Since these parameters are not in the GGP, we can only compare the true cross-covariance function between these variables pairs against the one indirectly estimated by GGP. 
	
	\begin{figure}[h]
		\centering
		% \hspace*{-1mm}\subfloat[Set 1A]{\includegraphics[scale=0.3, angle=0, trim={0 0 0 0},clip]{Figures/Cross_cov/crosscov_gem_gm.png}\label{fig:crosscorgem_gm}}
		\hspace*{-1mm}\subfloat[Set 1B]{\includegraphics[scale=0.3,trim={0 0 0 0},clip]{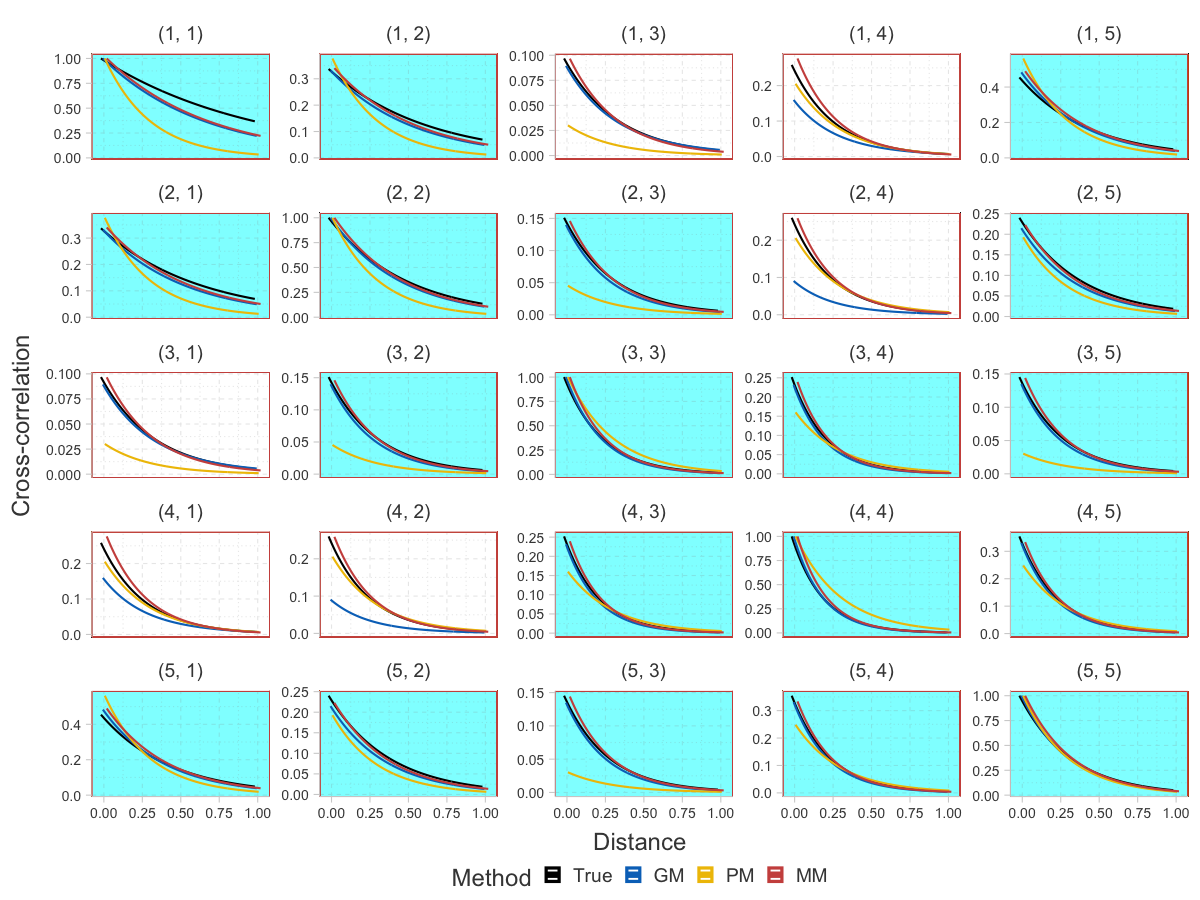}\label{fig:crosscorgem_mm}}
		\caption{Estimates of cross-correlation functions (GM, PM, MM) compared to the truth in Set~1B. The grids correspond to specific pair of the cross-correlations. The sky blue shaded grids correspond to edges in the gem graph assumed for GM.} 
		\label{fig:crosscorgem}
	\end{figure} 
	
	For the misspecified case of Set~1B, Figure \ref{fig:crosscorgem} shows that GM estimates the cross-correlation function pretty well for the assumed edges (blue background) and show bias for some of the variable pairs not included in $\calG_\calV$ (white background). The accurate estimation for the covariance functions (diagonal plots) is attributable to the GGP exactly preserving the marginal distributions of the multivariate Mat\'ern. Similarly, the estimates of the cross-covariance parameters for $(i,j) \in E_\calV$ is expectedly accurate (as is concluded in Proposition \ref{lemma: ub-ee}). 
	
	The bias observed in estimates of the cross-correlation for some pairs of $(i,j) \notin E_\calV$ is also unsurprising. The multivariate Mat\'ern used to generate the data does not follow any graphical model or any other low-rank structure, and any form of dimension-reduction (like modelling dependencies with a sparse graph) will lead to some tradeoff in terms of accuracy and scalability. For analyzing highly multivariate spatial data, even if the variables truly doesn't conform to any graphical model, the MM is not a feasible option due to its high-dimensional parameter space and computing requirements (Table \ref{tab:comp}) and hence dimension-reduction is necessary. Hence, using GGP with a reasonably chosen graphical model that does not exclude important variable pairs is a necessary dimension-reduction step. While it is challenging to establish a bound for the bias for excluded edges in GGP, we have proved that the marginal-preserving GGP is the information-theoretically optimal approximation of a full GP among the class of all GGP (Theorem \ref{th:exists}) . 
	
	We see from Figure that \ref{fig:crosscorgem} that the bias from GM is worse than that of PM for some $(i,j) \notin E_\calV$ (e.g., $(1,4)$ or $(2,4)$). On the other hand, estimates for PM are worse for some $(i,j) \notin E_\calV$ (e.g., $(1,3)$) and for a majority of the pairs $(i,i)$ and $(i,j) \in E_\calV$. PM achieves dimension-reduction by imposing simplifying parameter constraints which degrades its estimation accuracy substantially for most parameters. 
	%and is the trade-off for the dimension-reduction achieved by the graph. For analyzing highly multivariate setting, even if the variables truly doesn't conform to any graphical model, the MM is not an option due to its high-dimensional parameter space and computing requirements. The PM achieves dimension-reduction by imposing simplifying parameter constraints which degrades its estimation accuracy substantially as observed in Figure \ref{fig:cross}. 
	Moreover, PM cannot even be implemented in the truly highly multivariate settings (like sets 3A and 3B) due to requiring $O(q^3)$ for likelihood evaluation (see Table \ref{tab:comp}). %The GGP offers excellent estimates for all marginal parameters as it exactly retains the marginal distributions. It estimates all cross-correlation parameters of the MM included in the graph (Proposition \ref{lemma: ub-ee}). 
	%For cross-correlations not included in the model, for a well-chosen $\calL$, the GGP produced from stitching is a close approximation of the GGP guaranteed by Theorem \ref{th:exists} (b) to be the information theoretically optimal GGP for MM. Hence, one expects this bias to be low unless excluding edges corresponding to highly correlated variables. In fact, we see in \ref{fig:crosscorgem} that the GGP 
	%is generally less biased than the PM for these parameters. 
	The GGP offers drastic improvement in scalability over these alternatives, and for highly multivariate settings, maybe the only viable option guaranteeing accurate estimation of a large subset of the full model parameters. Additionally, we see that exclusion of edges does not severely impact the prediction quality of GM.} % among these choices.

\subsection{\black{Comparison with linear model of coregionalization}}\label{sec:simlmc}
\black{To compare relative performance of linear model of coregionalization and GGP in modelling low-rank processes, we consider the following simulation scenarios: (i) data is generated from an linear model of coregionalization; (ii) data is generated from a Graphical Mat\'ern (GM) respecting the graphical model that would arise from the linear model of coregionalization in scenario (i). For each simulation setting we fit GM and linear model of coregionalization with two factor processes (using \emph{spMisalignLM} function from the \textsf{R} package \emph{spBayes} for our setting of variable-specific locations). % and the Bayesian implementation of Graphical Mat\'ern (BGML).
	Since \emph{spMisalignLM} can be implemented only when the number of observed and latent processes are equal, for generating the data we considered two observed process based on two independent factor processes. This linear model of coregionalization leads to the graphical model from Figure~\ref{fig: graph-lmc-1}. Hence, for scenario (ii) we generate data from a graphical Mat\'ern using this graph to generate correlated factor processes. 
	
	Since the two models correspond to different sets of parameters, we cannot compare them directly. Instead, in Figure~\ref{fig:crosscor-gmlmc} we compare the estimates of the entire correlation and cross-correlation functions. We observe that the impact of misspecification is more pronounced for the linear model of coregionalization; when the true model is graphical Mat\'ern the estimate of the correlation function for the second variable by linear model of coregionalization is quite poor. In comparison, the graphical Mat\'ern estimates the correlation and cross-correlation functions reasonably well both in the correctly specified and in the misspecified case. % case but even when the true model is specified to be LMC. 
	
	\begin{figure}[h]
		\centering
		\hspace*{-1mm}\subfloat[True process: GM]{\includegraphics[scale=0.23, angle=0, trim={0 0 0 50},clip]{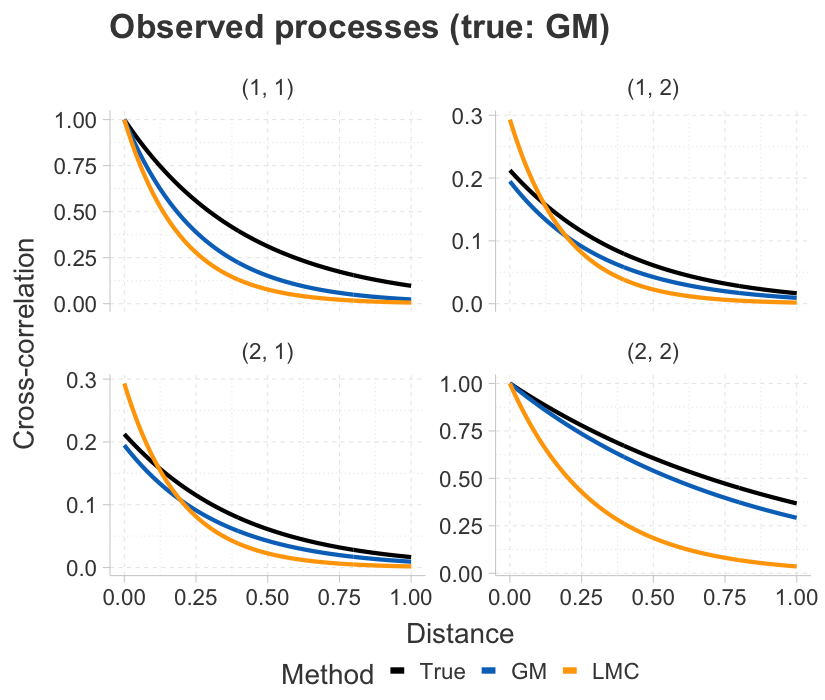}\label{fig:crosscorgm}}
		\hspace*{-1mm}\subfloat[True process: linear model of coregionalization]{\includegraphics[scale=0.23,trim={0 0 0 50},clip]{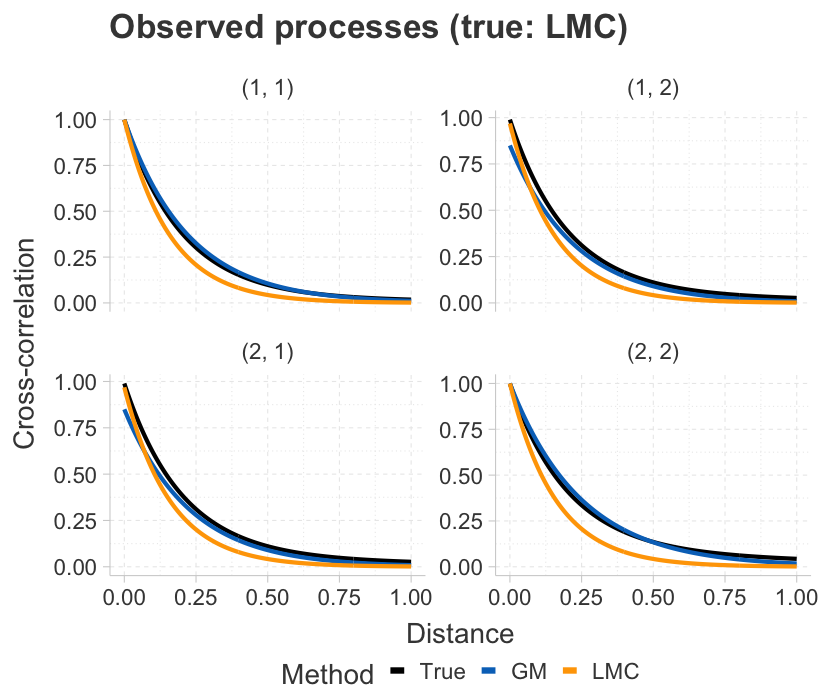}\label{fig:crosscorlmc}}
		\caption{Estimates of cross-correlation functions for the two observed processes. The grids correspond to specific pair of the cross-correlations.} 
		\label{fig:crosscor-gmlmc}
	\end{figure} 
	
	We also compare the predictive performance of the models. For all the simulations performed, we leave out $20\%$ of the data to create test sets in order to evaluate prediction accuracy of the models. Figure \ref{fig:pred-gmlmc} presents the comparison of the prediction and the true values for both models and both data generation scenarios. GM reports marginally improved root mean square prediction error than linear model of coregionalization in all of the situations.}

\begin{figure}[h]
	\centering
	\hspace*{-1mm}\subfloat[True process: GM]{\includegraphics[scale=0.23, angle=0, trim={0 0 0 50},clip]{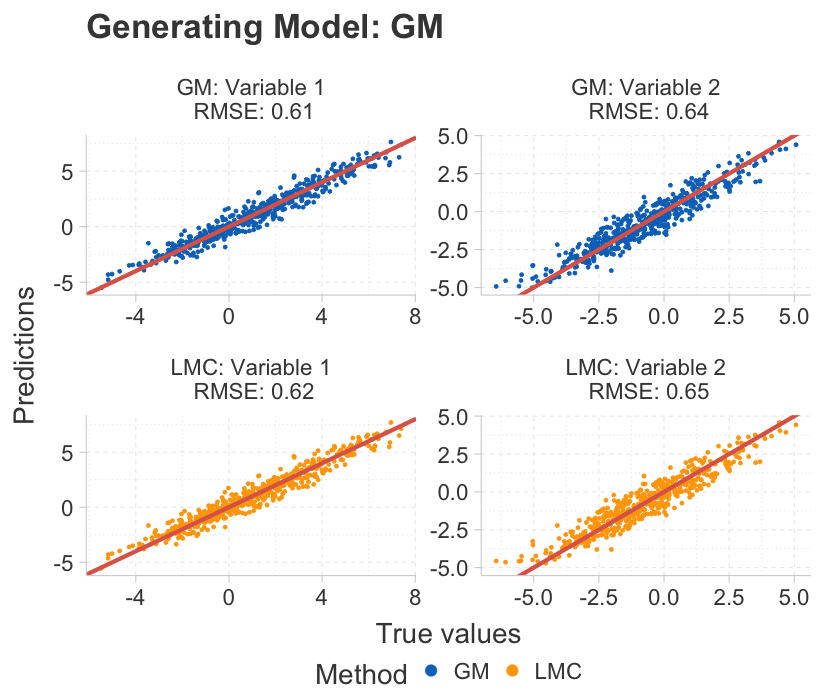}\label{fig:predgm}}
	\hspace*{-1mm}\subfloat[True process: linear model of coregionalization]{\includegraphics[scale=0.23,trim={0 0 0 50},clip]{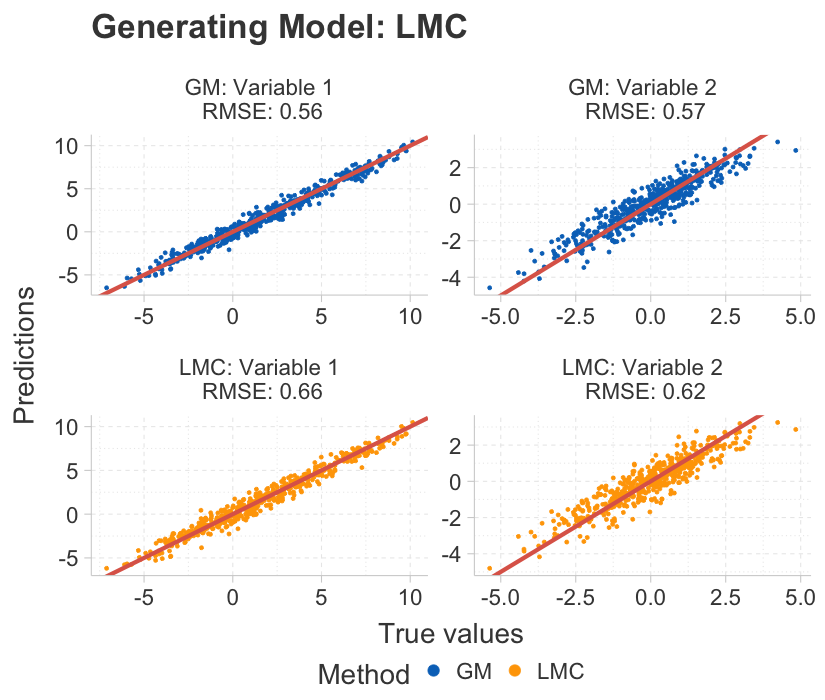}\label{fig:predlmc}}
	\caption{Truth vs prediction for test sets in different simulation scenarios with prediction RMSE reported.} 
	\label{fig:pred-gmlmc}
\end{figure}

%There was very little difference in the estimates from the latent model BGML and the response model BGMR.

\subsection{\black{Comparison with spatial dynamic linear models}}\label{sec:simdlm}
\black{The simulation set 3A corresponds to a highly multivariate setting ($q=100$), where the multivariate process truly follows a graphical model (path graph among the $100$ variables). None of the competing multivariate approaches besides the GGP can model such a graphical structure. These alternatives (PM and MM) also do not scale to our highly multivariate settings. In Section \ref{sec:time} we have illustrated how common univariate and multivariate spatial time-series correspond to decomposable graphical models among the variables and can be modelled using GGP. The path graph in setting 3A corresponds to the decomposable graph resulting from an AR(1) temporal structure. Hence, for this set we compare the performance of the GGP with a dynamic linear model (DLM) \citep{stroud2001,gel05,finley2012bayesian} commonly used for modelling spatial time-series with AR(1) temporal evolution. 
	
	In particular, we compare GGP with the spatial dynamic linear model ({\em SpDynLm}) of \cite{finley2012bayesian} which is set in the GP-based mixed-effect modelling setup similar to (\ref{eqn:mgp}), as opposed to spatial basis function based approach of \citep{stroud2001}. SpDynLM models the spatial process $w_t(\cdot)=w(\cdot,t)$ at time $t$ as 
	\begin{equation}\label{eq:dlm}
		w_t(s) = w_{t-1}(s) + \delta_t(s);\quad \delta_t(\cdot) \sim GP(0,C_{tt}),
	\end{equation}
	i.e., at each time-point the spatial process is a sum of the process at the previous time point and an independent time-specific GP. The rest of the model is the same as in our setup (Eq. \ref{eqn:mgp}) with SpDynLM enforcing an auto-regressive evolution model for the regression coefficients $\beta_t$ as well.
	
	Both SpDynLm and any GGP with a path graph between the time-specific variables model an AR(1) evolution over time. However, any DLM using an additive model of the type  (\ref{eq:dlm}) for the temporal evolution of the latent processes $w_t(c\dot)$, unfortunately, enforces the processes $w_t(\cdot)$ to have the same smoothness at all time-points $t$. Thus, even if the $\delta_t(\cdot)$'s are modelled using Mat\'ern GPs with time-specific smoothness, range and variance parameters, none of the processes $w_t(\cdot)$ will be Mat\'ern GPs and each will have the smoothness of the roughest of the independent processes $\delta_t(c\dot)$. 
	
	Another major restriction of the SpDynLM model is that (\ref{eq:dlm}) is the customary {\em random walk prior} \citep{stroud2001} for $w_t(\cdot)$ which imposes the assumption that $\mbox{Var}(w_t(s)) > \mbox{Var}(w_{t-1}(s))$ for all $t,s$, i.e., that the process variance is monotonically increasing over time. For most spatiotemporal processes this assumption is unlikely to hold. The GGP, on the other hand, ensures that the processes $w_t(\cdot)$ for each time $t$ can be modelled using a Mat\'ern GP with time-specific  variance parameters. 
	
	The dynamic model in (\ref{eq:dlm}) also implicitly assumes a constant (over time) auto-regression coefficient of $1$. While this can be easily relaxed by replacing $w_{t-1}(s)$ with $\rho_t w_{t-1}(s)$ in (\ref{eq:dlm}), the current implementation of SpDynLm does not allow modelling such a non-stationary auto-correlation coefficient $\rho_t$. In a GGP with a path graph, the cross-correlation parameter $b_{t,t-1}$ between the processes at two consecutive times is time-specific, thereby allowing it to capture non-stationary auto-regressive structures. % (as seen in the accurate estimation of the non-stationary autoregressive covariances in Figure \ref{fig:corset3mm}).
	
	%Here, we implement SpDynLm as a competing candidate to our methods as the path graph is the same as one arising from an AR(1) spatial time-series (see Section \ref{sec:time}).
	
	\begin{figure}[h]
		\centering
		\subfloat[Variance estimates (in log-scale).]{\includegraphics[scale=0.25]{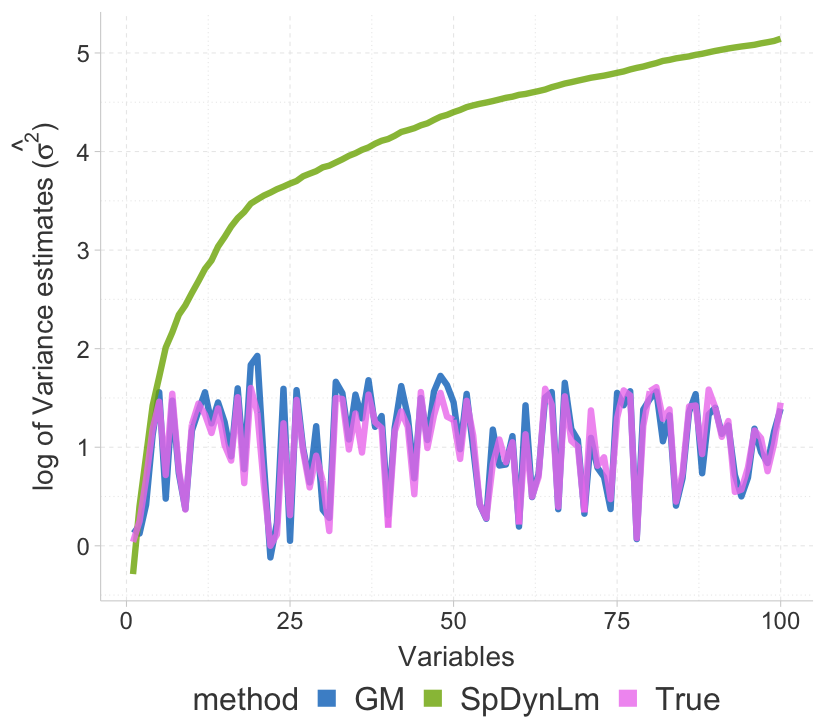}\label{fig:log-sigma-set3}}
		\subfloat[Median RMSPE over seeds for each variable (time-point).]{\includegraphics[scale=0.25]{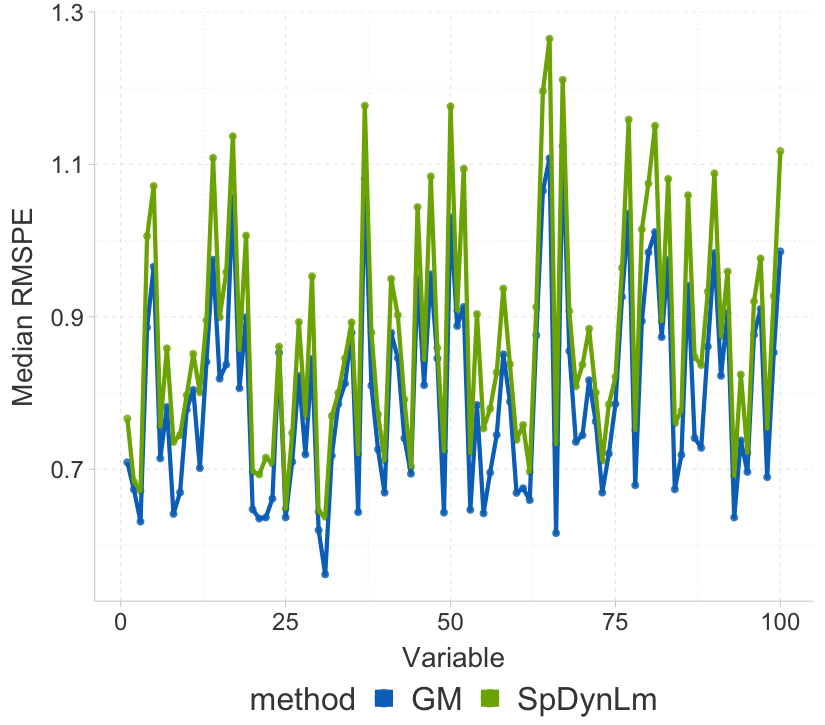}\label{fig:pred-set3}}
		%\hspace*{-1mm}\subfloat[3B]{\includegraphics[scale=0.25]{Figures/Set3/MM/Prediction_set3.png}\label{fig:predset3gm}}
		%\hspace*{-1mm}\subfloat[]{\includegraphics[scale=0.25]{Figures/Set 2/GM/phisigma_set2.png}\label{fig:phisigset2gm}}
		%\hspace*{-1mm}\subfloat[]{\includegraphics[scale=0.25]{Figures/Set 2/GM/Correlation2_set2.png}\label{fig:corset2gm}}
		%\hskip -1mm\subfloat[]{\includegraphics[scale=0.25]{Figures/Set 2/MM/Prediction_set2.png}\label{fig:predset2mm}}
		%\hskip -1mm\subfloat[]{\includegraphics[scale=0.25]{Figures/Set 2/GM/Prediction_set2.png}\label{fig:predset2gm}}
		\caption{Comparison between GGP and SpDynLm for modelling AR(1) spatial time series: (a) Variance estimates (in log-scale) for GM and SpDynLm compared to the true values for Set~3A. (b) Median RMSPE over seeds for each variable (time) for GM and SpDynLm for Set~3A.}\label{fig:ggpvsdlm}
	\end{figure} 
	
	%\begin{figure}[h]
	%    \centering
	%\includegraphics[scale=0.25]{Figures/Set3/GM/sigma_set3_log.png}
	%\hspace*{-1mm}\subfloat[3B]{\includegraphics[scale=0.25]{Figures/Set3/MM/Prediction_set3.png}\label{fig:predset3gm}}
	%\hspace*{-1mm}\subfloat[]{\includegraphics[scale=0.25]{Figures/Set 2/GM/phisigma_set2.png}\label{fig:phisigset2gm}}
	%\hspace*{-1mm}\subfloat[]{\includegraphics[scale=0.25]{Figures/Set 2/GM/Correlation2_set2.png}\label{fig:corset2gm}}
	%\hskip -1mm\subfloat[]{\includegraphics[scale=0.25]{Figures/Set 2/MM/Prediction_set2.png}\label{fig:predset2mm}}
	%\hskip -1mm\subfloat[]{\includegraphics[scale=0.25]{Figures/Set 2/GM/Prediction_set2.png}\label{fig:predset2gm}}
	%\caption{Log of variance estimates for BGML, BGMR and SpDynLm comapred to the true values for Set~3A.}% and Set~3B} 
	%    \label{fig:log-sigma-set3}
	%\end{figure}
	
	For Set~3A, the estimation accuracy of GM has already been demonstrated in Figure~\ref{fig:marg}(e) (for the marginal parameters) and in Figure 
	\ref{fig:supcross}(c) (for the cross-covariance parameters representing the auto-regression). In particular, Figure \ref{fig:supcross}(c) demonstrates the capability of GGP to successfully estimate non-stationary (time-specific) cross-covariance parameters. The competing model SpDynLm does not possess an autocovariance parameter that can been compared with these cross-covariances. However, we compare estimates of the marginal process variances from GGP and the SpDynLm function with the truth in Figure \ref{fig:log-sigma-set3}. We observe that while GGP accurately captures the marginal variances of the processes $w_t(\cdot)$ for each time $t$, the estimates from SpDynLm are monotonically increasing with time and far exceeding the true values. This demonstrates the detrimental implications of the model in \ref{eq:dlm} leading to variances exploding with time and, hence, prohibiting any meaningful insight regarding the underlying processes from these parameter estimates. 
	
	We also compare the models based on their predictive performance on hold-out data. We use the implementation of SpDynLm in the SpBayes R-package \citep{finley2013spbayes}. Figure \ref{fig:pred-set3} plots the median RMSPE for each variable (time-point). We see that SpDynLm produces higher predictive error (RMSPE=$0.868$) than GM (RSMPE=$0.8$) %and BGMR (RMSPE=$0.801$) 
	%produce similar RMSPE while 
	for most time-points. The numbers reported in parentheses are averaged across variables.
	
	Overall, while predictive performance between GGP and DLM is competitive, the GGP is  flexible and interpretable allowing estimation of spatial properties of the latent process $w_t(\cdot)$ for each time. The current implementation of SpDynLM imposes unnecessary constraints of monotonically increasing latent process variance with time leading to meaningless estimates of these parameters. More importantly, DLM assumes common smoothness over time, thereby offering no avenue to quantitatively study  smoothness of the process at each time which can reveal important scientific phenomenon, e.g., pollutant surfaces can be smooth on days where the pollutant is driven by regional sources, but will be much less smooth with high local variations on days where there are significant local sources of emission.}

\subsection{Comparison between different implementations of GGP}\label{sec:variants}

\begin{figure}[h]
	\begin{center}
		\hspace*{-1mm}\subfloat[Set 1B (Marginal parameters)]{\includegraphics[scale=0.215]{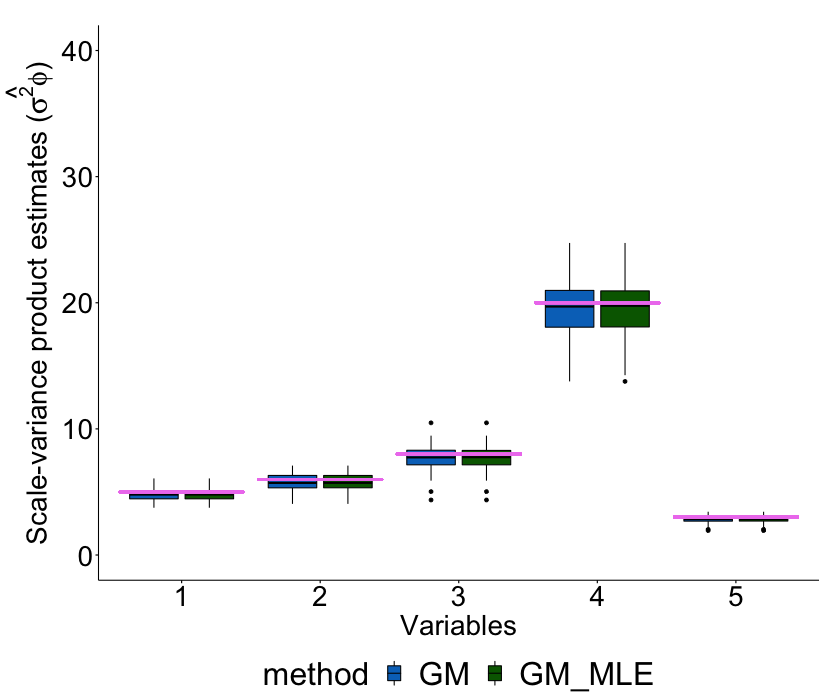}}
		\hspace*{-1mm}\subfloat[Set 1B (Cross-covariance parameters)]{\includegraphics[scale=0.215]{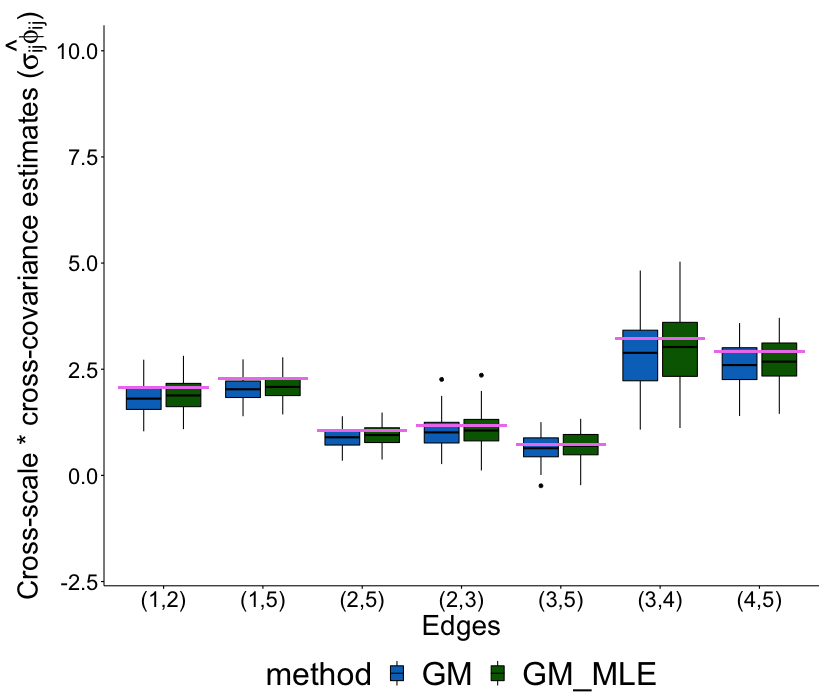}}
		%\hspace*{-1mm}\subfloat[3A]{\includegraphics[scale=0.25]{Figures/Set3/GM/Prediction_set3.png}\label{fig:predset3mm}}
		\caption{Comparison of performance of Graphical Mat\'ern (GM) and Graphical Mat\'ern frequentist (${\text{GM}}_{\text{MLE}}$) : (a) Estimates of the scale-covariance product parameters $\sigma_{ii}\phi_{ii}= i \in \calV$, (b) Estimates of the cross-covariance parameters $\sigma_{ij}\phi_{ij}=\Gamma(1/2)b_{ij}$, $(i,j) \in E_\calV$ for Set~1B. The horizontal pink lines in Figures (a) and (b) indicate true parameter values.} %Sub-figure (d): Median RMSPE across seeds for BGML, BGMR and SpDynLm for Set~3A.} %violet line to indicate truth. Three plots (a) Product of marginal scale and variance parameter estimates, (b) product of cross-scale and cross-covariance parameter estimates}
		\label{fig:mle}
	\end{center}
\end{figure}

\black{We have implemented 3 different variants of the GGP model. Besides the main focus on Bayesian model with GGP (GM) on the latent spatial processes (implementations details in Section \ref{sec: bayes-model}), we have also discussed GGP on the response process (GM$_{response}$) in Section \ref{subsec:bgmr} (implementation details in Section \ref{sec:gibbsmarg}), and have presented a frequentist estimation scheme for the parameters with maximum likelihood estimation (GM$_{MLE}$) using co-ordinate descent (see Section \ref{sec:freq}). In this Section we compare the performances of the 3 variants of GGP. 
	
	The MLE-based methods preclude misalignment among data locations for the different variables and excludes nugget processes $\eps_i(\cdot)$ in (\ref{eqn:mgp}). Set~1B conforms to such assumptions. Hence, we compare GM with GM$_{MLE}$ for this set. Since there is no nugget, GM and GM$_{response}$ are the same for this set. Figure \ref{fig:mle} plots the true covariance and cross-covariance parameters and their estimates from GM and GM$_{MLE}$ showing that the Bayesian and frequentist implementations yield similar estimates. } %which are similar for both the correctly specified setting 1A and the misspecified setting 1B. 

\begin{figure}
	\begin{center}
		\hspace*{-1mm}\subfloat[Set 2B (Marginal parameters)]{\includegraphics[scale=0.215]{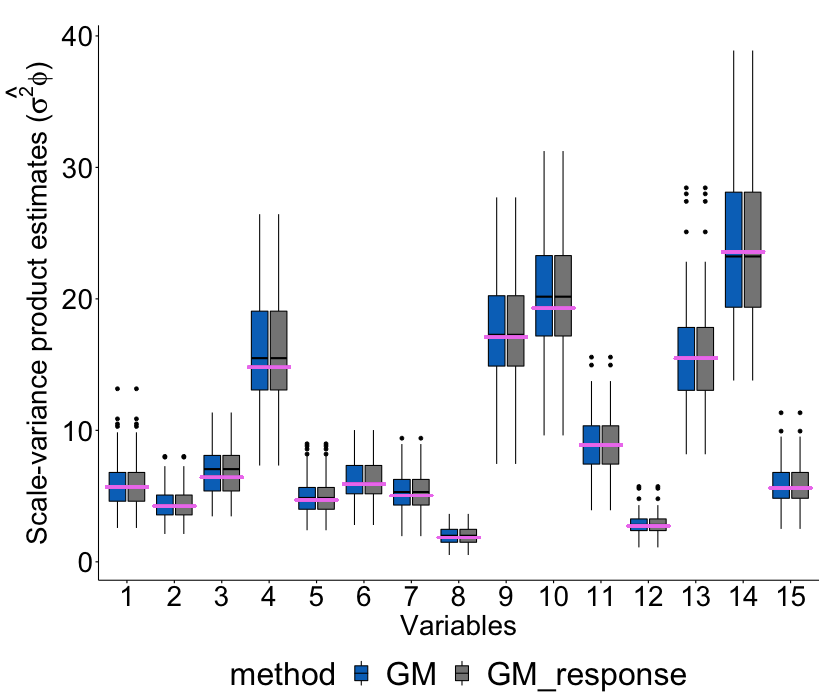}}
		\hspace*{-1mm}\subfloat[Set 2B (Cross-covariance parameters)]{\includegraphics[scale=0.215]{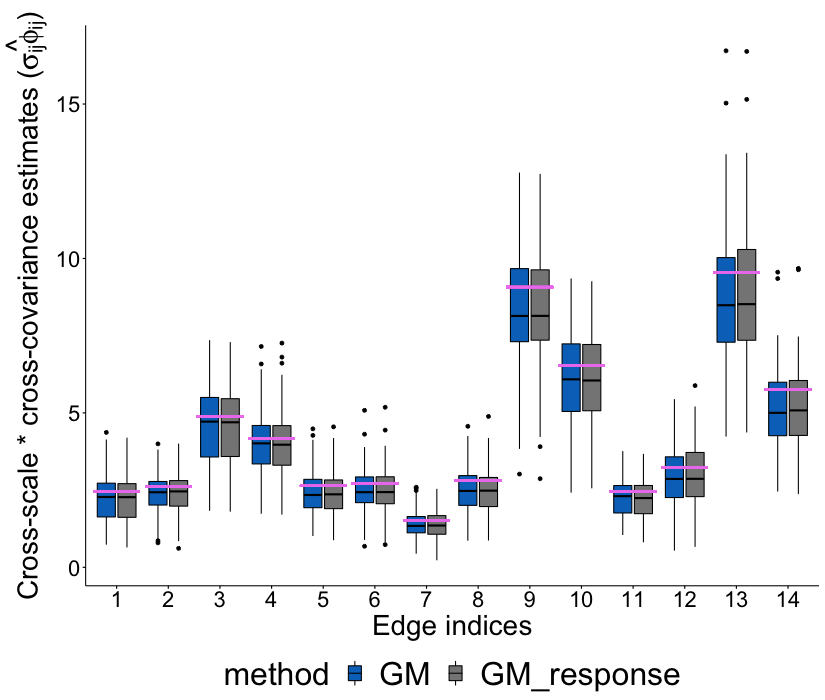}}\\
		\hspace*{-1mm}\subfloat[Set 2B (Predictions)]{\includegraphics[scale=0.215]{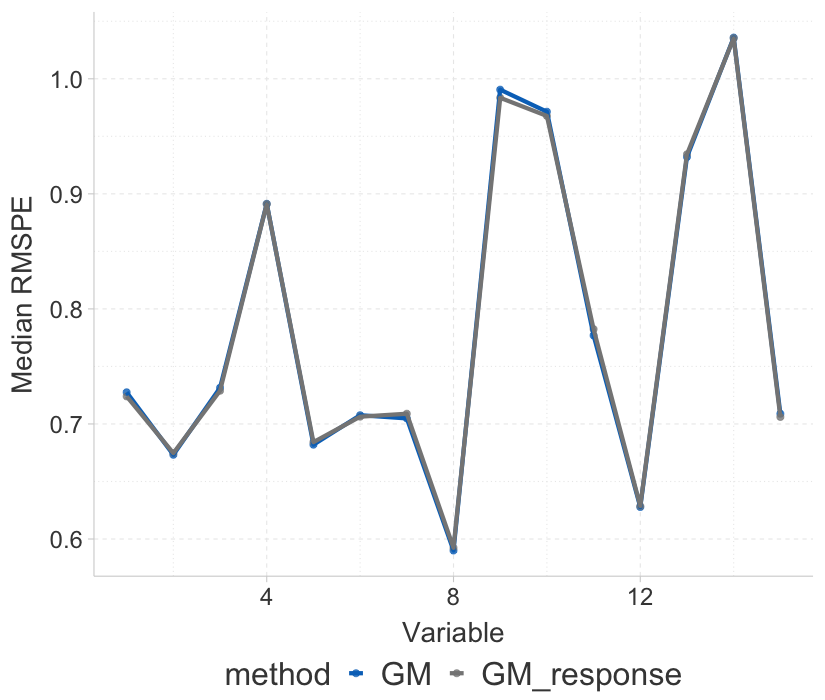}}
		%\hspace*{-1mm}\subfloat[3A]{\includegraphics[scale=0.25]{Figures/Set3/GM/Prediction_set3.png}\label{fig:predset3mm}}
		\caption{Comparison of performance of Graphical Mat\'ern (GM) and Graphical Mat\'ern response (${\text{GM}}_{\text{response}}$) : (a) Estimates of the scale-covariance product parameters $\sigma_{ii}\phi_{ii}, i \in \calV$, (b) Estimates of the cross-covariance parameters $\sigma_{ij}\phi_{ij}=\Gamma(1/2)b_{ij}$, $(i,j) \in E_\calV$ and (c) median RMSPE for Set~2B. The horizontal pink lines in Figures (a) and (b) indicate true parameter values.} %Sub-figure (d): Median RMSPE across seeds for BGML, BGMR and SpDynLm for Set~3A.} %violet line to indicate truth. Three plots (a) Product of marginal scale and variance parameter estimates, (b) product of cross-scale and cross-covariance parameter estimates}
		\label{fig:response}
	\end{center}
\end{figure}

\black{We then compare the two Bayesian implementations of GGP: GGP on the latent process (GM); and GGP on the response process (GM$_{response}$). Figure \ref{fig:response} plots the estimates of the covariance and cross-covariance parameters, and prediction RMSPE based on hold-out data for the two variants for Set 2B. We see that they produce similar estimates and predictive performance.}

%The BGML (average $RMSE = 0.78$) and BGMR (average $RMSE = 0.78$) performs better than SpDynLm (average $RMSE = 0.84$) in Set~$3B$, where all the candidate models are misspecified (Figure \ref{fig:pred-set3}).

%, which can be implemented using \textbf{SpBayes} \citep{finley2013spbayes} package in \textsf{R}. 

\section{Additional figures and tables}

\begin{figure}[h!]
	\centering
	\hspace*{-1mm}\subfloat[DAG for a univariate AR(2) model]{\includegraphics[scale=0.35]{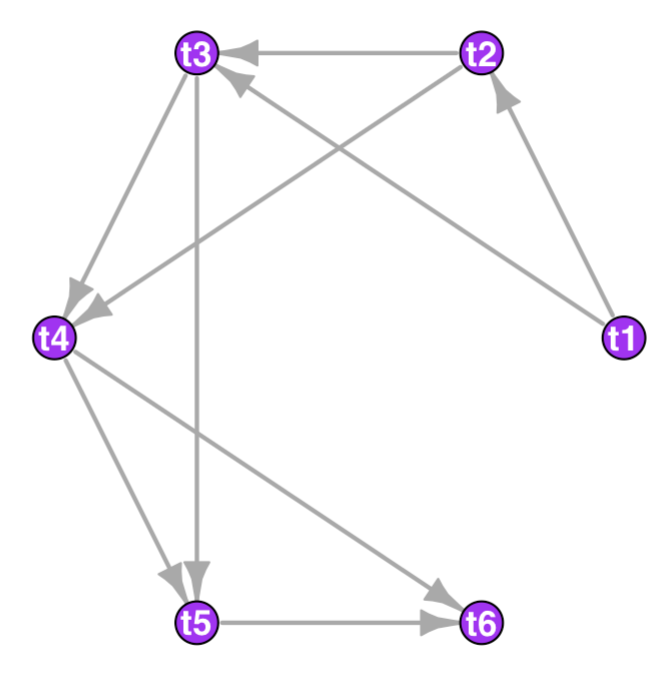}\label{fig:ardag}}
	\hspace*{10mm}\subfloat[Moralized $\calG_\calT$ for a univariate AR(2) model]{\includegraphics[scale=0.39]{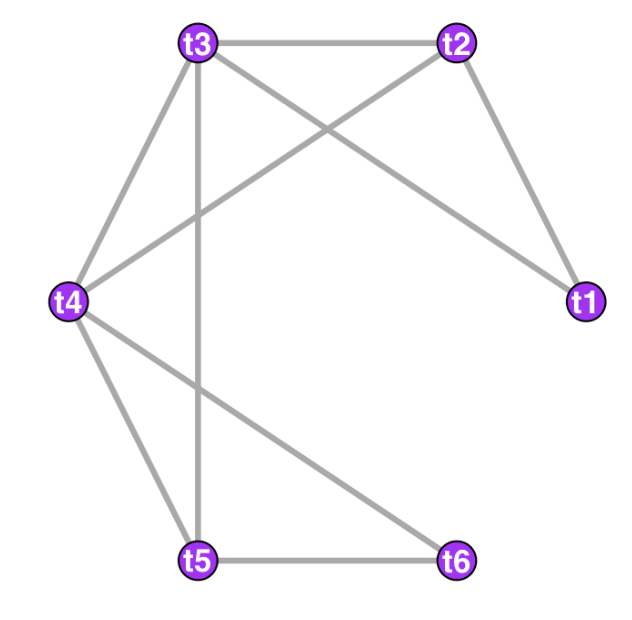}\label{fig:ar}}\\
	\hspace*{-1mm}\subfloat[DAG for the graphical VAR model example of Section \ref{sec:time} ]{\includegraphics[scale=0.28]{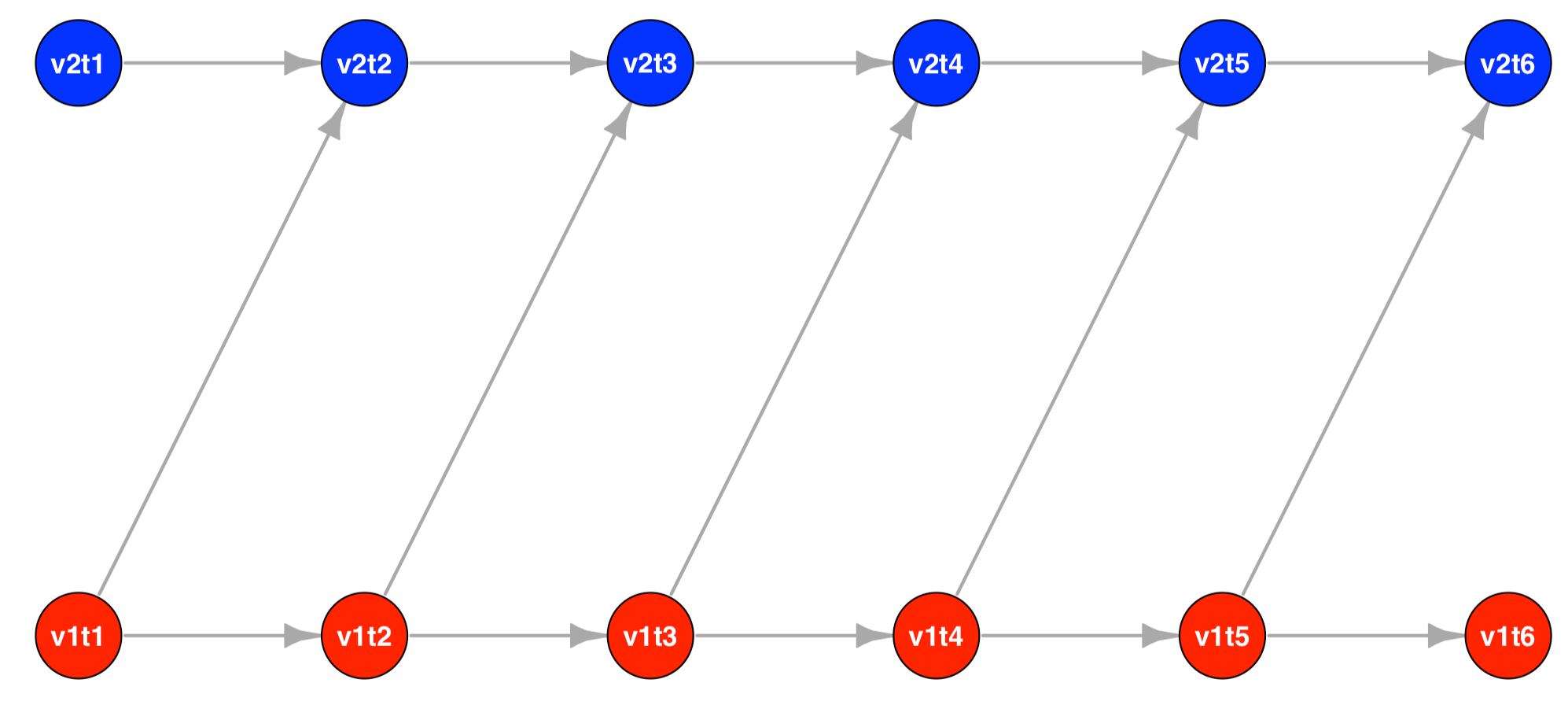}\label{fig:vardag}}\\
	\hspace*{-1mm}\subfloat[Moralized $\calG_{\calV\times\calT}$ for the graphical VAR model of Figure (c)]{\includegraphics[scale=0.3]{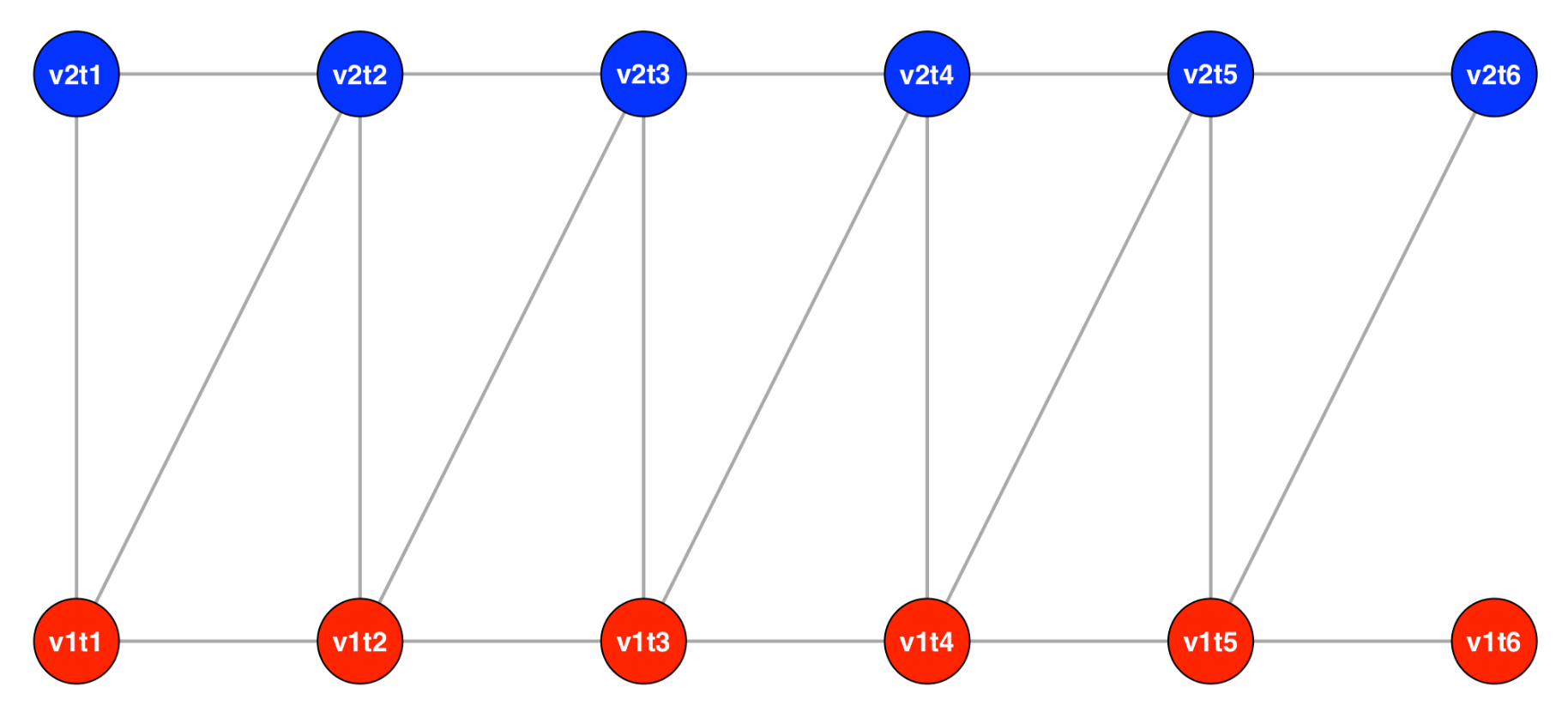}\label{fig:var}}
	\caption{Graphical models for autoregressive spatial time-series.}
	\label{fig:timeseries}
\end{figure}

\begin{figure}
	\begin{center}
		\hspace*{-1mm}\subfloat[Set 1A]{\includegraphics[scale=0.215]{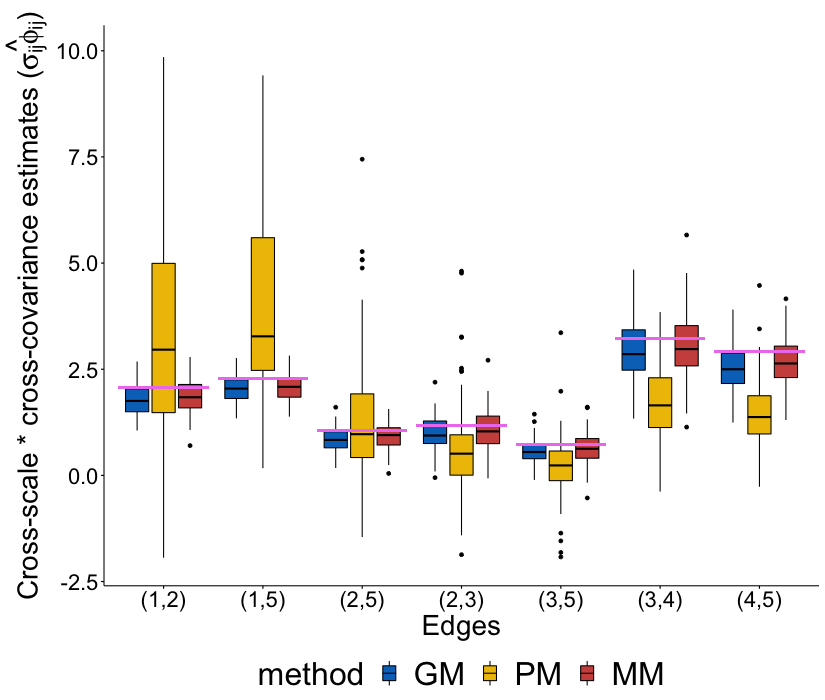}\label{fig:corset1A}}
		\hspace*{-1mm}\subfloat[Set 2A]{\includegraphics[scale=0.215]{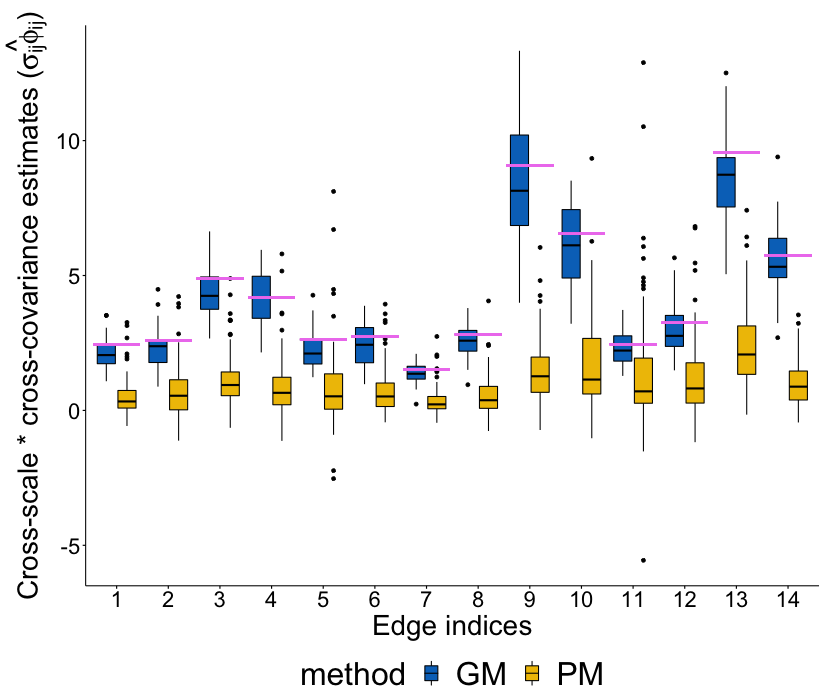}\label{fig:corset2A}}\\
		\hspace*{-1mm}\subfloat[Set 3A]{\includegraphics[scale=0.25]{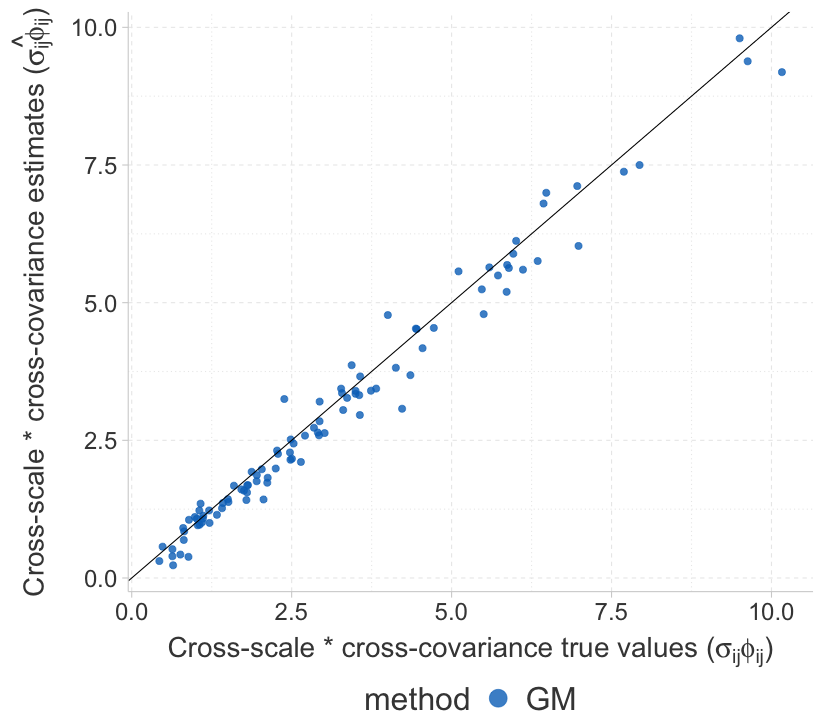}\label{fig:corset3mm}}
		%\hspace*{-1mm}\subfloat[3A]{\includegraphics[scale=0.25]{Figures/Set3/GM/Prediction_set3.png}\label{fig:predset3mm}}
		\caption{Estimation performance of graphical Mat\'ern in the correctly specified case: (a), (b) and (c): Estimates of the cross-covariance parameters $\sigma_{ij}\phi_{ij}=\Gamma(1/2)b_{ij}$, $(i,j) \in E_\calV$ for the 3 simulation sets (1A, 2A and 3A) where the graphical Mat\'ern is correctly specified. The horizontal pink lines in Figures (a) and (b) indicate true parameter values.} %Sub-figure (d): Median RMSPE across seeds for BGML, BGMR and SpDynLm for Set~3A.} %violet line to indicate truth. Three plots (a) Product of marginal scale and variance parameter estimates, (b) product of cross-scale and cross-covariance parameter estimates}
		\label{fig:supcross}
	\end{center}
\end{figure} 

\begin{figure}
	\begin{center}
		\subfloat[Posterior edge selection probabilities for Set 1A. ]{\includegraphics[scale=0.22,trim={0cm 0cm 0cm 1.2cm},clip]{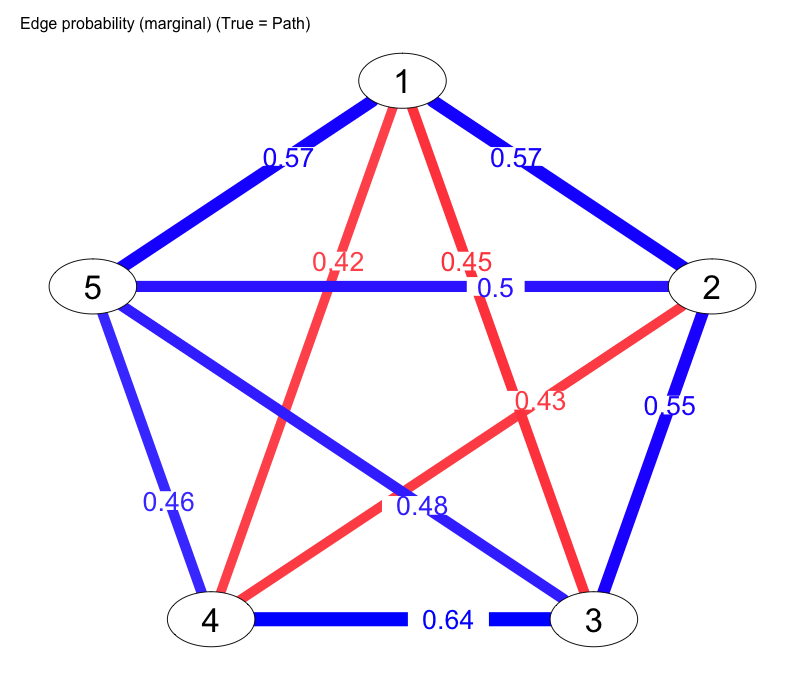}}\hspace{0 cm}
		\hspace*{-1mm}\subfloat[Cross-covariance parameter estimates for Set 1A while estimating the unknown graph]{\includegraphics[scale=0.23,trim={0 0 0 0},clip]{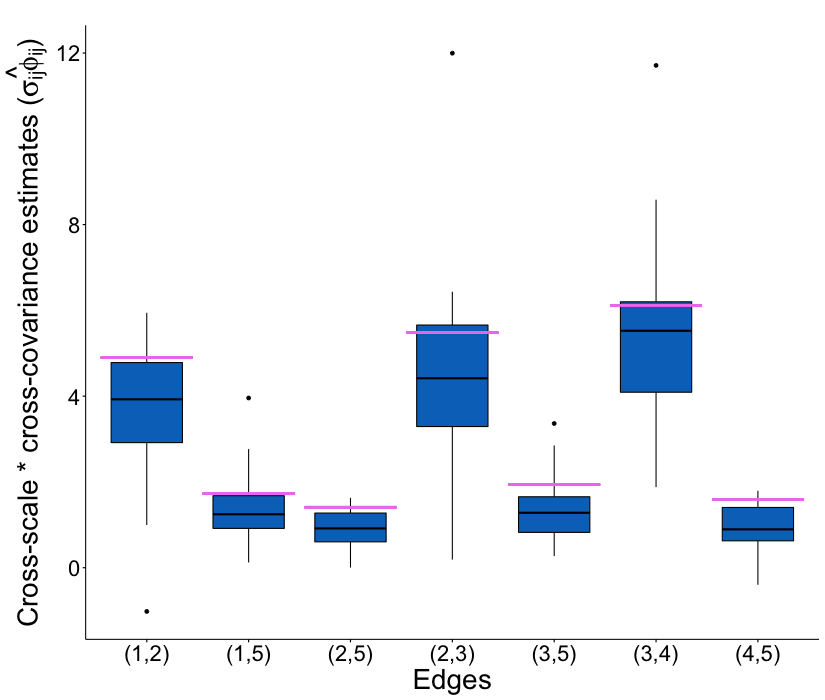}}
	\end{center}
	\caption{Performance of GGP with unknown graph for Set 1A: (a): Marginal edge probabilities estimated from the reversible jump MCMC sampler. Blue edges denote the true edges and red denotes the non-existent edges. Edges are weighted proportional to the estimated posterior selection probabilities. (b) GM estimates of cross-correlation parameters ($b_{ij}$) corresponding to true edges when the graph is unknown.}
	\label{fig: est-graph1A}
\end{figure}

\begin{table}[!h]
	\centering
	\begin{tabular}{|rr|rr|}
		\hline
		\multicolumn{2}{|r|}{Set 1A (True = Gem graph)} & \multicolumn{2}{|r|}{Set 2A (True = Path graph)}\\
		\hline
		Edges & Probability & Edges & Probability \\ 
		\hline
		\textbf{(3, 4)} & \textbf{0.64} & \textbf{(1, 2)} & \textbf{0.43} \\ 
		\textbf{(1, 2)} & \textbf{0.57} & \textbf{(11, 12)} & \textbf{0.42} \\ 
		\textbf{(1, 5)} & \textbf{0.57} & \textbf{(2, 3)} & \textbf{0.42} \\ 
		\textbf{(2, 3)} & \textbf{0.55} & \textbf{(14, 15)} & \textbf{0.41} \\ 
		\textbf{(2, 5)} & \textbf{0.50} & \textbf{(4, 5)} & \textbf{0.40} \\ 
		\textbf{(3, 5)} & \textbf{0.48} & \textbf{(7, 8)} & \textbf{0.38} \\ 
		\textbf{(4, 5)} & \textbf{0.46} & \textbf{(8, 9)} & \textbf{0.37} \\ 
		(1, 3) & 0.45 & \textbf{(13, 14)} & \textbf{0.34} \\ 
		(2, 4) & 0.43 & \textbf{(6, 7)} & \textbf{0.33} \\ 
		(1, 4) & 0.42 & \textbf{(10, 11)} & \textbf{0.32} \\ 
		& & \textbf{(9, 10)} & \textbf{0.31}\\ 
		&  & \textbf{(5, 6)} & \textbf{0.30} \\ 
		& & \textbf{(12, 13)} & \textbf{0.30} \\ 
		&  & \textbf{(3, 4)} & \textbf{0.27} \\ 
		&  & (1, 3) & 0.22 \\ 
		&  & (13, 15) & 0.21 \\ 
		& & (9, 11) & 0.20 \\ 
		&  & (7, 9) & 0.18 \\ 
		&  & (4, 6) & 0.18 \\ 
		&  & (10, 12) & 0.18 \\ 
		\hline
	\end{tabular}
	\caption{Posterior probabilities of including an edge when estimating the graph in a GGP. The rows of the table are ordered from highest to lowest. (a) Set 1A (all edges), (b) Set 2A (edges with the top 20 highest selection probabilities). Bold numbers indicate true edges.}
	\label{tab:edge-prob}
\end{table}

%\pagebreak

\begin{figure}[h]
	\centering
	\includegraphics[scale=0.3, angle=0, trim={0 0 0 0},clip]{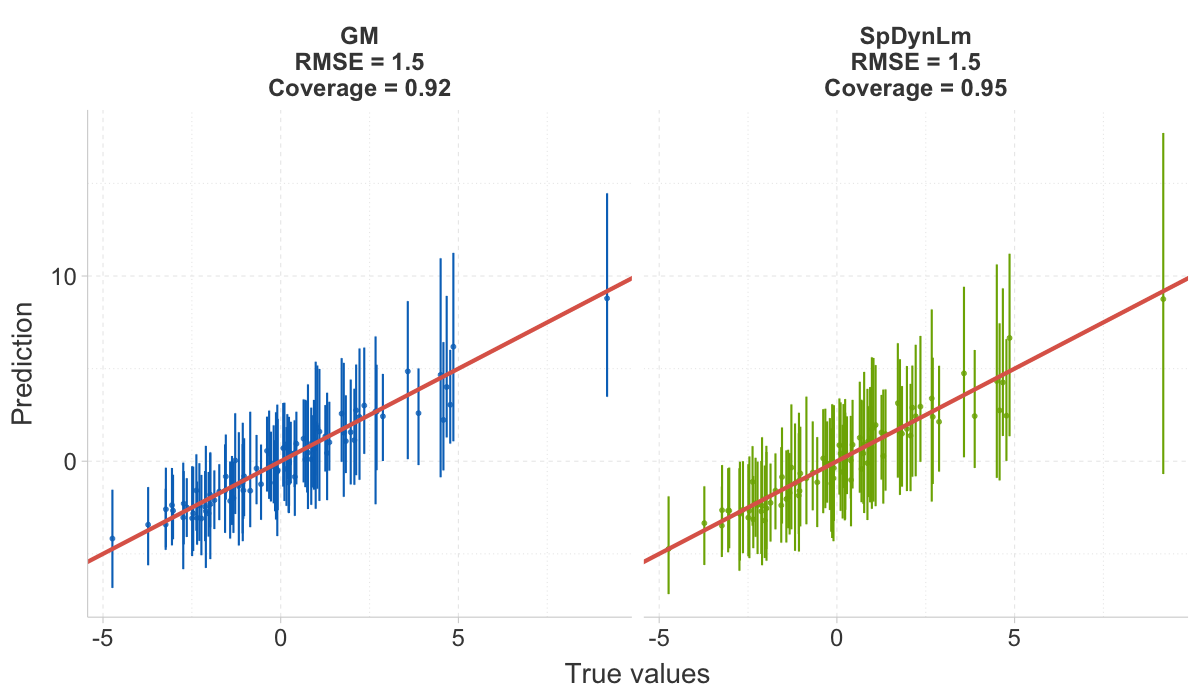}
	\caption{Truth vs prediction for test set data compared among GM and SpDynLM} 
	\label{fig:pred-data-ci}
\end{figure} 

\begin{figure}[h]
	\centering
	\includegraphics[scale=0.3, angle=0, trim={0 0 0 0},clip]{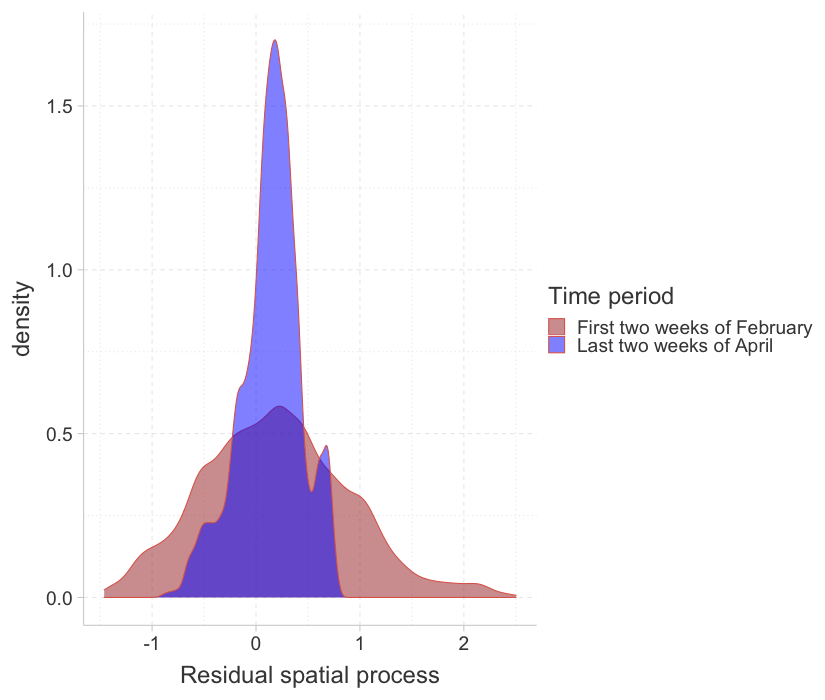}
	\caption{Density of residual spatial process values (across locations) for two different time periods - first two weeks of February and last two weeks of April} 
	\label{fig:pred-resid}
\end{figure}

\end{document}